%% file: main.tex
\documentclass[a4paper, fleqn]{cas-dc}
\usepackage[authoryear]{natbib}
\usepackage{graphicx}
\usepackage{tabularx}
\usepackage{multirow}
\usepackage{subfig}
\usepackage{epstopdf}
\usepackage{url}
\usepackage{amsmath}
\usepackage[dvipsnames]{xcolor}
\usepackage[flushleft]{threeparttable}
\usepackage{hyperref}
\usepackage{soul}
\usepackage{verbatim}
\usepackage{longtable}
\usepackage{pdflscape}
\usepackage{lscape}

\begin{document}

\let\WriteBookmarks\relax
\def\floatpaghttps://www.overleaf.com/projectepagefraction{1}
\def\textpagefraction{.001}

\def\tsc#1{\csdef{#1}{\textsc{\lowercase{#1}}\xspace}}
\newcommand{\referee}[1]{{\color{red}#1}} 
\newcommand{\mioblu}[1]{{\color{blue}#1}} 
\newcommand{\mioverde}[1]{{\color{green}#1}} 
\newcommand{\FIXBUG}[1]{{\em Check:  }{\color{red}{#1}}}
\newcommand{\FIXME}[1]{{\em Comment:  }{\color{blue}{#1}}}
\newcommand\gio[1]{{\bf \color{blue} #1}}
\newcommand\gionote[1]{{\bf \color{red} [GM:\,#1]}}

\def \astrima {ASTRI Mini-Array}

\newcommand{\hess}{H.E.S.S.}
\newcommand{\fermi}{\textit{Fermi}}
\newcommand{\astrisim}{\texttt{ASTRIsim}}
\newcommand{\chandra}{\textit{Chandra}}
\newcommand{\obsteide}{\textit{Observatorio del Teide}}

\defcitealias{scuderi21}{Paper~I}
\defcitealias{vercellone21}{Paper~II}
\defcitealias{saturni21}{Paper~IV}

\def \degmark{$^\circ$}
\def \ergsec{\hbox{erg s$^{-1}$}}
\def \ergcmsec{\hbox{erg cm$^{-2}$ s$^{-1}$}}
\def \phcmsec{\hbox{photons cm$^{-2}$ s$^{-1}$}}
\def \phcmsectev{\hbox{photons cm$^{-2}$ s$^{-1}$ TeV$^{-1}$}}
\def \ferg {erg cm$^{-2}$ s$^{-1}$}
\def \arcmin {\hbox{$^\prime$}}
\def \arcsec {\hbox{$^{\prime\prime}$}}
\def \chisq {$\chi ^{2}$}
\def \rchisq {$\chi_{\nu} ^{2}$}
\def \gray {$\gamma$-ray}
\def \grays {$\gamma$-rays}
\def \gammapy {\textsc{Gammapy}}
\def \ctools {\textsc{ctools}}
\def \xspec {\textsc{xspec}}
\def \heasoft {\textsc{heasoft}}
\def \aip {AIP Conf. Ser.}
\def \msun {$M_{\odot}$}
\def \ATel {ATel} 
\def \aapr {A\&A Rev.}
\def \aip {AIP Conf. Ser.}
\def \ATel {ATel}
\def \aj {The Astronimical Journal}
\def \apj {ApJ}
\def \apjl {ApJL}
\def \apjs {ApJS}
\def \aap {A\&A}
\def \aapl {A\&AL}
\def \aara {A\&A Rev.}
\def \araa {ARA\&A }
\def \asr {AdSpR}
\def \baas {BAAS}
\def \bcao {BCAO}  
\def \gcn {GCN Circ.} 
\def \jcap {JCAP}
\def \iaucirc {IAU Circ.}
\def \mnras {MNRAS}
\def \nat {Nature}
\def \pasj {PASJ}
\def \prd {PRD}
\def \prl {PRL}
\def \ssr {SSRv}
\def \spie {SPIE Conf. Ser.}

\setlength{\tabcolsep}{5pt}
\renewcommand{\arraystretch}{1.2}

\def \tev2032 {TeV~J2032\,+\,4130}
\def \lat {\textit{Fermi}-LAT}

\let\WriteBookmarks\relax
\def\floatpagepagefraction{1}
\def\textpagefraction{.001}

\shorttitle{Galactic Observatory Science with \astrima}
\shortauthors{A. D'Aì et~al.}
\title [mode = title]{Galactic Observatory Science with the \astrima\ at the \obsteide}
\author[1]{A. D'Aì}[type=editor, orcid=0000-0002-5042-1036]
\cormark[1]
\ead{antonino.dai@inaf.it}
\author[2]{E. Amato}[orcid=0000-0002-9881-8112]  
\author[2]{A. Burtovoi}[orcid=0000-0002-8734-808X]  
\author[1]{A.~A. Compagnino}[orcid=0000-0003-4727-9136] 
\author[3]{M. Fiori}[orcid=0000-0002-7352-6818] 
\author[4]{A. Giuliani}[orcid=0000-0002-4315-1699] 
\author[4]{N. {La Palombara}}[orcid=0000-0001-7015-6359] 
\author[4]{A. Paizis}[orcid=0000-0001-5067-0377] 
\author[5]{G. Piano}[orcid=0000-0002-9332-5319] 
\author[6,7]{F.~G. Saturni}[orcid=0000-0002-1946-7706] 
\author[1,8]{A. Tutone}[orcid=0000-0002-2840-0001] 
\author[4]{A. Belfiore}[orcid=0000-0002-2526-1309]  
\author[5]{M. Cardillo}[orcid=0000-0001-8877-3996] 
\author[4]{S. Crestan}[orcid=0000-0002-8368-0616] 
\author[1]{G. Cusumano}[orcid=0000-0002-8151-1990] 
\author[9,10]{M. {Della Valle}}[orcid=0000-0003-3142-5020] 
\author[1]{M. {Del Santo}}[orcid=0000-0002-1793-1050] 
\author[1]{A. {La Barbera}}[orcid=0000-0002-5880-8913] 
\author[1]{V. {La Parola}}[orcid=0000-0002-8087-6488] 
\author[6,7]{S. Lombardi}[orcid=0000-0002-6336-865X] 
\author[4]{S. Mereghetti}[orcid=0000-0003-3259-7801] 
\author[2]{G. Morlino}[orcid=0000-0002-5014-4817] 
\author[1]{F. Pintore}[orcid=0000-0002-3869-2925] 
\author[11]{P. Romano}[orcid=0000-0003-0258-7469] 
\author[11]{S. Vercellone}[orcid=0000-0003-1163-1396] 
\author[6]{A. Antonelli}[orcid=0000-0002-5037-9034]  
\author[12]{C. Arcaro}[orcid=0000-0002-1998-9707]  
\author[6,7]{C. Bigongiari}[orcid=0000-0003-3293-8522] 
\author[13]{M. B\"oettcher}[orcid=0000-0002-8434-5692] 
\author[14]{P. Bruno}[orcid=0000-0003-3919-9611] 
\author[15]{A. Bulgarelli}[orcid=0000-0001-6347-0649] 
\author[15]{V. Conforti}[orcid=0000-0002-0007-3520]  
\author[14]{A. Costa}[orcid=0000-0003-0344-8911] 
\author[16]{E. {de Gouveia Dal Pino}}[orcid=0000-0001-8058-4752]
\author[15]{V. Fioretti}[orcid=0000-0002-6082-5384] 
\author[17]{S. Germani}[orcid=0000-0002-2233-6811] 
\author[18]{A. Ghedina}[orcid=0000-0003-4702-5152] 
\author[15]{F. Gianotti}[orcid=0000-0003-4666-119X] 
\author[14]{V. {Giordano}}[orcid=0000-0001-8865-5930] 
\author[14]{F. {Incardona}}[orcid=0000-0002-2568-0917] 
\author[14]{G. {Leto}}[orcid=0000-0002-0040-5011] 
\author[19,20]{F. Longo}[orcid=0000-0003-2501-2270] 
\author[21]{A. {L\'opez Oramas}}[orcid=0000-0003-4603-1884] 
\author[6,7]{F. Lucarelli}[orcid=0000-0002-6311-764X] 
\author[22]{B. Olmi}[orcid=0000-0001-6022-8216 ] 
\author[1]{A. Pagliaro}[orcid=0000-0002-6841-1362] 
\author[15]{N. Parmiggiani}[orcid=0000-0002-4535-5329] 
\author[14]{G. Romeo}[orcid=0000-0003-3239-6057] 
\author[6]{A. Stamerra}[orcid=0000-0002-9430-5264] 
\author[6]{V. Testa}[orcid=0000-0003-1033-1340] 
\author[15,17]{G. Tosti}[orcid= 0000-0002-0839-4126] 
\author[14]{G. Umana}[orcid=0000-0002-6972-8388] 
\author[5]{L. Zampieri}[orcid=0000-0002-6516-1329] 
\author[4]{P. Caraveo}[orcid=0000-0003-2478-8018] 
\author[11]{G. Pareschi}[orcid=0000-0003-3967-403X] 
\address[1]{INAF -- Istituto di Astrofisica Spaziale e Fisica Cosmica, via Ugo La Malfa 153, I-90123, Palermo, Italy} 
\address[2]{INAF -- Osservatorio Astrofisico di Arcetri, Largo E. Fermi 5, I-50125, Florence, Italy} 
\address[3]{INAF -- Osservatorio Astronomico di Padova, Vicolo dell'Osservatorio 5, I-35122, Padova, Italy} 
\address[4]{INAF -- Istituto di Astrofisica Spaziale e Fisica Cosmica, Via Alfonso Corti 12, I-20133, Milan, Italy} 
\address[5]{INAF -- Istituto di Astrofisica e Planetologia Spaziale, Via Fosso del Cavaliere 100, I-00133, Rome, Italy} 
\address[6]{INAF -- Osservatorio Astronomico di Roma, Via Frascati 33, Monte Porzio Catone, I-00040, Rome, Italy} 
\address[7]{ASI -- Space Science Data Center, Via del Politecnico s.n.c., I-00133, Rome, Italy} 
\address[8]{Università degli studi di Palermo, Dipartimento di Fisica e Chimica, Via Archirafi 36, I-90123, Palermo, Italy} 
\address[9]{INAF -- Osservatorio Astronomico di Capodimonte Astronomical, Salita Moiariello 16, I-80131, Naples, Italy}
\address[10]{Icranet -- Piazza della Repubblica 10, I-65122 , Pescara, Italy} 
\address[11]{INAF -- Osservatorio Astronomico di Brera, Via Emilio Bianchi 46, I-23807, Merate, Italy} 
\address[12]{Universit\`a degli Studi di Padova and INFN, I-35131, Padova, Italy}
\address[13]{Centre for Space Research, North-West University, Potchefstroom, South Africa} 
\address[14]{INAF -- Osservatorio Astronomico di Catania, Via Santa Sofia 78, I-95123, Catania, Italia} 
\address[15]{INAF -- Osservatorio di Astrofisica e Scienza dello Spazio di Bologna, via Gobetti 93/3, I-40129, Bologna, Italy} 
\address[16]{Univ. de S{\~a}o Paulo, Inst. de Astronomia, Geof{\'i}sica e Ci{\^e}ncias Atmosf{\'e}ricas, Cid. Universitaria, R. do Mat{\~a}o 1226, BR-05508-090, S{\~a}o Paulo (SP), Brazil}
\address[17]{Università degli studi Di Perugia, Via Alessandro Pascoli, I-06123, Perugia, Italy} 
\address[18]{INAF -- Fundaci{\'o}n Galileo Galilei, Rbla. J. A. Fern{\'a}ndez P{\'e}rez 7, ES-38712, San Antonio de Bre{\~n}a (TF), Spain}
\address[19]{Universit{\`a} degli Studi di Trieste, Dip. di Fisica, Via A. Valerio 2, I-34127, Trieste, Italy}
\address[20]{INFN -- Sezione di Trieste, Via A. Valerio 2, I-34127, Trieste, Italy}
\address[21]{Instituto de Astrof{\'i}sica de Canarias, C/ V{\'i}a L{\'a}ctea s/n, E-38205, La Laguna (Tenerife), Spain}
\address[22]{INAF -- Osservatorio Astronomico di Palermo, P.zza del Parlamento 1, I-90134, Palermo, Italy}

\cortext[cor1]{Corresponding author}
\begin{abstract}
The ASTRI (Astrofisica con Specchi a Tecnologia Replicante Italiana) Mini-Array will be composed of nine imaging atmospheric Cherenkov telescopes at the \obsteide\ site. The array will be best suited for astrophysical observations in the 0.3--200 TeV range with an angular resolution of few arc-minutes and an energy resolution of 10-15\%. A core-science programme in the first four years will be devoted to a limited number of key targets, addressing the most important open scientific questions in the very-high energy domain. At the same time, thanks to a wide field of view of about 10\degmark, \astrima\ will observe many additional field sources, which will constitute the basis for the long-term observatory programme that will eventually cover all the accessible sky. In this paper, we review different astrophysical Galactic environments, e.g. pulsar wind nebulae, supernova remnants, and gamma-ray binaries, and show the results from a set of \astrima\ simulations of some of these field sources made to highlight the expected performance of the array (even at large offset angles) and the important additional observatory science that will complement the core-science program.
\end{abstract}
\begin{keywords}
Telescopes \sep \grays: general \sep $\gamma - $rays: stars \sep  \sep © 2022. This manuscript version is made available under the CC-BY-NC-ND 4.0 license \href{https://creativecommons.org/licenses/by-nc-nd/4.0/}{https://creativecommons.org/licenses/by-nc-nd/4.0/}
\end{keywords}
\maketitle
\tableofcontents
\input{sect_intro}
\input{sect_cygnusregion}
\input{sect_snr}
\input{sect_ic443}
\input{sect_pwn}
\input{sect_j1813}
\input{sect_j2032}
\input{sect_pulsars}
\input{sect_gammabinaries}
\input{sect_ss433}
\input{sect_ls5039}
\input{sect_dmgc}
\input{sect_terzan5}

\input{sect_conclusions}
\section*{Acknowledgments}
\noindent
This work was conducted in the context of the ASTRI Project. This work is supported by the Italian Ministry of Education, University, and Research (MIUR) with funds specifically assigned to the Italian National Institute of Astrophysics (INAF). We acknowledge support from the Brazilian Funding Agency FAPESP (Grant 2013/10559-5) and from the South African Department of Science and Technology through Funding A\-gree\-ment 0227/2014 for the South African Gamma-Ray Astronomy Programme. This work has been supported by H2020-ASTERICS, a project funded by the European Commission Framework Programme Horizon 2020 Research and Innovation action under grant agreement n. 653477. IAC is supported by the Spanish Ministry of Science and Innovation (MICIU).
The ASTRI project is becoming a reality thanks to Giovanni ``Nanni'' Bignami, Nicol\`{o} ``Nichi'' D'Amico two outstanding scientists who, in their capability of INAF Presidents,  provided continuous support and invaluable guidance. While Nanni was instrumental to start the ASTRI telescope, Nichi transformed it into the Mini Array in Tenerife. Now the project is being built owing to the unfaltering support of Marco Tavani, the current INAF President. Paolo Vettolani and Filippo Zerbi, the past and current INAF Science Directors, as well as Massimo Cappi, the Coordinator of the High Energy branch of INAF, have been also very supportive to our work. We are very grateful to all of them. Nanni and Nichi, unfortunately, passed away but their vision is still guiding us. 

This research made use of \ctools\ \citep{knoedlseder16}, a community-developed analysis package for Imaging Air Cherenkov Telescope data. \ctools\ is based on GammaLib, a community-developed  toolbox for the scientific analysis of astronomical \gray\ data \citep{knoedlseder_gammalib, knoedlseder_gammalib_ctools}.

This research made use of \gammapy,\footnote{\url{https://www.gammapy.org}} a community-developed core Python package for TeV \gray\ astronomy \citep{deil17, nigro19}.
\newpage

\section{Appendix}
This Appendix presents results for two bright PWNe, which were originally studied in the context of the ACDC science project \citep[see details in][]{pintore20}. They well illustrate scientific aspects of notable interest in the PWN field and the expected performance of the \astrima\  in this context: the Vela X PWN is a prototype of a bright extended nebula and we will show how the VHE morphology could be matched with template morphology maps obtained in the radio and  in the X-ray bands; the moderately bright and extended PWN HESS J1303-631 has an energy dependent morphology, 
which is indicative of the path the NS and its associated nebula has travelled since the SN explosion.

\input{sect_velax}
\input{sect_j1303}

\bibliographystyle{mnras}
\bibliography{biblio}

\end{document}

%% file: sect_intro.tex
\section{Introduction} \label{sect:intro}
The Italian National Institute for Astrophysics (INAF), together with international partners from South Africa and Brazil, in the last decade has led the construction of one class of the Small-Size Telescopes (SSTs) in the context of the Cherenkov Telescope Array (CTA) for its  Southern site: the  \textit{Astrofisica con Specchi a Tecnologia  Replicante Italiana}  (ASTRI) telescope \citep{Pareschi_2016}. An ASTRI prototype (\textit{ASTRI-Horn}) was built on Mt. Etna (Sicily) in 2014. The ASTRI collaboration adopted an end-to-end  approach that comprised all aspects from design, construction and management of the entire hardware and software system up to final scientific products.  
Following the successful validation of all the engineering aspects of the telescope and the successful matching of the expected scientific performance by the \textit{ASTRI-Horn} prototype, INAF has financed a larger project, that will lead to the construction of an array of nine SST ASTRI-like telescopes, implemented on the base of the prototype's camera and structural design: the \astrima\, that will be built at the \obsteide\footnote{\url{https://www.iac.es/en/observatorios-de-canarias/teide-observatory}}, located in the Canary island of Tenerife at 2390 m of altitude. 

After the calibration phase and the validation of the expected performances, in the first years, the array will be run as an experiment and the \astrima\ Collaboration has defined an ambitious observational plan focused on key scientific questions (\emph{Science Pillars}) anchored to a corresponding set of celestial objects to be observed extensively. Moreover, thanks to its large field of view (FoV) of $10^{\circ}$ in diameter and good spatial resolution, the \astrima\ will be able to observe large portions of the sky in a single observation, allowing deep monitoring of multiple targets at the same time and a great opportunity for catching transient and serendipitous events. 

The work presented in this paper has three companion papers that will present each different aspects of the project. \citet[][hereafter Paper I]{scuderi21} will detail the current status of the ASTRI Project, namely the \astrima\ technological solutions, the array lay-out and site characteristics. \citet[][hereafter Paper II]{vercellone21} will describe the \emph{Science Pillars} targets that will be observed during the first years of operation. The aim of this paper (paper III) is to present the potentially interesting very high energy (VHE) Galactic targets that might be observed together with the \emph{pillar} targets, because present in the same FoV, or  from a long-term  planning. \citet[][hereafter Paper IV]{saturni21} will similarly present the base of the scientific long-term planning for the extragalactic science. 

\subsection{Scientific simulation setups} \label{sect:setups}
The instrument response function (IRF) used to produce the simulated event list files, and the subsequent high-level scientific analyses presented in this work, was obtained from dedicated Monte Carlo (MC) simulations and MC reconstruction analysis which are fully detailed in Sect. 2.1 and 2.2 of \citetalias{vercellone21}.

Here, we briefly summarise the most important characteristics of this IRF (optimised for a 50 hours exposure) regarding energy and spatial resolution and sensitivity
\citepalias[see Sect. 2.3 of ][for a comprehensive discussion]{vercellone21}. For an on-axis source, in the 1--200 TeV band, the energy resolution is in the range 10-15\%; the angular resolution is $\sim$\,8\arcmin\ at 1 TeV, with a minimum value of $\sim$\,3\arcmin\ at 10 TeV, degrading very little up to 100 TeV. The differential sensitivity curve for 50 hours has its minimum value of 7\,$\times$\,10$^{-13}$ erg cm$^{-2}$ s$^{-1}$ between 5 and 8 TeV. The dependence of this performance on the offset angle is contained within a factor $\sim$\,1.5 degradation from the best nominal values up to 3\degmark; the degradation reaches a factor $\sim$\,2 for an offset angle of $\sim$\,5\degmark. 

In this paper we simulated the event list files associated with each source, or field, using the same simulation setup and tools detailed in \citetalias[see Sect. 2.4][]{vercellone21}; briefly, we used \ctools\ v.1.7.2  \citep{knoedlseder_gammalib_ctools} and the \gammapy\ package \citep[v.0.17,][]{deil17} to generate event lists, sky maps, spectra and to perform maximum likelihood fitting. For every source model, we have generally produced 100 independent realisations and then averaged the best-fitting results~\citep{2018MNRAS.481.5046R, 2020MNRAS.494..411R}. We did not take into account the energy dispersion of the IRF and any systematic error introduced at the IRF level. In the spectral fit, the background is treated using the IRF model. We did not take into account any model for the diffuse \gray\ emission. We will detail on data analysis of single observations in the \textit{Feasibility and Simulations} paragraph of each section. We also used \heasoft\ v.6.26\footnote{\url{https://heasarc.gsfc.nasa.gov/docs/software/lheasoft/}} and \textsc{SAOImage ds9}\footnote{\url{http://ds9.si.edu/site/Home.html}} for general data extraction and analysis of spectra and images.

Additionally, in the Appendix of this paper, we will also show results for two pulsar wind nebulae (Vela X and HESS J1303-631), which were originally simulated within the ASTRI Comprehensive Data Challenge (ACDC) project \citep{pintore20}. Spectra for these sources were simulated with an IRF corresponding to an array of nine ASTRI-like telescopes for the Paranal site. This IRF is sufficiently close (in terms  of effective area, angular and spectral resolution) to the Teide IRF (differences in sensitivity less than 20\% in the 2--100 TeV range), so that the analysis carried out on these two sources can be confidently used to illustrate the expected \astrima\ performance for similarly extended pulsar wind nebulae in the Northern Hemisphere.
\section{Overview of the Galaxy in the TeV band} \label{sect:intro_galaxy}

\begin{figure*}
	\centering
	\includegraphics[width=0.9\textwidth]{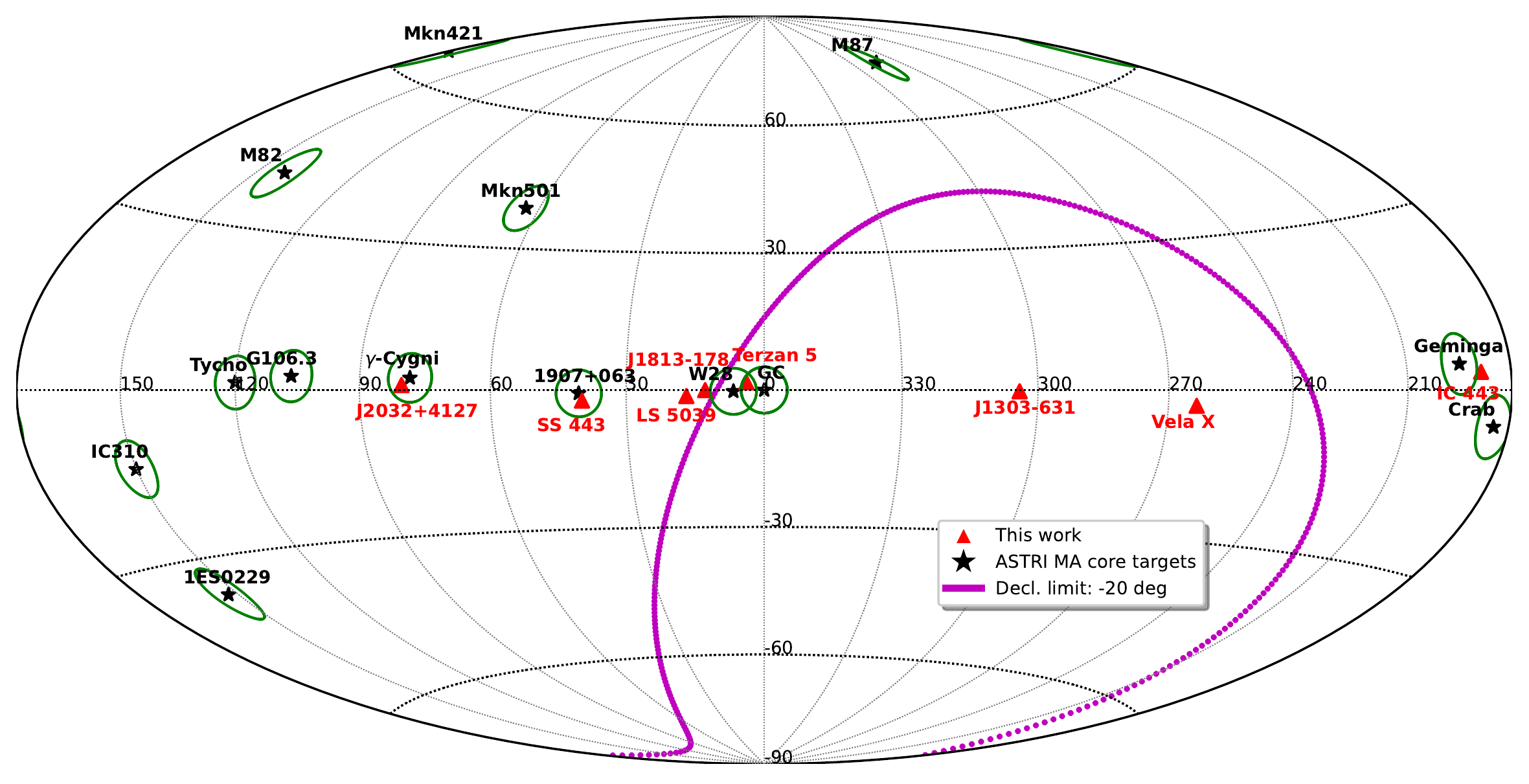}
	\caption{Sky distribution, in Galactic coordinates and Hammer-Aitoff projection, of the \astrima\ Galactic targets presented in this paper (red triangles). The position in Galactic coordinates of the core-science targets described in \citetalias{vercellone21} (black stars) is also shown inside a 6\degmark\ radius circle (green solid circles). The assumed limit on source declination for the objects visible by the \astrima\ is 20\degmark\ (purple line).}
	\label{fig:galactic_map}
\end{figure*}

The energy coverage, effective area and expected performances of the \astrima\ make it  an ideal observatory of the bright Galactic TeV sources. In this section, we briefly summarise the state-of-the-art population of Galactic TeV sources. A recent census of the source classes and relative size of the populations is provided by the Galactic Plane survey (GPS) obtained with the  High-Energy Stereoscopic System \citep[\hess,][]{hess18_gps}. While based on a scan of the plane best observed from the Southern hemisphere, it is reasonable to assume that the HESS GPS is representative  of the source population also for the Northern sky. The \hess\ survey gathers $\sim$\,2700 hours of observations covering the 65--250 deg longitude range for latitudes |b| $<$ 3 deg, with 5\arcmin\ angular resolution. With a sensitivity reaching 1.5\% Crab, the survey yielded in a catalogue of 78 discrete sources. 31 out of 78 sources have secure counterparts at other wavelengths and their nature is well established: 12 sources are pulsar wind nebulae (PWN), 16 sources are supernova remnants (SNRs, 8 with shell-like morphology and 8 with composite morphology) and 3 sources are related to $\gamma$-ray binary systems, most likely hosting a neutron star (NS) as compact object. 36 out of 78 objects are likely associated either to a PWN or to a SNR, or to an energetic $\gamma$-ray pulsar. 11 of the remaining objects do not have a convincing counterpart, with one source of possible extra-galactic origin \citep[HESS J1943+213,][]{archer18}. Apart from the 3 $\gamma-$ray binaries, which appear as point-like sources, all other sources show spatially extended emission.
A view of the observable sky from the \obsteide\ together with the sky  position of the sources analysed in this paper, is presented in Fig.~\ref{fig:galactic_map}.
For the Northern hemisphere, we find a similar distribution of sources among the different classes. Using the second TeVCat catalogue\footnote{\url{http://tevcat2.uchicago.edu/}} we report in Table~\ref{tab:sources_list}  all the Galactic TeV point sources observed from at least one of the three most important VHE IACT Northern observatories: MAGIC \citep{magic16}, HAWC \citep{hawc17}, and VERITAS \citep{staszak15}. We note that the \astrima\ can likely observe many other known sources, in addition to those listed in Table~\ref{tab:sources_list}: for instance, many sources from the \hess\ Galactic survey will be observable, although at higher zenith angles.

\onecolumn
\begin{scriptsize}
\begin{center}
\captionsetup{width=17cm}
\begin{longtable}{m{0.25\textwidth}crrccccc}
	\caption{\normalsize List of identified TeV sources which were reported by at least one of the following observatories in the Northern hemisphere: MAGIC, VERITAS or HAWC. Sky positions,	source identifications and fluxes reported according to the TeVCat \citep{TevCat08} catalogue. The 0\degmark$-$\,45\degmark and 45\degmark$-$\,60\degmark columns indicate the maximum available hours of visibility from the \obsteide\ site, in moonless conditions, for two zenith angle ranges.\label{tab:sources_list}}\\
  \toprule
    \hline
        \multicolumn{1}{l}{Name} & \multicolumn{1}{c}{{RA}} & \multicolumn{1}{c}{{Dec.}} &  \multicolumn{1}{c}{Flux} & \multicolumn{1}{c}{0\degmark$-$\,45\degmark} &
        \multicolumn{1}{c}{45\degmark$-$\,60\degmark} & \multicolumn{1}{l}{Notes} & \multicolumn{1}{l}{Section}  
        \\ 
        \multicolumn{1}{c}{} & \multicolumn{1}{c}{deg} & \multicolumn{1}{c}{deg} &
        \multicolumn{1}{c}{\% Crab}&
        \multicolumn{1}{c}{hr}&
        \multicolumn{1}{c}{hr}
        & \multicolumn{1}{c}{}
        & \multicolumn{1}{c}{}\\
\midrule
\endfirsthead
\multicolumn{8}{c}%
{{\bfseries \tablename\ \thetable{} -- continued from previous page}} \\
\toprule 
        \multicolumn{1}{c}{Source Name} & \multicolumn{1}{r}{{RA}} & \multicolumn{1}{r}{{Dec}} &  \multicolumn{1}{c}{Flux} & \multicolumn{1}{r}{0-45\,deg} &
        \multicolumn{1}{r}{45-60\,deg} & \multicolumn{1}{l}{Notes} 
        \\ 
        \multicolumn{1}{c}{} & \multicolumn{1}{c}{{(deg)}} & \multicolumn{1}{c}{{(deg)}} &
        \multicolumn{1}{c}{Crab \%}&
        \multicolumn{1}{c}{hr}&
        \multicolumn{1}{c}{hr}
        & \multicolumn{1}{c}{}
        & \multicolumn{1}{c}{}\\
\midrule
\endhead
\hline \multicolumn{8}{|r|}{{Continued on next page}} \\ \hline
\endfoot
\hline
\endlastfoot
\midrule
\multicolumn{7}{c}{Pulsar Wind Nebulae}\\
\midrule
\vspace{0.2cm}
{HESS J1857+026} & 284.30 & 2.67  & 16\%  & 370 & 170 & &\\ 
{2HWC J1953+294} & 298.06 & 29.42 & 10\%  & 495 & 170 & &\\ 
\midrule
\multicolumn{7}{c}{Shell Supernova Remnants} \\
\midrule
{Tycho} & 6.34   & 64.14  & 1\% & 410  & 340 & & \citetalias[][]{vercellone21}\\   
{Cas A}          & 350.85 & 58.81 & 2\% & 470  & 280 & &\\
{HESS J1912+101} &  288.00 & 10.10 & 0.1\%      & 420 & 160 & &\\
{SNR G106.3+02.7}& 336.9958  & 60.8769 & 5\% & 460  & 300 &  & \citetalias[Sect.  4.1,][]{vercellone21} \\
\midrule
\multicolumn{7}{c}{Other Supernova Remnants} \\
\midrule
{\textbf{IC\,443}}        & 94.21 & 22.50 & 3\% & 480  & 170 & & Sect.~\ref{sect:ic443} \\ 
{W\,51} & 290.96 & 14.10 & 3\%             &  440 & 160 & Cloud &  \citetalias[][]{vercellone21}\\ 
{SNR G015.4+00.1} &  274.02 & -15.20 & 23\% &  100 & 300 & Composite SNR & \\
\midrule
\multicolumn{7}{c}{Mixed PWN/SNR}\\
\midrule
{CTA 1}       & 1.65    & 72.78   & 4\%         & 70  & 690 & &\\
{3C 58}       & 31.40 & 64.83  & 0.65\%     & 400 & 360 & &\\
{Crab Nebula} & 83.633 & 22.0145    & 100\%     & 470 & 170 & & \citetalias[Sect. 4.3 ,][]{vercellone21}\\
{Geminga}     & 98.48 & 17.77 & 23\%          & 400 & 170 & also TeV Halo & \citetalias[Sect. 4.3 ,][]{vercellone21}\\ 
{\textbf{HESS J1813-178}} & 273.36 & -17.85 & 6\%      & 0   & 370 & & Sect.\ref{sect:j1813}\\
{HESS J1825-137} & 276.55 & -13.58 & 54\%     & 150 & 260 & also TeV Halo\textsuperscript{1}\\
{HESS J1831-098} & 277.85 & -9.90   & 4\%     & 230 & 220 & &\\ 
{HESS J1837-069} & 279.51 & -6.93  & 53\%     & 275 & 210 & &\\ 
{IGR J18490-0000} & 282.26 & -0.02  & 1.5\%   & 346 & 180 & &\\ 
{SNR G054.1+00.3} & 292.63 & 18.87 & 2.5\%    & 460 & 160 & &\\ 
{MGRO J2019+37} & 305.02 & 36.72  & 67\%      & 510 & 180 & &\\ 
{Boomerang} & 337.25 & 61.2 &  44\%           & 450 & 300 & & \citetalias[][]{vercellone21}\\
\midrule
\multicolumn{7}{c}{Gamma-ray binaries} \\
\midrule
{LS 61+303} & 40.13 & 61.23 & 16\% & 440 & 310    &  &\\
{HESS J0632+057} & 98.25 & 5.80 & 3\% & 400 & 170 & &\\
{\textbf{LS 5039}} & 276.56 & -14.85 & 3\% & 130 & 280 & & Sect.~\ref{sect:ls5039}\\
{\textbf{SS 433}} & 287.96 & 4.98  & 1\% & 390 & 170 & & Sect.~\ref{sect:ss433}\\
{\textbf{PSR J2032+4127}} & 308.05 & 41.46  & 1\% &  510 & 190 & also PWN & Sect.~\ref{sect:j2032}\\
\midrule
\multicolumn{7}{c}{TeV halos} \\
\midrule
{HAWC J0635+070} & 98.71  & 7.00   &   &  400  & 170  & &\\ 
{HAWC J0543+233} & 85.78 & 23.40   &   &  480 & 170  & &\\ 
{2HWC J0700+143} & 105.12 & 14.30  &   &  450 & 170  & &\\ 
\bottomrule
        \multicolumn{8}{l}{\textsuperscript{1} The interpretation of HESS J1825-137 
        as a TeV Halo is discussed in \citet{sudoh19} and \cite{aharonia04, aharonian13b}}.
\end{longtable}
\end{center}
\end{scriptsize}
\twocolumn
In the following we illustrate the \astrima\ expected performance on representative objects taken from all the main classes of sources identified in the \hess\ GPS. We will start with the Cygnus region, an extended field crowded of multiple TeV objects, ideal for showing the \astrima\ capabilities at detecting known and serendipitous sources at different flux levels (Sect.~\ref{sect:cygnusregion}).\\
We will then discuss the SNR class in Sect.~\ref{sect:snr}, PWNe in Sect.~\ref{sect:PWN}, \gray\ pulsars and \gray\ binaries in Sect.~\ref{sect:gammaraypsr} and in Sect.~\ref{sect:binaries}, respectively. Other scientific cases, like search for dark matter signal in the Galactic Centre, possible detection of Novae, and VHE emission from a candidate Globular Cluster, are discussed in Sect.~\ref{sect:dm_gc}.

%% file: sect_cygnusregion.tex
\section{A Survey of the Cygnus Region} \label{sect:cygnusregion}

\paragraph{Scientific Case}
The \astrima\ wide FoV will cover large parts of the $\gamma$-ray sky in one observing night. Here, we present the expected outcome of such extensive coverage of the sky, as a possible mapper of both the persistent and transient Galactic population of VHE sources. For the Northern sky, the richest and most interesting extended region to look at is the Cygnus region, a region of the Galaxy which extends from 64\degmark to 84\degmark in Galactic longitude $l$ and from -3\degmark to 3\degmark in Galactic latitude $b$. The region comprises the nearest and most massive star-forming regions of the Galaxy, with a wealth of possible cosmic accelerators, among the many SNRs and PWNe. 

\begin{figure*}
	\centering
	\begin{tabular}{c}
	\includegraphics[width=2\columnwidth, height=5cm]{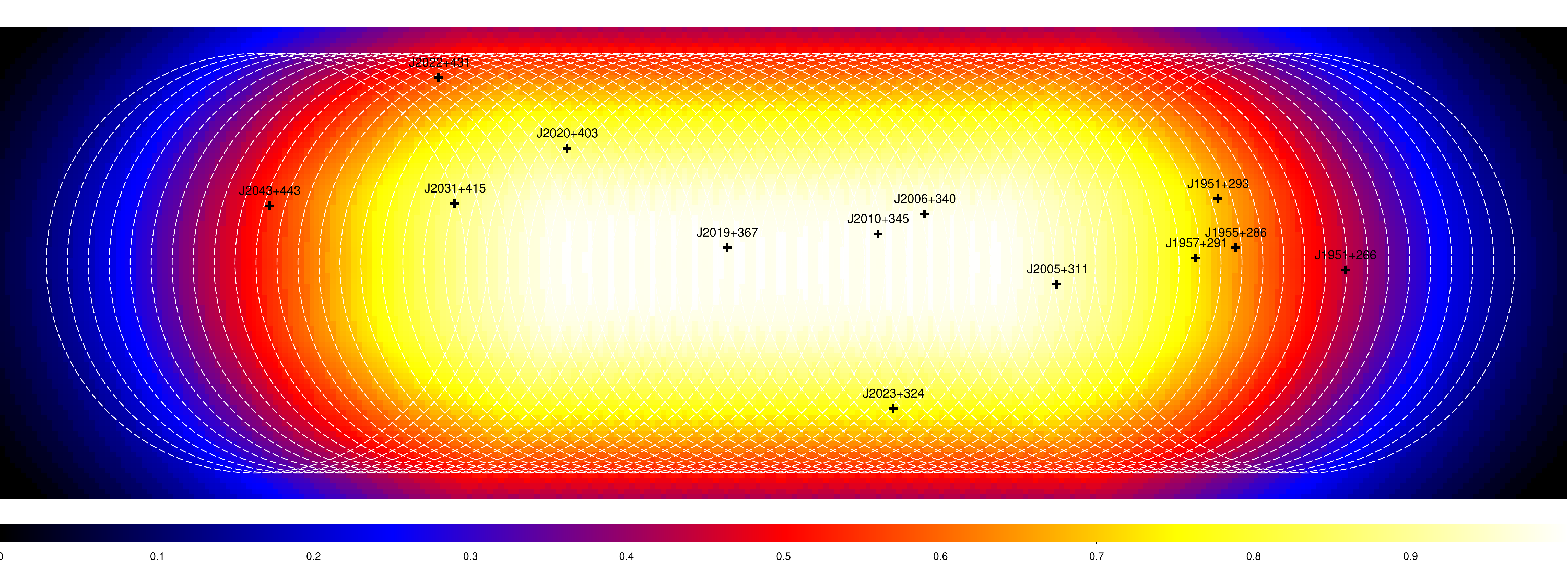} \\
		\includegraphics[width=2\columnwidth, height=5cm]{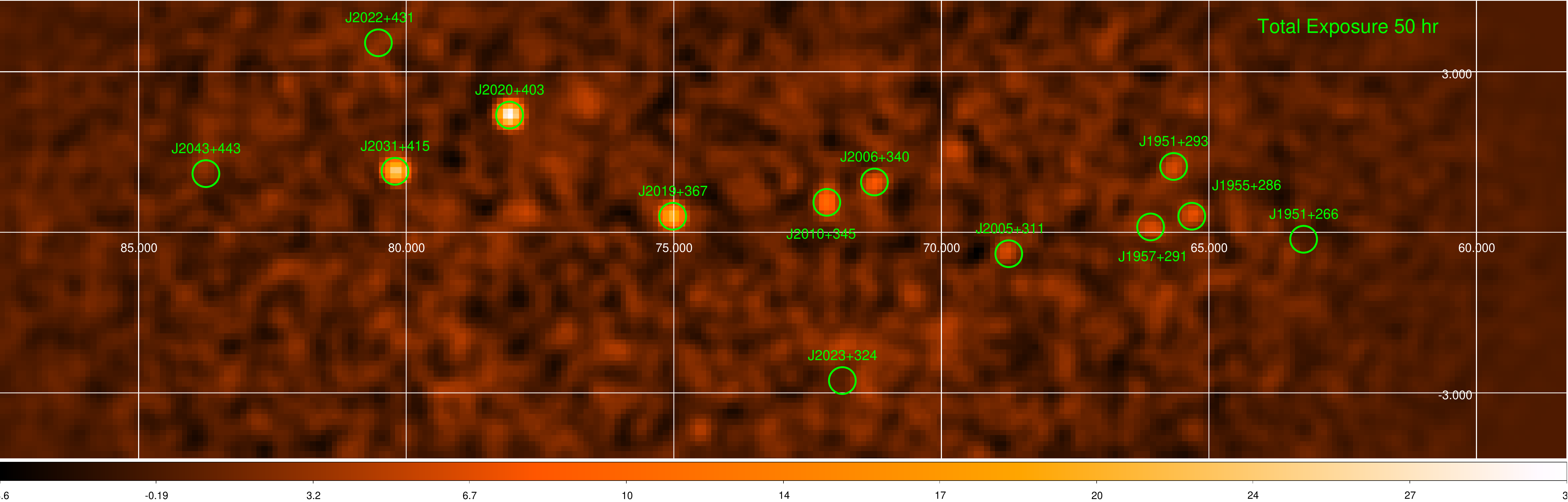} \\
		\includegraphics[width=2\columnwidth, height=5cm]{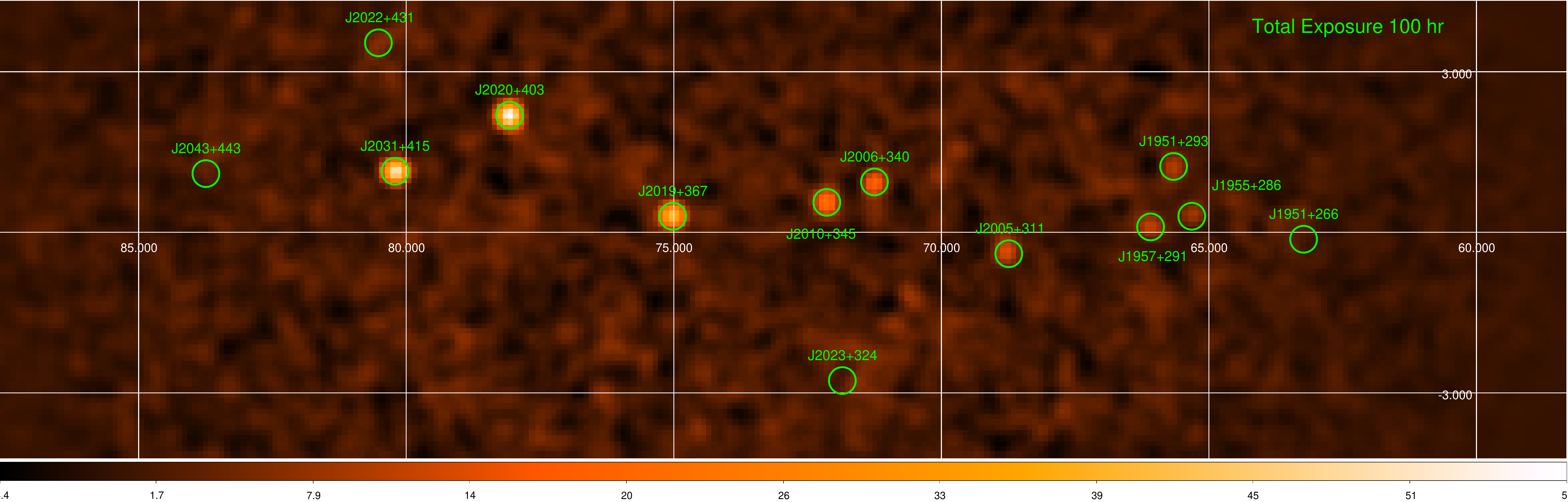}\\
		\includegraphics[width=2\columnwidth, height=5cm]{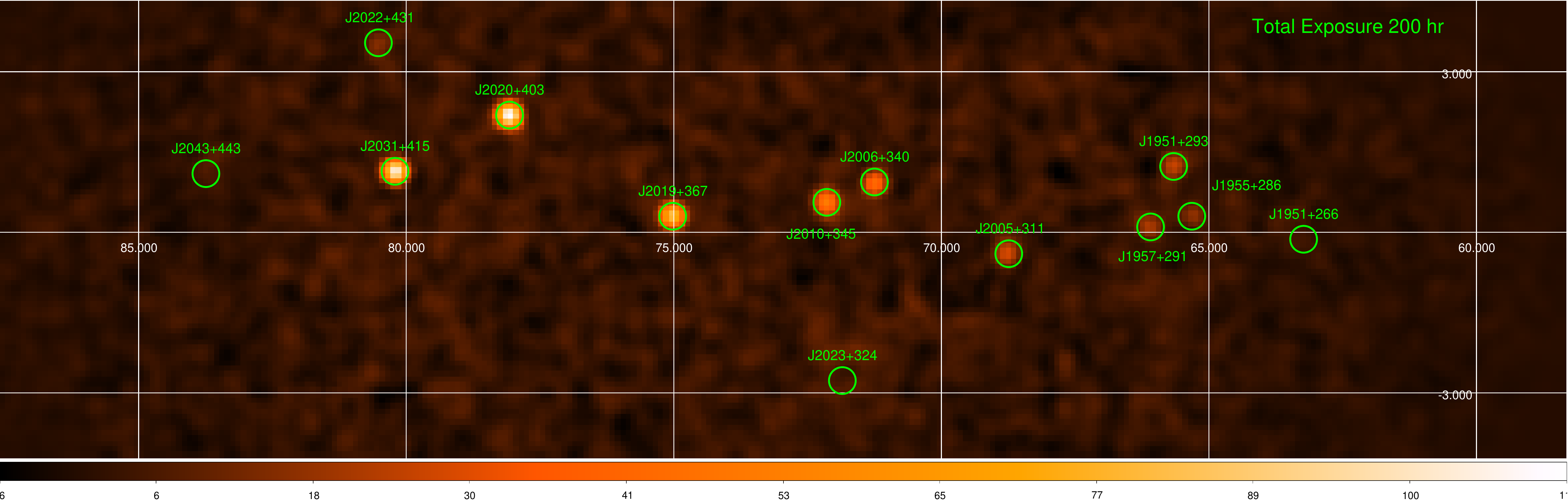}
\end{tabular}
	\caption{Sky maps of the Cygnus region, showing the position of the simulated sources of Table \ref{tab:hawcsources}. From top to bottom: normalised exposure map of the simulated field: each pointing shown as a circle of radius 4\degmark, uniformly spanning the Galactic longitude from $l$\,=\,64 deg to $l$\,=\,84 deg; count maps assuming for each pointing an exposure of 1 hr (second panel), 2 hr (third panel)and 4 hr (fourth panel). Sky map units are counts/pixel.}
	\label{fig:cygnus_skymaps}
\end{figure*}

\paragraph{Feasibility and Simulations}
We simulated a possible survey of the Cygnus region assuming 50 different pointings, at the same Galactic latitude and spaced by 0.4\degmark in Galactic longitude, from ($l,b$) = (64,0) to ($l,b$) = (84,0). We tested three different exposures (1, 2, and 4 hours) for each pointing, to assess the detection efficiency and parameter constraints as a function of the total observing time, which therefore resulted in a global (sum of all the pointings exposures) observing time of 50, 100, and 200 hours, respectively. This strategy maximises the exposure at the centre of the field, whereas the boundary regions of the survey result much less covered. An exposure map of the region (normalised at 1 for the central regions) computed for a reference energy of 10 TeV is shown in Fig.~\ref{fig:cygnus_skymaps}.
For this simulation the list of TeV sources, with their spectra and morphology, are taken from the most recent HAWC catalogue  \citep{2020ApJ...905...76A}. The spatial and spectral parameters are comprehensively reported in Table \ref{tab:hawcsources}. Two sources from this list have been studied in much more detail: the PWN/\gray-binary PSR J2032+4127 (Sect.\ref{sect:j2032} of this paper) and the SNR $\gamma$-Cygni \citepalias{vercellone21}. 
For the sake of consistency, in the present analysis we adopted only the spectral and morphological parameters given by HAWC although most of these sources have been also investigated with facilities with better angular and spectral resolution than HAWC. However, we note that the flux measured by HAWC tends to be generally higher than the other measurements: this might be due to the fact that HAWC uses a power-law spectral model to fit the spectra, whereas most of these sources show a cut-off above 10 TeV. To avoid this high-energy flux bias in our detection estimates, we maintained a more conservative approach and performed an unbinned likelihood analysis in the restricted energy range 0.5--10 TeV. 
From the event list and best-fit models, we produced sky maps from the whole set of simulations and for the three overall exposures (50\,hr, 100\,hr and 200\,hr) with the \textsc{ctskymap} tool using the IRF subtraction method (Fig.\ref{fig:cygnus_skymaps}). 

\begin{table*} [width=2.0\linewidth,cols=7,pos=htp!]
\caption{List of HAWC  TeV sources detected in the Cygnus region \citep{2020ApJ...905...76A} together with spatial and spectral model parameters: RA and Dec. positions, radius (in case of extended sources), power-law spectral photon index and differential flux ($prefactor$) at the reference energy of 7 TeV.}
\label{tab:hawcsources}
\begin{tabular}{ l c c c c c c}     
\hline
\multirow{2}*{Name}     & \multirow{2}*{RA} & \multirow{2}*{Dec.} &       Radius          & \multirow{2}*{Index} &   Prefactor                        \\
                        &   deg    &   deg     & deg      &                      & [$10^{-21}$ MeV$^{-1}$ cm$^{-2}$ s$^{-1}$]  & Notes\\
\hline                                                                                              
\textbf{3HWC J1951+266} &      $297.9$      &      $26.61$       &         $0.5$         &        $2.36$        &           $8.5$ & \\
\textbf{3HWC J1951+293} &      $297.9$      &      $29.40$       &         $0.0$         &        $2.47$        &           $7.9$ & \\
\textbf{3HWC J1954+286} &      $298.7$      &      $28.63$       &         $0.0$         &        $2.42$        &           $6.4$ & \\
\textbf{3HWC J1957+291} &      $299.4$      &      $28.63$       &         $0.0$         &        $2.54$        &           $6.2$ & \\
\textbf{3HWC J2005+311} &      $301.5$      &      $31.17$       &         $0.0$         &        $2.58$        &           $5.6$ & \\
\textbf{3HWC J2006+340} &      $301.7$      &      $34.00$       &         $0.0$         &        $2.56$        &           $8.5$ & \\
\textbf{3HWC J2010+345} &      $302.7$      &      $34.55$       &         $0.0$         &        $2.91$        &           $5.4$ & \\
\textbf{3HWC J2019+367} &      $304.9$      &      $36.80$       &         $0.0$         &        $2.04$        &           $34.7$ & \\
\textbf{3HWC J2020+403} &      $305.2$      &      $40.37$       &         $0.0$         &        $3.11$        &           $11.4$ &  $\gamma$-Cygni (\citetalias{vercellone21})  \\
\textbf{3HWC J2022+431} &      $305.5$      &      $43.16$       &         $0.0$         &        $2.34$        &           $6.0$   &   \\
\textbf{3HWC J2023+324} &      $305.8$      &      $32.44$       &         $1.0$         &        $2.70$        &           $13.8$  &  Sect.\ref{sect:j2032} \\
\textbf{3HWC J2031+415} &      $307.9$      &      $41.51$       &         $0.0$         &        $2.36$        &           $30.7$  &    \\
\textbf{3HWC J2043+443} &      $310.9$      &      $44.30$       &         $0.5$         &        $2.33$        &           $9.7$   &    \\
\hline
\end{tabular}
\end{table*}

\paragraph{Analysis and Results}

We report  in Table~\ref{tab:cygnus_fits} the best-fitting values and associated uncertainties according to the three global exposures of 50, 100 and 200 hr. Among the 13 simulated field sources, 10 sources are always significantly detected even at the shortest 50 hr exposure. In some cases, we obtained a Test Statistic (TS)\,$<$\,9 (e.g. for 3HWC J1951+266, 3HWC J2022+431, and 3HWC J2043+443). The source 3HWC J1951+266 is never well detected even with 200 hr exposure. This is because of the survey strategy: it can be 
noted also from the upper panel of Fig.~\ref{fig:cygnus_skymaps} that these sources are at the edges of the exposure map. For detected sources, the relative errors on their positions are of the order of 1\arcmin\ or less, thus indicating that a census of a VHE population at the sensitivity presently reached by HAWC can be obtained using just few tens of hours of exposure.
\onecolumn
\begin{scriptsize}
\begin{center}
\captionsetup{width=17cm}
\begin{longtable}{m{0.25\textwidth}r c c c c c c c} 
\caption{Best-fitting model parameters for the Cygnus sources according to the exposures (50 hr, 100 hr and 200 hr).\label{tab:cygnus_fits}}\\
\hline
\hline
\multirow{2}*{Name} & Exposure  & Radius        & RA & Dec. & \multirow{2}*{Index} & Norm    & TS \\
                    &   hours   &  deg & deg& deg  &                      &$10^{-21}$ MeV$^{-1}$ cm$^{-2}$ s$^{-1}$ & \\
\hline
\multirow{3}*{\textbf{3HWC J1951+293}} &    50     &          --           &$297.988 \pm 0.013$& $29.391 \pm 0.010$ &   $2.58 \pm 0.25$    &         $7.24 \pm 2.6$                    & $59$\\
                                       &    100    &          --           &$297.994 \pm 0.008$& $29.399 \pm 0.007$ &   $2.46 \pm 0.17$    &         $8.2 \pm 1.9$                    &$123$\\
                                       &    200    &          --           &$297.994 \pm 0.006$& $29.397 \pm 0.005$ &   $2.44 \pm 0.11$    &         $9.1 \pm 1.4$                    &$291$\\
\hline
\multirow{3}*{\textbf{3HWC J1954+286}} &    50     &          --           &$298.699 \pm 0.010$& $28.629 \pm 0.011$ &   $2.47 \pm 0.24$    &         $8.50 \pm 2.8$                    & $59$\\
                                       &    100    &          --           &$298.698 \pm 0.011$& $28.627 \pm 0.010$ &   $2.4 \pm 0.2$    &         $6.5 \pm 1.8$                    & $77$\\
                                       &    200    &          --           &$298.689 \pm 0.007$& $28.627 \pm 0.006$ &   $2.35 \pm 0.16$    &         $6.5 \pm 1.3$                    &$140$\\
\hline
\multirow{3}*{\textbf{3HWC J1957+291}} &    50     &          --           &$299.362 \pm 0.011$& $29.176 \pm 0.011$ &   $2.58 \pm 0.25$    &         $6.9 \pm 2.5$                    & $62$\\
                                       &    100    &          --           &$299.370 \pm 0.008$& $29.170 \pm 0.008$ &   $2.30 \pm 0.18$    &         $9 \pm 2$                    &$128$\\
                                       &    200    &          --           &$299.362 \pm 0.007$& $29.178 \pm 0.005$ &   $2.46 \pm 0.14$    &         $6.7 \pm 1.2$                    &$197$\\
\hline
\multirow{3}*{\textbf{3HWC J2005+311}} &    50     &          --           &$301.465 \pm 0.018$& $31.208 \pm 0.011$ &   $2.70 \pm 0.27$    &         $3.9 \pm 1.6$                    & $34$\\
                                       &    100    &          --           &$301.461 \pm 0.011$& $31.167 \pm 0.010$ &   $2.73 \pm 0.20$    &         $3.7 \pm 1.1$                    & $72$\\
                                       &    200    &          --           &$301.466 \pm 0.007$& $31.170 \pm 0.005$ &   $2.53 \pm 0.13$    &         $5.4 \pm 1.0$                    &$193$\\
\hline
\multirow{3}*{\textbf{3HWC J2006+340}} &    50     &          --           &$301.736 \pm 0.014$& $34.011 \pm 0.011$ &   $2.60 \pm 0.23$    &         $6.0 \pm 2.0$                    & $65$\\
                                       &    100    &          --           &$301.738 \pm 0.008$& $34.001 \pm 0.006$ &   $2.73 \pm 0.14$    &         $6.4 \pm 1.3$                    &$191$\\
                                       &    200    &          --           &$301.720 \pm 0.005$& $33.999 \pm 0.004$ &   $2.50 \pm 0.09$    &         $9.6 \pm 1.2$                    &$464$\\
\hline
\multirow{3}*{\textbf{3HWC J2010+345}} &    50     &          --           &$302.694 \pm 0.012$& $34.552 \pm 0.010$ &   $3.11 \pm 0.18$    &         $4.0 \pm 1.3$                    &$103$\\
                                       &    100    &          --           &$302.695 \pm 0.008$& $34.546 \pm 0.006$ &   $2.95 \pm 0.13$    &         $5.2 \pm 1.1$                    &$225$\\
                                       &    200    &          --           &$302.693 \pm 0.006$& $34.558 \pm 0.005$ &   $3.18 \pm 0.09$    &         $3.7 \pm 0.6$                    &$456$\\
\hline
\multirow{3}*{\textbf{3HWC J2019+367}} &    50     &          --           &$304.937 \pm 0.005$& $36.800 \pm 0.004$ &   $2.12 \pm 0.10$    &        $31 \pm 4$                    &$472$\\
                                       &    100    &          --           &$304.939 \pm 0.004$& $36.805 \pm 0.003$ &   $2.02 \pm 0.08$    &        $34.8 \pm 3.2$                    &$965$\\
                                       &    200    &          --           &$304.943 \pm 0.002$& $36.802 \pm 0.002$ &   $2.02 \pm 0.05$    &        $35.7 \pm 2.2$                   &$2032$\\
\hline
\multirow{3}*{\textbf{3HWC J2020+403}} &    50     &          --           &$305.155 \pm 0.006$& $40.376 \pm 0.004$ &   $3.21 \pm 0.09$    &        $11 \pm 4$                    &$643$\\
                                       &    100    &          --           &$305.160 \pm 0.005$& $40.372 \pm 0.003$ &   $3.11 \pm 0.06$    &        $12.9 \pm 1.4$                   &$1169$\\
                                       &    200    &          --           &$305.161 \pm 0.003$& $40.373 \pm 0.002$ &   $3.14 \pm 0.05$    &        $12.2 \pm 1.0$                   &$2279$\\
\hline
\multirow{3}*{\textbf{3HWC J2022+431}} &    50     &          --           &         --        &         --         &          --          &             --                             & $<9$\\
                                       &    100    &          --           &$305.525 \pm 0.023$& $43.135 \pm 0.018$ &   $2.5 \pm 0.3$    &        $4.4 \pm 1.9$                     & $29$\\
                                       &    200    &          --           &$305.531 \pm 0.015$& $43.153 \pm 0.010$ &   $2.48 \pm 0.20$    &        $5.6 \pm 1.4$                     & $84$\\
\hline
\multirow{3}*{\textbf{3HWC J2023+324}} &    50     &   $0.94 \pm 0.11$   &$305.35 \pm 0.17$& $32.56 \pm 0.14$ &   $3.2 \pm 0.5$    &         $8.2 \pm 7.8$                    & $12$\\
                                       &    100    &   $1.10 \pm 0.08$   &$306.01 \pm 0.12$& $32.32 \pm 0.10$ &   $2.6 \pm 0.3$    &         $26 \pm 8$                   & $24$\\
                                       &    200    &   $0.96 \pm 0.06$   &$305.28 \pm 0.10$& $32.54 \pm 0.08$ &   $2.8 \pm 0.3$    &         $11.2 \pm 5.2$                   & $31$\\
\hline
\multirow{3}*{\textbf{3HWC J2031+415}} &    50     &          --           &$307.930 \pm 0.006$& $41.506 \pm 0.004$ &   $2.46 \pm 0.09$    &        $27 \pm 4$                    &$574$\\
                                       &    100    &          --           &$307.930 \pm 0.004$& $41.506 \pm 0.003$ &   $2.37 \pm 0.06$    &        $31 \pm 3$                   &$1226$\\
                                       &    200    &          --           &$307.929 \pm 0.002$& $41.510 \pm 0.002$ &   $2.32 \pm 0.04$    &        $34.7 \pm 2.1$                   &$2706$\\
\hline
\multirow{3}*{\textbf{3HWC J2043+443}} &    50     &          --           &         --        &         --         &          --          &             --                             & $<9$\\
                                       &    100    &    $0.32 \pm 0.06$  &$310.59 \pm 0.10$& $44.33 \pm 0.08$ &   $2.4 \pm 0.5$    &         $7 \pm 5$                    & 9 \\
                                       &    200    &    $0.50 \pm 0.07$  &$310.85 \pm 0.12$& $44.42 \pm 0.09$ &   $2.2 \pm 0.4$    &        $10 \pm 5$                    &11\\
\hline
\hline
\end{longtable}
\end{center}
\end{scriptsize}
\twocolumn

%% file: sect_snr.tex
\section{\bf Supernova Remnants} \label{sect:snr}

Supernova Remnants (SNRs) are sources of primary interest for High Energy Astrophysics: not only they are bright high energy radiation sources but they are also thought to be the primary Cosmic Ray (CR) accelerators in the Galaxy \citep[see e.g.][for some recent reviews]{2013A&ARv..21...70B, 2014IJMPD..2330013A}. A SNR is a system composed of the debris of a Supernova (SN) explosion and of the material, either interstellar or circumstellar, that the blast wave associated to the explosion progressively sweeps up depending on the type of explosion and on the evolutionary stage of the source. SNRs are characterised by the presence of at least two shock waves: the forward shock, associated with the supersonic expansion of the SN ejecta in the surrounding medium, and the reverse shock, propagating through the ejecta and back towards the explosion site. The processes of plasma heating and particle acceleration associated with these shock waves make SNRs bright sources throughout the entire electro-magnetic spectrum, from radio to VHE $\gamma$-rays. The latter are privileged messengers when it comes to probe the SNR-CR connection (\citet{aharonian13}; see also \citet{Amato21} for a very recent review of this subject).

Indeed, while the paradigm associating galactic CRs with SNRs has been in place for about 80 years, observational evidence of the connection is still only partial. X-ray observations of young SNRs reveal the presence of electrons accelerated to energies of several tens of TeV \citep{Vink_2012}, suggesting that protons could be accelerated beyond $\sim$\,100 TeV through Diffusive Shock Acceleration \citep[DSA, see e.g.][]{Malkov_2001}. However these high energy protons can only show their presence through nuclear collisions, ultimately producing $\pi^0$-decay $\gamma$-rays. This emission is difficult to disentangle from leptonic Inverse-Compton scattering (ICS) radiation contributing in the same energy band. Solid evidence of the presence of relativistic protons has only been found in middle-aged SNRs, as W44 and IC443 \citep{Giuliani_2011, Ackermann_2013, 2014A&A...565A..74C, 2017hsn..book.1737F}, where the $\pi^0$-decay signature is easier to recognise but where we do not find, nor expect, very high energy particles: the emission in these sources is cut off at energies consistent with protons below 100 TeV and with a steep spectrum. In addition, these particles could in principle be reaccelerated, rather than freshly accelerated at the SNR shock \citep{2016A&A...595A..58C}.

For the vast majority of \gray\ emitting SNRs, and most importantly in the young and most efficient ones, the \gray\ emission can be modelled as hadronic or leptonic just as well, in spite of the fact that the two interpretations usually imply two very different scenarios in terms of the underlying properties of the source. A striking example in this sense is the remnant RX J1713.7-3946 \citep{2009MNRAS.392..240M, 2010ApJ...708..965Z}. 

The nature of the emission becomes however clear as soon as one moves in photon energy beyond a few tens of TeV: at these energies the ICS radiation is suppressed due to the transition to the Klein-Nishina scattering cross-section, so that emission beyond this energy must come from hadronic processes. At these very high energies one would be observing protons of energy close to 1 PeV, namely in the so-called {\it knee} region where a break in the CR spectrum is observed, presumably indicating the maximum achievable proton energy in galactic sources. Investigation of SNRs at energies around 100 TeV is then particularly topical, especially in light of the recent theoretical developments casting doubts on the ability of SNRs to accelerate protons up to the {\it knee} \citep{2015APh....69....1C,Cristofari20}. 

It is possible to classify SNRs on the basis of their morphology. Traditionally, this classification comprises three main classes: shell-type SNRs, plerions, and composite SNRs \citep{Vink_2012}. Shell-type SNRs are clearly characterised by a patchy ring of X-rays emission around the centre of the explosion. This emission is usually observed with the same shape also at other frequencies (e.g. in radio or in \gray) and is produced at the shock front of the blast wave.
 
At the moment, 14 distinct sources are classified as shell-type SNRs in the TeVcat catalogue\footnote{\url{http://tevcat2.uchicago.edu/}} (note though that the north-east and south-west limbs of SN 1006 are listed as separate entries). These sources are all young remnants with negligible fluxes at GeV energies. At the present date, only Cas A \citep{Abdo_2010_casa}, Tycho \citep{Giordano_2012}, RX J1713.7-3946 \citep{Abdo_2011}, RX J0852-4622 \citep{Tanaka_2011}, and RCW86 \citep{Renaud_2012} have been detected at GeV energies with \textit{Fermi-LAT}. The spectral shape of these sources is usually well-fitted with a power-law of photon-index in the range 2.3--3.0. Only three sources show a clear TeV cut-off in the spectrum: RCW 86 at 3.5 TeV, Vela Junior at 6.7 TeV, and RX J1713.7-3946 at 12.9 TeV. RX J1713.7-3946, being very bright and extended, is one of the best targets for detailed morphological studies. Recently, for the first time, \hess\ observations have shown that the TeV emitting regions are not perfectly coincident with the X-ray emitting regions \citep{hess18_j1713}. The TeV region appears more radially extended, suggesting  the existence of a leakage of the most energetic particles from the shock front. A quantitative observational assessment of the particle escape from SNRs, in terms of their spectrum and transport properties, would be of the utmost importance in view of probing the SNR-CR connection.\\

Plerions constitute another class of SNRs. They result from a core-collapse SN event and are characterised by a centre-filled morphology. The nebular emission is mainly due to a young and fast spinning NS, which releases its rotational energy in the form of a relativistic wind mainly made of electron-positron pairs and magnetic fields. This nebula of accelerated particles is named Pulsar Wind Nebula (PWN). The PWN energetics is mainly driven by the pulsar's activity rather than the SNR blast wave, and, in most cases, it is difficult to clearly disentangle and assess the contribution of the primeval SNR explosion. PWNe constitute the majority of the firmly established Galactic TeV emitters \citep{hess18_gps} and, given their importance, we will focus on this class in the next section.\\
Finally, the composite SNRs show distinctively the contributions from the plerion and from the ring-shaped blast wave.

The HGPS survey lists eight firmly identified shell-like SNRs and eight plerion-composite SNRs. In \citetalias{vercellone21}, we showed how a deep observation of the prototypical shell-like SNR, Tycho, with \astrima\ would conclusively prove or disprove the PeVatron nature of this source. Similarly, other two SNRs suggested as possible PeVatron candidates are examined: the SNR G106.3\,+\,2.7 and SNR G40.4\,-\,0.5, associated to the VHE sources VER J2227\,+\,608 and VER J1907\,+\,062, respectively.
 
Another topic of special interest in the context of the SNR class, and their connection with galactic CRs, is how particles escape into the Galactic medium once they are accelerated at the shock front. This can be investigated through detailed spectral and morphological studies of middle-aged SNRs.
\citetalias{vercellone21} shows how outstanding issues regarding this topic can be successfully investigated with the \astrima\ in two key-target objects: $\gamma$-Cygni and W\,28. 


Among the SNRs listed in Table~\ref{tab:sources_list}, we will take as possible case-study the middle-aged SNR IC\,433, as this source is in the field of two targets of the {\it Science Pillars}. For the long-term observatory programme, two bright SNRs will certainly receive particular attention: HESS  J1912+101 and the SNR G015.4+0.1. 
The former \citep{2018A&A...612A...8H}, also seen by HAWC \citep[2HWC J1912+099,][]{2017ApJ...843...40A}, is a candidate shell SNR hosting the radio pulsar J1913+1011 \citep{2008ApJ...682.1177C}. 
Recent detection of extended GeV emission \citep{2020ApJ...889...12Z} close to this TeV source led to speculation that this GeV/TeV emission is originated by a TeV halo. 
However this scenario is incompatible with the shell-like TeV morphology hypothesis advanced by \hess. 
It is also possible that the TeV and GeV emission have distinct origins, from the SNR and the PWN, respectively, with the TeV emission from the PWN eclipsed by the SNR. 
The VHE source HESS J1818-154 \citep{2014A&A...562A..40H}, also seen by HAWC \citep[2HWC J1819-150][]{2017ApJ...843...40A}, has been observed in the centre of the shell-type SNR G 15.4+0.1. 
This TeV emission, coincident with the diffuse X-ray emission, is compatible with a PWN scenario classifying the G 15.4+0.1 as a composite SNR. 
G 15.4+0.1 is, therefore, one of the few examples of this category seen at TeV energies  \citep[as G\,0.9+0.1 and G\,21.5-0.9, see][]{2016ApJ...821..129A, hess18_pwn}.

The \astrima\ will give an important contribution to the study of these interesting sources. Thanks to their high flux, the \astrima\ would allow the study of their spectral properties (such as the presence of a  cut-off) up to tens of TeV with moderate exposure times ($<$\,100 hrs). Moreover, both remnants are close to other sources of  potential interest for the \astrima:  HESS J1912+101 is at < 5\degmark\ from the SNRs W51 and W49B, while HESS J1818-154 is at < 3\degmark\ from the PWNe HESS J1813-178 (Sect.~\ref{sect:j1813}), HESS J1825-137 and from the \gray\ binary LS 5039 (Sect.~\ref{sect:ls5039}).

%% file: sect_ic443.tex
\subsection{VHE emission from a middle-age SNR: IC\,443}  \label{sect:ic443} \vspace{0.3cm}
\paragraph{Scientific Case}
IC\,443 (also known as G189.1\,+\,3.0) is a SNR located at 1.5 kpc with a $\sim 20'$ angular radius~\citep{2003A&A...408..545W}. The age of the SNR is still uncertain, with a possible range between $\sim 3$ kyr~\citep{2008A&A...485..777T} and $\sim 30$ kyr~\citep{2008ApJ...676.1050B}. 
Recent 3D hydrodynamical simulations suggest an age of $\sim 8.4$  kyr~\citep{2021A&A...649A..14U}. It is classified as a mixed-morphology SNR (MMSNR,~\citealt{1998ApJ...503L.167R}), i.e. a remnant with a shell-like morphology visible in the radio band and a centrally filled thermal X-ray emission. The environment around the remnant is rather complex: a dense molecular cloud in the northwest and southeast region~\citep{1977A&A....54..889C} forms a semi-toroidal structure that encloses IC 443~\citep{2006ApJ...649..258T, 2014ApJ...788..122S} and an atomic cloud in the northeast~\citep{1978MNRAS.183..187D} confines the remnant. The remnant has been observed through radio~\citep{2004AJ....127.2277L, 2008AJ....135..796L, 2012ApJ...749...34L}, infrared~\citep{2014ApJ...788..122S} and X-rays~\citep{2006ApJ...649..258T, 2008A&A...485..777T, 2018A&A...615A.157G}. Strong $\gamma$-ray emission is associated with the interaction of the SNR with the nearby molecular cloud at both HE~\citep{2010ApJ...710L.151T, Abdo_2010_ic443, Ackermann_2013} and VHE~\citep{2007ApJ...664L..87A, Acciari_2009}. Spectral features strictly related with the characteristic pion-bump~\citep{Ackermann_2013} strongly suggests that IC\,443 is a CR proton accelerator.

\paragraph{Feasibility and Simulations}
IC 433 is observable from the Teide site for about 470 hours per year with a zenith angle between 0\textdegree  and 45\textdegree\ and 165 hr per year with a zenith angle between 45\textdegree\ and 60\textdegree\, in moonless conditions. In a region with 10\textdegree\ radius around IC\,443, two key target sources of the {\it Science Pillars} are also located: the Crab PWN ($\sim$ 10\textdegree\ angular separation) and the TeV halo of Geminga ($\sim$ 6\textdegree\ angular separation). The vicinity to these promising sources on one hand and the large FoV of the \astrima\, on the other hand, will guarantee a long exposure for IC\,443 in the first years of operation. Therefore, to understand what spectral and morphological constraints are achievable for reasonable observational exposures, we simulated the source for 100 and 200 hours. We adopted an extended spatial model (\textit{RadialDisk} in \textsc{ctools}) assuming a radius of 0.16\textdegree~\citep{Acciari_2009} and a position 3\textdegree\ off-axis from the pointing position; the spectrum is described as a power-law with an index of 2.99 \citep{Acciari_2009}. 
We fitted the data using a binned analysis likelihood in the 0.5--200 TeV range, according to standard procedures.

\begin{figure}
	\centering
		\includegraphics[scale=.5]{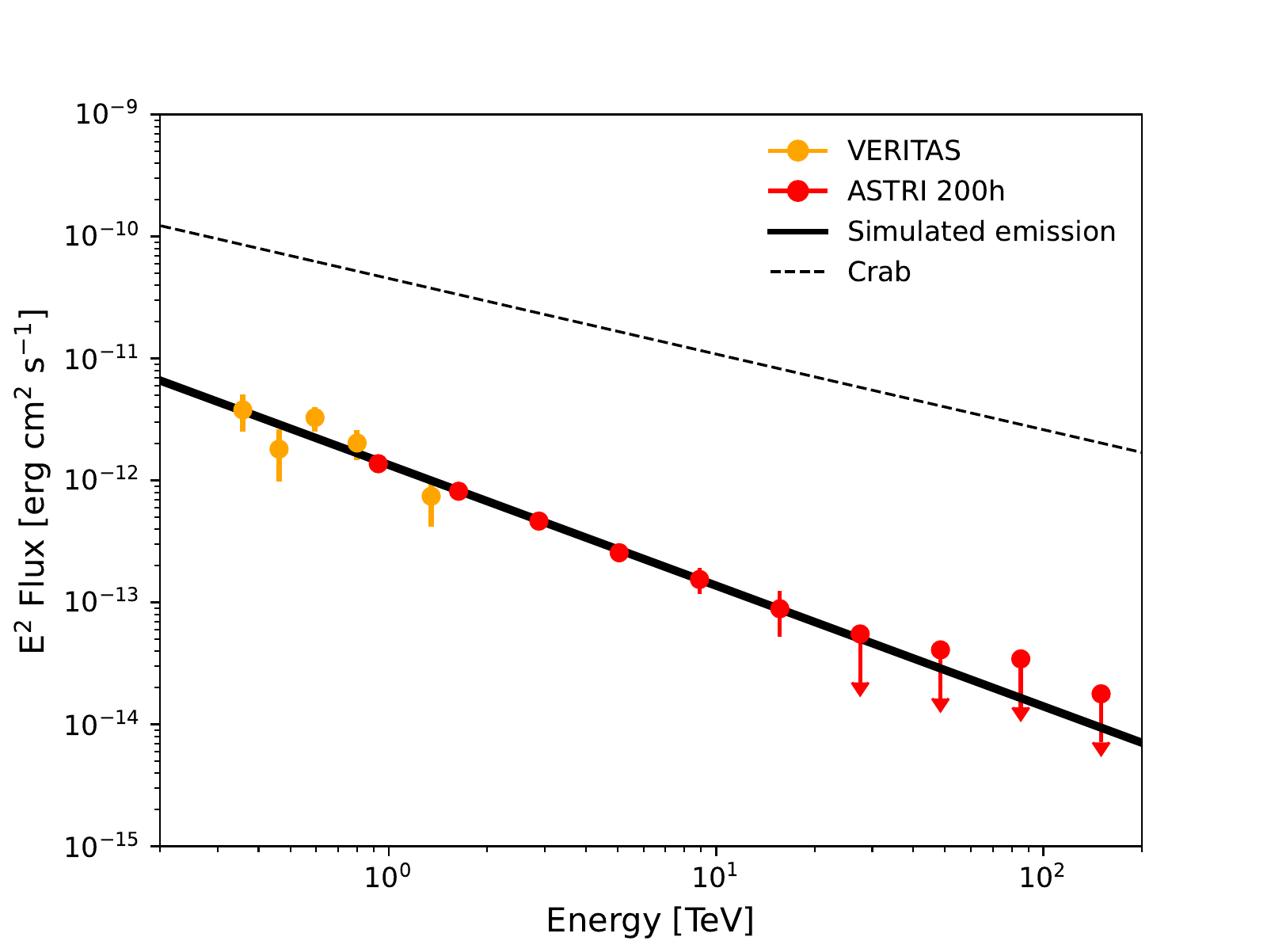}
	\caption{IC\,443: \gray\ data from VERITAS~\citep{Acciari_2009} (orange dots) and \astrima\ (red dots).
	The solid black line shows the best-fit model for the combined data-sets. We show for illustrative purposes  the Crab spectrum as a dotted line.}
	\label{fig:ic433}
\end{figure}

\paragraph{Analysis and Results}
As a first step, we checked if it is possible to constrain the extended nature of the source. To this aim we compared the TS values computed using a point source (our null-hypothesis) and an uniform disk spatial model for the fit. With 100 hr of exposure, we found a $\Delta$TS\,=\,334  implying a significant improvement ($>18 \sigma$)  that increases to $\sim 25 \sigma$ (TS\,=\,664) for 200 hours of exposure. In particular, for 200 hr of exposure, we might be able to constrain the photon index with a relative uncertainty of  2\%  ($3.08 \pm 0.07$); a similar relative error is obtained for the disk radius  ($0.166 \pm 0.004^{\circ}$). Moreover, the last significant energy bin is filled at energies of $\sim 20$ TeV, thus extending the current available SED (see Fig. \ref{fig:ic433}) and allowing us to detect the presence of any possible cut-off in the 1--10 TeV range.

%% file: sect_pwn.tex
\section{\bf Pulsar Wind Nebulae} \label{sect:PWN}
The majority of core-collapse SN events give birth to a fast spinning, highly magnetised neutron star, which then spins down by emitting a highly relativistic magnetized wind, mainly composed of electron-positron pairs, efficiently produced via cascades in the pulsar magnetosphere. For several thousands of years after the SN explosion, the wind is confined by the surrounding SN ejecta, which are expanding at non-relativistic speed, and therefore it has to slow down. This occurs at a termination shock where the wind kinetic energy is dissipated and a large fraction of it ($\sim 20\%$ in the Crab Nebula) is converted into particle acceleration \citep{Amato14}. The Pulsar Wind Nebula (PWN) is the bubble of magnetised plasma that develops downstream of the termination shock \citep{gaensler06}. Particles are highly relativistic, with a non-thermal distribution, and emit synchrotron and Inverse Compton radiation over a very wide range of frequencies: synchrotron is typically the dominant emission process from radio to hard X-rays/soft gamma-rays, while Inverse Compton is the primary process at VHE. The size of the PWN is typically observed to shrink with increasing frequency from radio to X-rays, a fact that is easily understood in terms of synchrotron losses. However, since the VHE emission is mostly contributed by the particles that are responsible for radio/optical synchrotron emission, the nebular size is larger at TeV energies than in the X-rays.

Taking into account the results from the recent \hess\ survey of PWNe, comprising 19 well-established sources and about 10 \textit{bona fide} candidates \citep{hess18_pwn},  significant correlations have been found between the properties of the wind generating pulsar and the most important properties observed at TeV energies. PWNe emit a significant fraction of their power in the VHE band: the efficiency, defined as the ratio of the 1--10 TeV luminosity and the present spin-down power, is 0.1--10\%, thus resulting in a range of TeV inferred luminosities $\log_{10}(L_{\rm TeV}[{\rm erg/s}]$ = 32.3--35.9. PWNe are spatially extended sources by nature, although at TeV energies some of them could be difficult to resolve either because of their young age, or because of a large distance from Earth. For the \hess\ PWN catalogue, this translates into inferred dimensions in the few pc, up to $\sim$\,40 pc range; for young PWNe, the nebular radius ($R_{\rm PWN}$) correlates with the age of the source and, as expected, anti-correlates with the spin-down power of the pulsar. Because of the interaction of the expanding bubble with the SNR reverse shock, and of the kick imparted to the NS at the time of the explosion, the centre of the PWN and the position of the NS show sometimes an offset, which becomes larger with age. In two PWNe,  HESS J1303-631 and HESS J1825-137, this is also accompanied by a different, spatially-dependent, TeV spectrum along the path described by the pulsar in the nebula, with the most energetic photons more concentrated at the present position of the NS. From a spectral point of view, PWNe typically show a power-law spectrum with a cut-off, generally below 10 TeV; the averaged spectral index is 2.3, although this value is possibly biased by the fact that softer indices are more difficult to constrain and the corresponding sources are more difficult to detect. Since the primary mechanism for the TeV emission is IC scattering, the cut-off is due to the onset of the Klein-Nishina regime for the IC scattering cross-section. The presence of a contribution from p-p processes is also possible \citep{2003A&A...402..827A}, although it may require special environmental conditions to become detectable \citep[see e.g.][]{horns06}. How well deep observations of the Crab PWN with the \astrima\ might constrain this additional hadronic contribution is discussed in \citetalias{vercellone21}.

In recent years, detection of extended TeV emission around two PWNe showed that high-energy electrons and positrons can escape from the PWN and become trapped in a larger region of suppressed diffusion; whether these sources constitute a separate class or a natural outcome of the PWN evolution is still a subject of debate \citep{giacinti20}. Detailed spectral-morphological studies above 10 TeV will help to shed light on the matter, as was shown in \citetalias{vercellone21} for the case of Geminga.

Among the many PWNe observable from Teide, we will discuss in detail the case of the moderately bright HESS J1813-178 in Sect.~\ref{sect:j1813} and of TeV J2032+4130, a PWN powered by a young NS likely belonging to a binary \gray\ system in Sect.~\ref{sect:j2032}.  We also present in the appendix two PWNe, originally studied in the context the ACDC project \citep{pintore20}. They offer a good benchmark for illustrating two characteristics of the analysis of this class of sources affecting in general the source emission of PWN: their multi-wavelength (MWL) emission (as in the case of Vela X, see Sect. \ref{sect:velax}) and the possible energy-dependent morphology, shaped by their past emission (as in the case of the PWN HESS J1303\,-\,631, see Sect.~\ref{sect:j1303}). 

In addition to the simulated spectra of PWNe analysed in the \citetalias{vercellone21} and in this work, the PWN HESS J1825-137 deserves a special mention. This bright nebula is powered by the energetic pulsar PSR J1826-1334 and, similarly  to the case of PWN HESS J1303-631 (Sect. \ref{sect:j1303}), the pulsar position is offset with respect to the centre of the nebula, because of the pulsar proper motion. The spectral shape shows a dependence on the distance from the pulsar, thus suggesting that the highest energy electrons trace the pulsar proper motion \citep{aharonian06}. The source is likely associated with the source 2HWC 1825-134 \citep{2017ApJ...843...40A}, although HAWC angular resolution does not allow to clearly disentangle the contribution from another close-by TeV source HESS J1826-130 \citep{2017arXiv170804844A, 2019A&A...623A.115D}. This latter  source shows a hard spectrum above 1 TeV ($\Gamma$ < 2) and a possible cut-off around 13 TeV. The combined spatial and spectral resolution of \astrima\ will prove to be especially important in crowded regions like this one.

%% file: sect_j1813.tex
\subsection{A moderately bright PWN: HESS J1813-178} \label{sect:j1813}
\paragraph{Scientific Case}
We consider here the candidate PWN HESS J1813-178, a source of moderate brightness (6\% of Crab flux) but with a hard spectrum ($\Gamma$ $\sim$\,2). The \astrima\ improvement in the effective area at $>10$ TeV, with respect to existing IACTs, will be essential.\\
HESS J1813-178 was discovered in 2005 by \hess\ \citep{2005Sci...307.1938A} at TeV energies and then associated via positional coincidence with the source SNR G\,12.82--0.02, discovered in the radio and X-ray bands \citep{2005ApJ...629L.105B, 2005ApJ...629L.109U}. A few years later, \citet{2009ApJ...700L.158G} discovered the highly energetic pulsar PSR J1813-1749 within SNR G12.82-0.02. This pulsar has a spin-down luminosity $\dot{E} = (6.8\pm 2.7) \times 10^{37} \text{erg} \text{ s}^{-1}$, a characteristic age $\tau_c = 3.3-7.5$ kyr, and it is certainly capable of powering the TeV emitting particles. The 4.7 kpc distance has been determined by the association of SNR G12.82-0.02 with a nearby young stellar cluster \citep{2008ApJ...683L.155M}. The age of the system is unknown but it is believed to be young due to the small radius of the SNR shell that is expanding in a regular interstellar medium \citep{2005ApJ...629L.105B}. 

Even before the pulsar discovery, \citet{2007A&A...470..249F} proposed a composite SNR scenario for the multi-wavelength emission of HESS J1813-178, where the $\gamma$-rays could arise either from a young PWN in the core or from the SNR shell, or both. The radius of the SNR shell -- and subsequently the radius of the PWN -- is relatively small ($\sim 3'$) and, given the typical angular resolution of IACT observatories, it is not possible to understand where the emission comes from. \cite{2007A&A...470..249F} and \cite{2010ApJ...718..467F} tried to model its MWL emission considering two different scenarios, one where the TeV emission mostly comes from leptons accelerated in the PWN and one where this emission mostly comes from hadrons in the SNR shell. However, neither of the two studies could point on a preferred scenario.

A possible way to make progress on the subject is through observations in the GeV band, where the emission properties are expected to be different for hadrons and leptons. 
\citet{acero13} made a detailed search for GeV emission with \textit{Fermi}-LAT in the direction of HESS J1813-178 but could only put upper limits. These upper limits were used to constrain the models by \citet{2007A&A...470..249F} and \citet{2010ApJ...718..467F}, but no conclusion could be reached on the nature of HESS J1813-178.
In a more recent work, \citet{2018ApJ...859...69A} analysed more years of Fermi-LAT data in the direction of the source and found an extended GeV emission (uniform radial disc with R$_\text{GeV}=0.6$ deg) plus a small excess near the centre of the TeV source. However, the GeV emission turned out to be much more extended than the TeV source and might be associated with the nearby star-forming region \citep{2018ApJ...859...69A}. As a result, still no conclusion can be drawn for the TeV emission of HESS J1813-178. 

A complementary strategy to unveil the nature of the source is to look in the highest-energy part of the TeV spectrum. From the models of \cite{2007A&A...470..249F} and \cite{2010ApJ...718..467F}, it is clear that a population of protons is expected to have a lower energy cut-off with respect to a population of electrons. The quality of the currently available data from \hess\ above 10 TeV is not sufficient to discriminate between the models.

The \astrima\ could give a major contribution in clarifying the nature of this source, by improving its detected spectrum up to, and above, 100 TeV. To test this expectation we have used for our simulations the models from \citet{2007A&A...470..249F} and \citet{2010ApJ...718..467F}. 

For the PWN scenario, both studies used a time-dependent leptonic model \citep[for more detail see e.g.][]{fiori2020} to describe the MWL emission, with the difference that \citet{2007A&A...470..249F} considered a power-law shape for the injection spectrum, while \citet{2010ApJ...718..467F} used a Maxwellian plus a power-law tail injection spectrum.
Furthermore for the SNR scenario the two studies used a similar approach, where the TeV emission originates from the interaction of protons accelerated in or near the SNR shell. An additional difference is in the underlying particle distribution: in \citet{2007A&A...470..249F} a power-law with a spectral index of 2.1 is considered, whereas in \citet{2010ApJ...718..467F} the protons spectrum is computed from a semi-analytical non-linear model.  

\paragraph{Feasibility and simulations}
HESS J1813-178 is observable from Teide for $\sim$\,370 hr/yr only at large zenith angle (between 40\degmark\ and 60\degmark). This will lead to an increase of the lower energy threshold for the observations, but also an increase of the effective area of the array in the highest energy band \citep[see e.g. the studies made for the MAGIC telescopes; ][]{magic16_bis}.

We simulated with \ctools\ the observations with the \astrima\ in the energy range $0.5-200$ TeV. As spatial model for the source, we used the best-fit radial Gaussian with $\sigma$\,=\,0.049\degmark\ as reported in \cite{hess18_gps}. We simulated 4 different cases using the spectral models described in the previous section. 

To derive the minimum amount of time needed to obtain statistically significant spectral bins above 10 TeV, we performed the simulations with 50, 100 and 200 hours of observing times. The latter, very large duration of the observations, takes into account the fact that the source is located at about $\sim6$\degmark\, from W28, one of the sources that will be extensively observed during the first years of the activities of the \astrima\ as part of the core science program \citepalias{vercellone21}. 

\paragraph{Analysis and results}
We performed an unbinned maximum likelihood analysis on the simulated data with \ctools\ to obtain the best-fit models and from the latter we extracted the spectral data points.
\begin{figure*}
\centering
    \includegraphics[width=\textwidth]{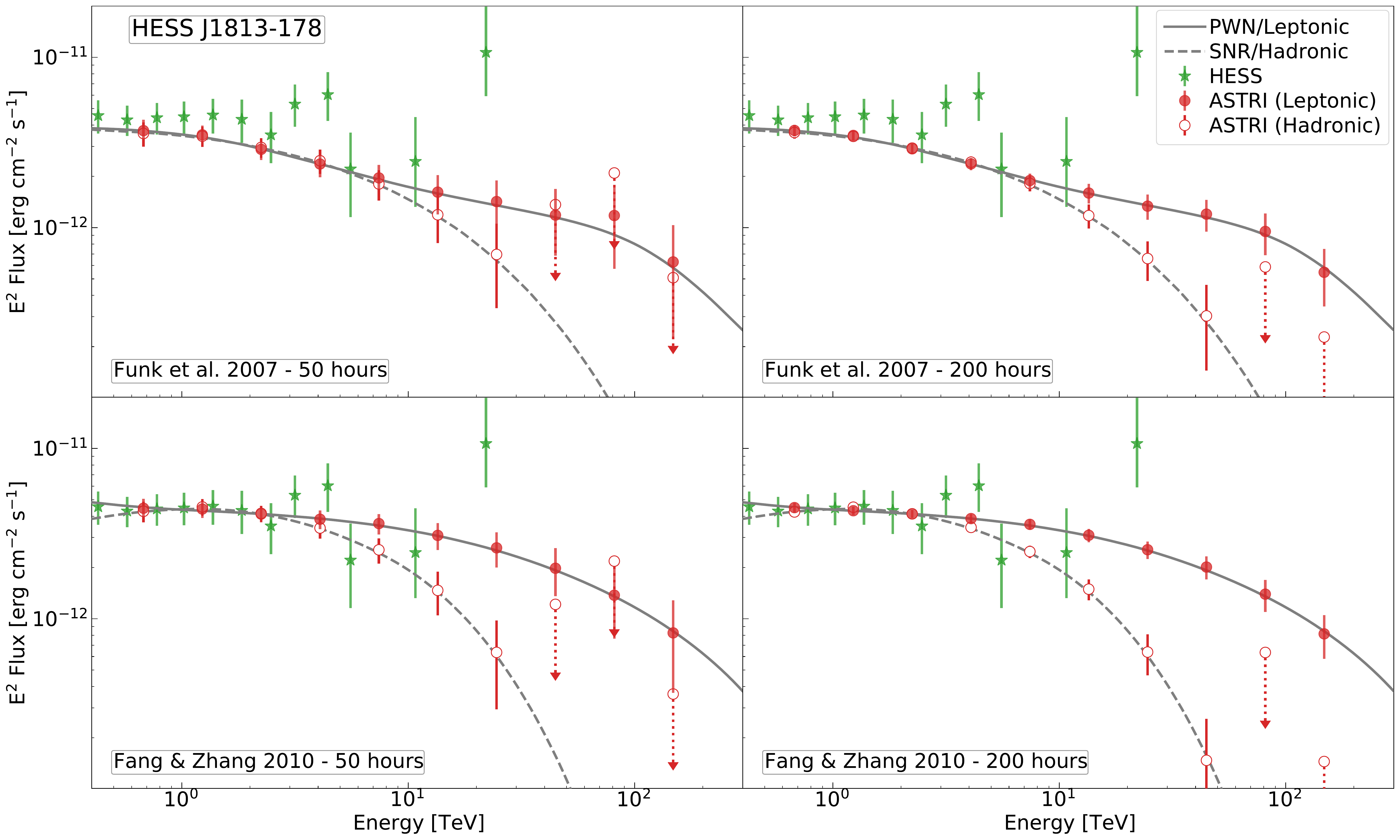}
	\caption{Simulated \astrima\ spectral points (red filled circles for leptonic models, open red circles for hadronic models) for 50 hours (left panels) and 200 hours (right panels) of observing times, together with the input models (solid lines for leptonic models, dotted lines for hadronic models) and the available data from \hess\ \citep[green stars; ][]{2005Sci...307.1938A}, for the leptonic and hadronic models from \cite{2007A&A...470..249F} (upper panels) and \cite{2010ApJ...718..467F} (lower panels). The points with the dashed red error-bars are the upper limits found in the case of the hadronic models.}
\label{fig:hessj1813}
\end{figure*}

In all the simulated cases, we found that the quality of data is always better than the currently available \hess\ data; we obtain meaningful results already with 50 hours of observations. Data points simulated with the two models by \citet{2010ApJ...718..467F} -- that predict slightly higher fluxes compared to the \cite{2007A&A...470..249F} models -- are already well separated with 50 hours, and it would be possible to select a preferred scenario. In the case of the two different models by \cite{2007A&A...470..249F}, 50 hours are not sufficient to distinguish them but we need at least of 100 hours of observations. Interestingly, we found that the source can be detected above 100 TeV already with 50 -- 100 hours of observations in the case of leptonic origin of the emission, while in the hadronic scenario the maximum energy is well below this value (even when increasing the observing time up to 500 hours).

In Fig.~\ref{fig:hessj1813} we report the simulated \astrima\ spectral points for 50 and 200 hours of exposure, together with the template models and the available data from \hess\ \citep{2005Sci...307.1938A}. We note that they always follow the corresponding input models and allow us to discriminate between the different scenarios, while this is not possible taking into account only the \hess\ data.

We conclude that for HESS J1813-178  an observation made with the \astrima\ will be crucial. Moreover, a clear preference among the proposed models can be obtained within a reasonable amount of time.

%% file: sect_j2032.tex
\subsection{A \gray-binary in a PWN: the strange case of TeV J2032+4130} \label{sect:j2032}

\paragraph{Scientific Case}  \tev2032 \ is an extended source, located in the Cygnus OB2 stellar association ($d$\,=\,1.7 kpc). 
It was discovered as a steady extended TeV source by HEGRA \citep{aharonian02}, and later detected  by Whipple \citep{lang04}, Milagro \citep{abdo07}, MAGIC \citep{albert08}, VERITAS \citep{weinstein09}, ARGO-YBJ \citep{bartoli12}, and HAWC \citep{2020ApJ...905...76A}. It partially overlaps with the position of the binary pulsar PSR J2032+4127 \citep[a young and energetic NS with a period of $P = 143$ ms,][]{abdo09}, that likely powers the PWN. The companion star, MT91 213, is a B0Ve star with a mass of $M_{\star} = 15 M_{\odot}$. During the last periastron passage, which occurred on 2017 November 13, the X-ray flux increased by a factor of $\sim$\,20 compared to 2010 and $\sim$\,70 compared to 2002 \citep{ho17}. The GeV emission from the system, as detected by \lat, was found to be almost steady during the periastron passage, possibly because the interaction of the binary members is hidden behind the dominant magnetospheric emission from the pulsar \citep{li18}. MAGIC and VERITAS observed the system during the periastron 
\citep{abeysekara18b}, discovering significant excess point-like emission from the location of the pulsar/Be-star system, at the boundary of the extended emission from \tev2032 . The MAGIC and VERITAS observations led to independent spectral analyses of the emission before the periastron (``baseline'', pre-2017) and during and immediately after the periastron. The results show a simple power-law emission, possibly related to the persistent and steady PWN, far from the periastron. For the periastron phase the spectra were fitted adding another component  to the PWN model, possibly related to the interaction between the pulsar wind and the Be star wind. The best-fitting spectral shape of this new component was found to have an exponential cut-off around 1 TeV, for both MAGIC and VERITAS datasets. 

The observations of \tev2032 \ with the \astrima\ will provide crucial information on the VHE \gray\ emission of this extended source, especially when the pulsar is far from the periastron and the VHE emission could be ascribed entirely to the PWN. The observation above 10 TeV will shed new light on the emission spectrum, constraining the presence and nature of a possible cut-off: this could reflect the maximum energy of the  accelerated particles in the nebula and/or the onset of the Klein-Nishina regime for the IC cross-section. The source exhibits a very hard spectrum, and a cut-off at energy of tens of TeV is expected in a PWN scenario \citep{aliu14b}. Moreover, the spectral shape at the highest energy could provide important constraints on the emission models for this source, and help discriminate the nature of the VHE \gray \ emission. In addition to an in-depth  analysis of the source spectrum, a continuous monitoring of \tev2032 \ over the years, along its eccentric orbit, could put constraints  on the variability of the VHE \gray \ flux far from the periastron. 

\paragraph{Feasibility and Simulations}
According to existing TeV observations, the extended emission from this source represents one of brightest VHE \gray \ signals from the Cygnus region at energies above 1 TeV \citep{popkow15}. Nevertheless, in order to properly investigate the angular and spectral characteristics of the source, long exposure times are still required (see Table~\ref{tab:sources_list} for its maximum visibility from Teide).

We carried out our simulations with \textsc{ctools} with 200 hours exposure. 
We assumed as template spectral and spatial model for \tev2032 \ in our simulations the MAGIC best-fit model \citep{abeysekara18b}. 
The spectral shape is a simple power law with a photon index $\Gamma$\,=\,2.06, a normalisation $N_0 = 2.63 \times 10^{-20}$ photons cm$^{-2}$ s$^{-1}$ MeV$^{-1}$ at a pivot energy $E_0 = 3.5$ TeV.  For the morphology, we assumed an elliptical disk morphology centred at (R.A., Dec) = (307$^{\circ}$.92,  41$^{\circ}$.57), major semi-axis = $0^{\circ}.125$,  minor semi-axis = $0^{\circ}.080$, and position angle = 326$^{\circ}$ (counterclockwise from North).

\paragraph{Analysis and Results}

\begin{figure}
\centering
\includegraphics[width=\columnwidth]{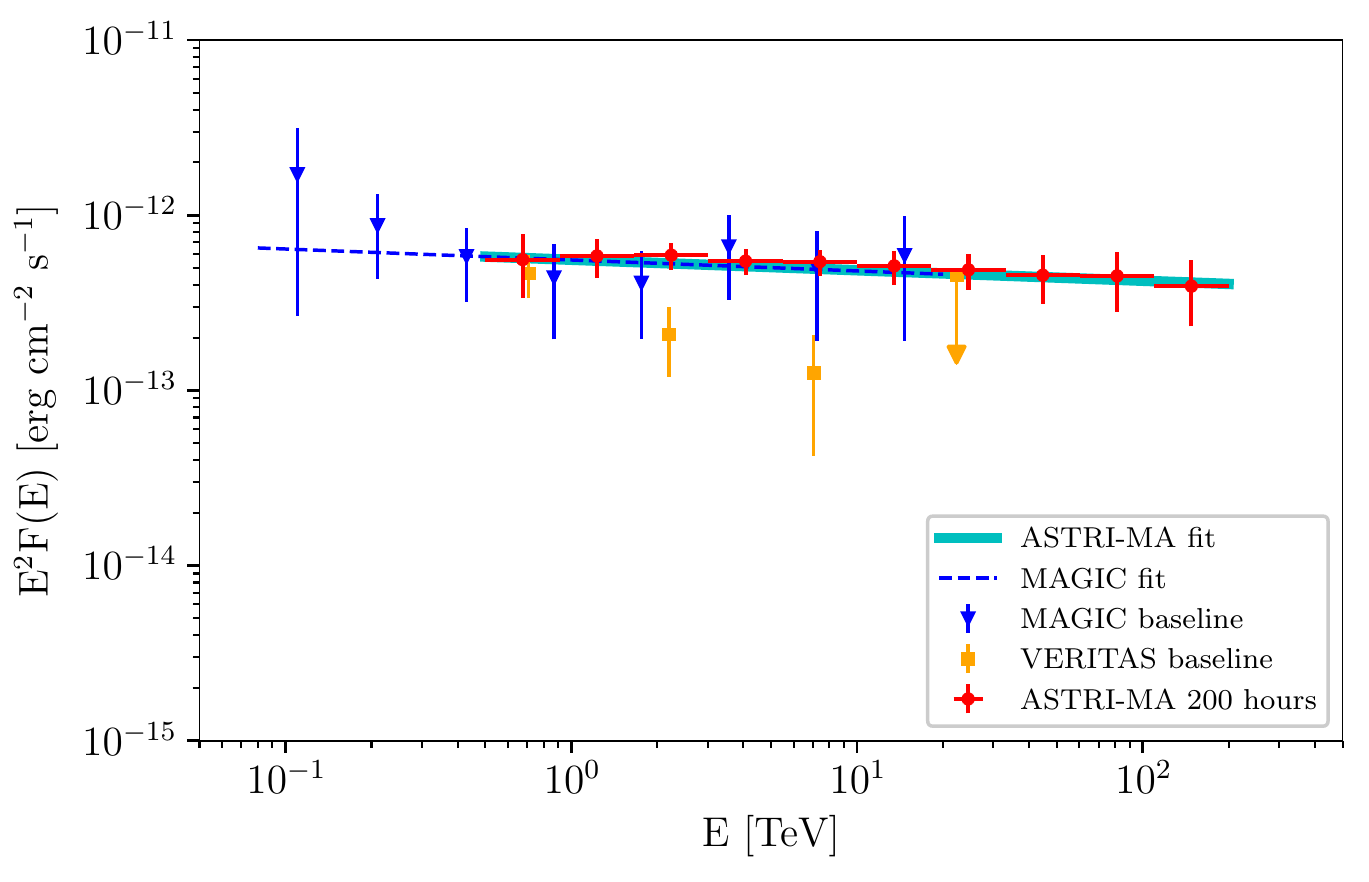} 
    \caption{SED for TeV J2032+4130. Baseline spectrum, as observed by MAGIC (blue triangles and blue dotted curve) and VERITAS (yellow squares) arrays \citep{abeysekara18b}. Spectral points from a 200 hr \astrima \ simulation (values averaged over 100 independent realisations) are shown in red and best-fit curve in cyan.}
    \label{fig:j2032_spectrum}
\end{figure}

We performed an unbinned analysis in the energy band 0.5--200 TeV for 100 independent realisations. 
We left  both the spectral and morphological parameters free to vary, and, using the template model, we found an average TS\,=\,362 $\pm$ 47, which corresponds to a detection significance of  (19 $\pm$ 7) $\sigma$. The best-fitting mean values for the spectral parameters are: $\Gamma = 2.06 \pm 0.09$, $N_0 = (2.6 \pm 0.2) \times 10^{-20}$ photons cm$^{-2}$ s$^{-1}$ MeV$^{-1}$. The mean morphological parameters for the extended source are: R.A. = $307^{\circ}.919 \pm 0^{\circ}.009$ , Dec. = $41^{\circ}.572 \pm 0^{\circ}.003$, major semi-axis = $0^{\circ}.082 \pm 0^{\circ}.008$,  minor semi-axis = $0^{\circ}.078  \pm 0^{\circ}.008$, while the position angle was kept fixed in the analysis. In Fig.~\ref{fig:j2032_spectrum} we show the resulting spectral points, obtained dividing the whole energy range in 10 logarithmically-spaced bins. The \astrima\  spectral bins are shown together with real data obtained by the MAGIC and VERITAS Collaborations \citet{abeysekara18b}.

\begin{figure}
\centering
\includegraphics[width=\columnwidth]{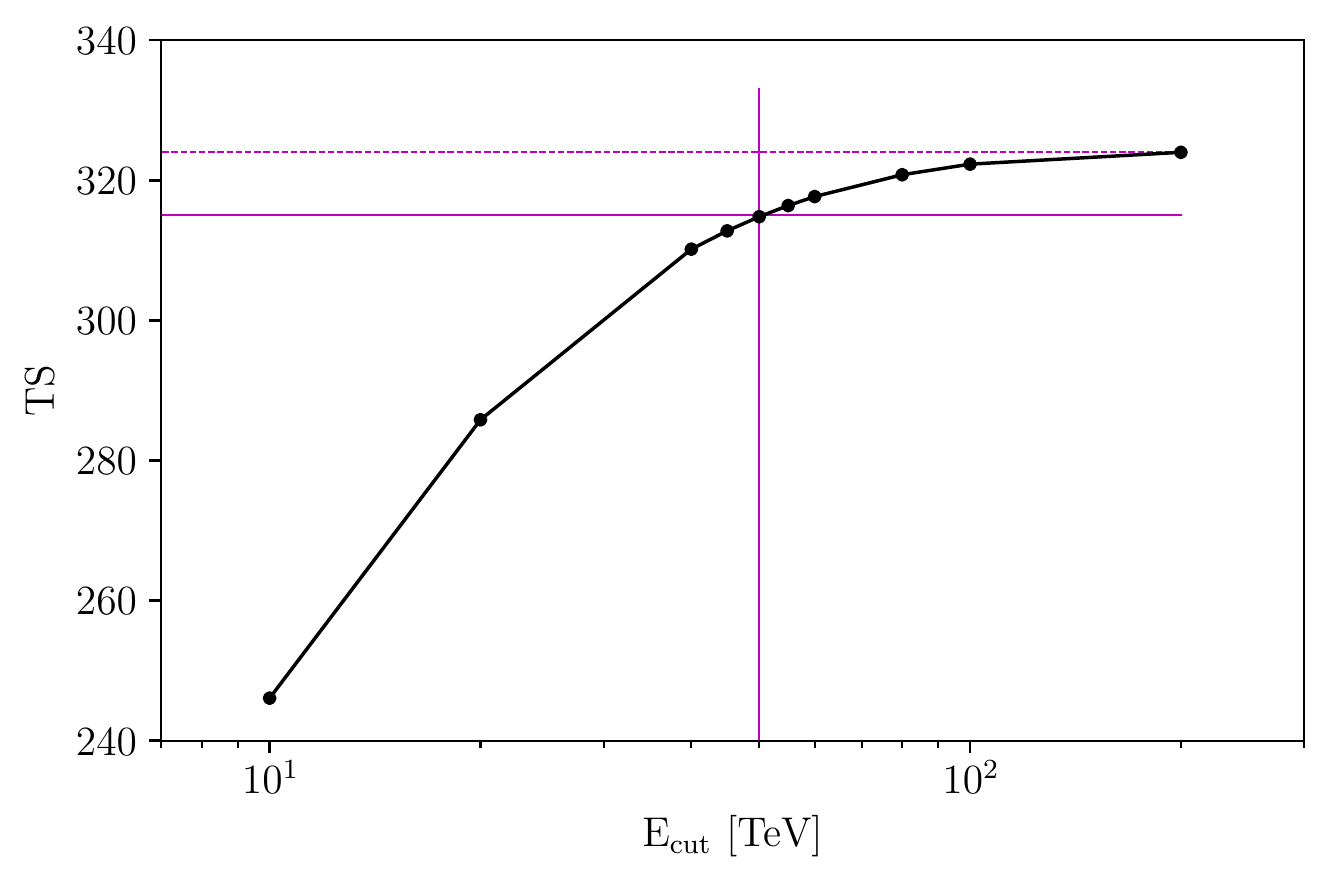} 
    \caption{TS versus $E_{\textrm{cut}}$ for a given simulation. The magenta horizontal dashed line marks the maximum TS value for the simulation (in this case, TS$_{\textrm{max}}$\,=\,324). The magenta horizontal solid line indicates the value of  TS$_{\textrm{max}}$ - 9. The vertical magenta line is the corresponding value of $E_{\textrm{cut}}$ (in this case, $E_{\textrm{cut}}$\,=\,50 TeV).}
    \label{fig:j2032_cutoff}
\end{figure}

We assumed that the emission has no spectral cut-off up to $\sim$\,200 TeV (the pulsar potential drop corresponds to $\sim$\,500 TeV). However, a spectrum with no cut-off appears unrealistic; therefore, to assess the capability of detecting a possible high-energy cut-off, we adopted the following method (see also Sect.~\ref{sect:ss433}). For a given spectrum generated using a power-law template model, we found the corresponding associated TS$_{\rm max}$; then, we fitted the same data with an exponential cut-off spectral model, where the $E_{\textrm{cut}}$ parameter is fixed on a grid of values, while the other parameters were left free to adjust to new values; by varying the $E_{\textrm{cut}}$ value (generally in the range 20--80 TeV), we kept note of each new TS value. As shown in Fig.~\ref{fig:j2032_cutoff}, we obtained the $E_{\textrm cut}$ vs. TS trend and the $\Delta$TS value corresponding to a decrease by 9 from its maximum. This provides a 3\,$\sigma$ upper limit on the possible lowest detectable energy cut-off given this exposure time.

We repeated this procedure for the 100 different realisations and computed the mean of this threshold value for the cut-off energy, obtaining $E_{\textrm{cut}} = (60 \pm 13)$ TeV, which is well above the expected limit usually observed for PWNe (see Sect.~\ref{sect:PWN}). Therefore it appears very likely that \astrima \ will be able to detect a possible curvature in the spectrum of this source. \\
Moreover, a monitoring of the source over the years will allow us to investigate a possible low-level flux instability of the VHE \gray \ flux along the orbit of the binary system. Repeated observations of the Cygnus field with the \astrima \ will allow a study of its flux variability and provide precious information about the nature of the pulsar-stellar wind interaction, possibly emerging over the PWN emission.

We finally note that recently the LHAASO Collaboration reported on a significant detection of VHE emission above 100 TeV from the direction of this source \citep{lhaaso21};  if the emission is of hadronic origin, as seems likely, the highest energy photons detected come from $\sim$\,15 PeV protons. The derived position of the LHAASO source is about half degree from TeV J2032+4130 (but still consistent with it). The authors propose the OB2 Cygnus association as the likely cradle for this highly energetic hadron population, but the region encompasses several potential accelerators.
Given the \astrima\ spatial resolution of $\sim$0.08\degmark\ above 100 TeV, with a deep exposure of the region we expect to be able to eventually disentangle the contributions from the different candidate sources (PSR, PWN and OB stars), thus confirming or rejecting such hypothesis.

%% file: sect_pulsars.tex
\section{TeV Pulsars}\label{sect:gammaraypsr}
At the centre of most PWNe, energetic pulsars produce \textit{pulsed} radiation in a broad energy range: from radio up to extremely energetic gamma-rays. Electrons and positrons accelerated to relativistic energies within the pulsar magnetosphere are believed to produce high-energy radiation \citep{Arons1983, Cheng1986}. To date, more than 200 \gray\ pulsars have been detected above 100 MeV with the \textit{Fermi}-LAT \citep{Abdo2013}, and significant emission from four of them has been recently observed with the currently operating IACTs \citep{Aliu2011, ansoldi16, HESS2018_Vela, SpirJacob2019arx, MAGIC2020_Geminga}. The Crab and Vela pulsars show the VHE gamma-ray emission up to TeV energies. Detection of pulsed emission in this energy range provides evidence for possible acceleration even outside the NS magnetosphere \citep{2012Natur.482..507A}.

\begin{figure}
	\centering
	\includegraphics[width=0.49\textwidth]{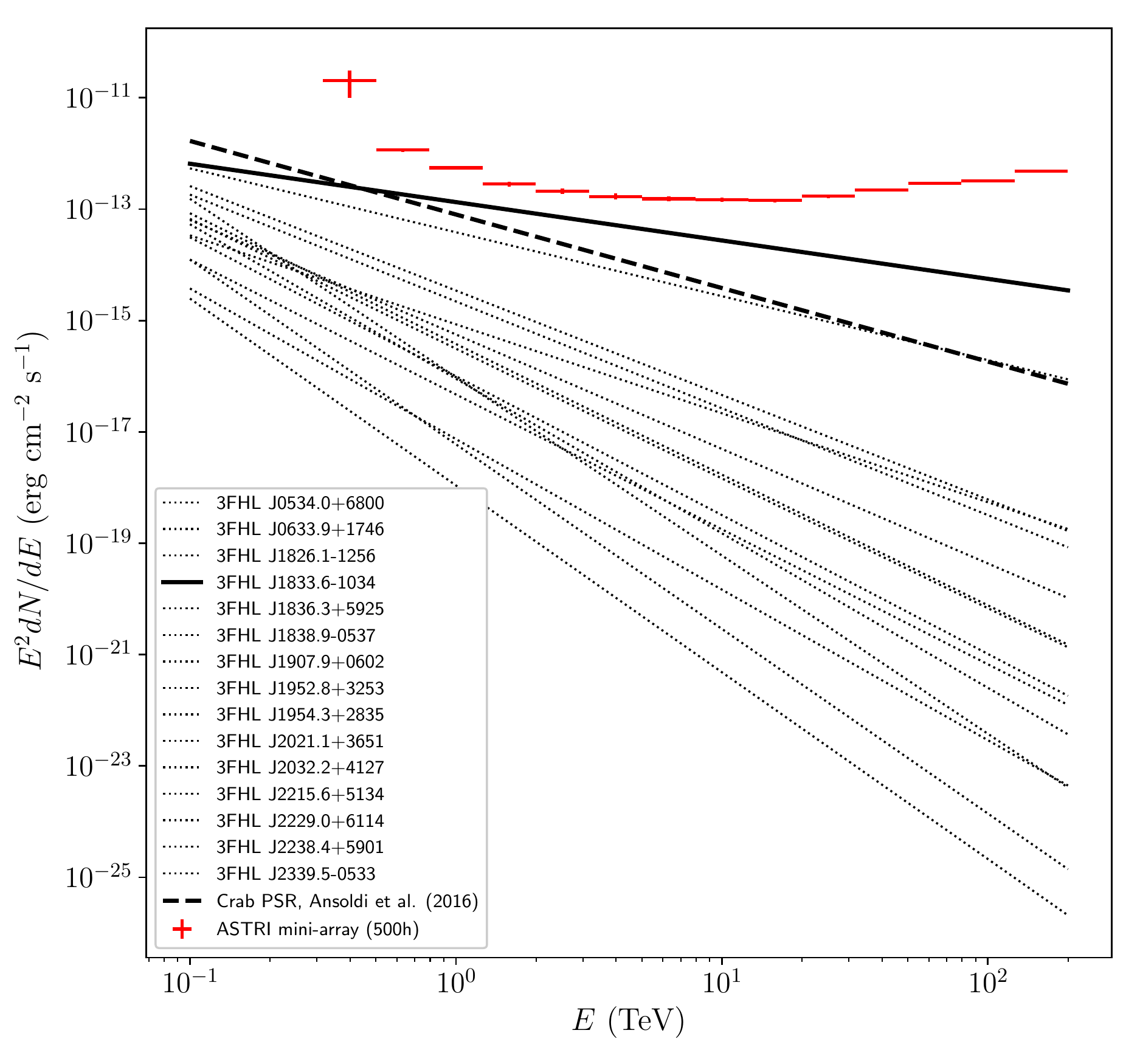} \\
	\caption{Spectral energy distribution of pulsars observable from the northern hemisphere over-imposed to the \astrima\ 500 h sensitivity curve. Pulsar spectra are taken from the Third Catalogue of Hard \textit{Fermi}-LAT Sources (3FHL) and extrapolated up to 200 TeV.}
	\label{fig:sed_pulsars}
\end{figure}

In this section, we explore and discuss the prospects for the \astrima\ to detect pulsed emission from the most promising \gray\ pulsars visible from Teide. 
\paragraph{Scientific Case}
The detection of pulsations from the Crab pulsar at TeV energies \citep{ansoldi16} strongly suggests IC scattering as the main emission mechanism. In this interpretation, low-energy photons are upscattered by the relativistic magnetospheric electrons and positrons in regions close to the light cylinder of the pulsar. The VHE spectrum has a power-law shape with a spectral index of about 3.3, which can be smoothly connected with the spectrum measured above 10 GeV with the \textit{Fermi}-LAT. 
Recently detected TeV pulsations from the Vela pulsar most probably correspond to a different spectral component of unknown origin. The detailed properties of this component are yet to be published.
The detection of pulsed emission  above 1 TeV from pulsars  will allow us to determine the maximum energy of the accelerated electrons and positrons, and to  also put constraints to the geometry of the emission region and to the physical mechanism producing pulses at VHE. Additionally, new discoveries will clarify whether the VHE components discovered so far in the Crab and Vela pulsars are common also to other \gray\ pulsars \citep[see e.g.][]{burtovoi17}.
\paragraph{Feasibility and simulations}
To investigate the prospects for the detection of $\gamma$-ray pulsars with the \astrima, we selected pulsars from the Third Catalogue of Hard \textit{Fermi}-LAT Sources \citep[3FHL,][]{Ajello2017} visible from the Northern hemisphere with the highest energy events probably coming from the source $E_{\rm pulsar}>25$ GeV (for details see 3FHL). Assuming a Crab-like VHE spectral component, we extrapolated 3FHL spectra\footnote{For the Crab Pulsar we extrapolated the spectrum obtained with MAGIC \citep{ansoldi16}.} up to 200 TeV and compared them with the \astrima\ 500 h sensitivity curve. As shown in Fig.~\ref{fig:sed_pulsars}, even in this very optimistic case when there is no cut-off in the pulsars spectra up to 200 TeV, 500 hr are not sufficient for a statistically significant detection of pulsars. 
We estimated the minimum exposure time necessary for the significant detection of the most favourable candidate targets, which have the hard spectra and high gamma-ray flux: PSR 3FHL J1833.6-1034 ($\Gamma=2.68$) and the Crab pulsar ($\Gamma=3.32$).
To this aim, we carried out a number of simulations\footnote{The simulations were repeated 100 times in order to take in account the statistical fluctuations of the simulated data.} with the \ctools, considering the energy range 0.9--199 TeV. Both pulsars were simulated as a point-like source which has a power-law spectra taken from the 3FHL catalogue -- for PSR 3FHL J1833.6-1034, and from \citet{ansoldi16} -- for the Crab pulsar. As a background we considered the contribution from the cosmic-rays and from the surrounding pulsar wind nebulae (HESS J1833-105, \citealt{HGPS2018}, and the Crab nebula, \citealt{Aleksic_2015}, respectively).
Performing the maximum likelihood analysis in the binned mode, we fit the simulated emission from the pulsar with a power-law function and obtained its detection significance. A 5-sigma detection of the pulsar 3FHL J1833.6-1034 above $\sim$1 TeV would be obtained in $\sim$\,2000 h.
In order to improve the detection sensitivity, we carried out similar simulations with a reduced background contamination to mimic the results expected from an analysis of the sole on-peak phase intervals\footnote{Phase intervals which correspond to prominent and sharp emission peak(s).}. With this, it is possible to take into account different duty cycles of the pulsars, normalising the background contribution to the on-peak phase interval. The duty cycle of the pulsar 3FHL J1833.6-1034 is taken to be equal to 0.35 \citep{Kerr2015}. In this case, the resulting time required for the 5-sigma detection would be $\sim$\,700 h. 
We note that PSR 3FHL J1833.6-1034 (RA\,=\,278.41, Dec.\,=\,$-$10.57) is visible from Teide for about 220 hr/yr for a zenith angle range 0--45 deg (in moonless conditions). A similar amount of hours is available for the higher zenith angle range 45--60 deg. Finally, we obtained that the Crab pulsar is not expected to be detectable with \astrima\ even in $\sim$\,10000 h.

%% file: sect_gammabinaries.tex
\section{\bf \gray\ Binaries} \label{sect:binaries}
Unlike X-ray and radio emitting binaries, \gray\ binaries are a small  group of sources \citep[known to date only nine persistent sources, see][]{chernyakova19} composed of a giant OB star and a compact object. Their essential characteristic is a peak in their spectral energy distribution above 1 MeV. Only for three sources is the compact object firmly identified as a NS thanks to the detection of their pulsed emission: PSR B1259-63, PSR J2032+4127 and, only recently, for LS I 61+303 \citep{weng22}., In the remaining sources the nature of the compact object is still debated. Besides the TeV emitting binaries mentioned above, there are also three microquasars (binaries with accretion and jet emission), that show $\gamma$-ray emission up to tens of GeV (Cyg X-1, Cyg X-3, and SS\,443). In the case of SS\,433, additional TeV emission has been detected from the interaction regions between the jet and the surrounding nebula, far from the central microquasar, which implies an emission scenario different from the rest of the sources \citep{abeysekara18}. Establishing the nature of the compact object is important, as different mechanisms can be envisaged for the VHE emission depending on the nature of the compact object. In the case of a NS, the wind of the pulsar interacts with the wind of the companion, producing two different shock regions. Particles at the shock fronts can be re-accelerated and transfer part of their kinetic energy to background photons via IC. If the compact object is a black-hole, some authors also proposed an extended microquasar scenario for the whole class of the \gray\ binaries. In this case, some mass transfer should be taking place and a jet should be present at the site for photon IC scattering \citep{dubus13}.
It is however extremely difficult to build a homogeneous scenario for these systems, because the emitting processes (be they leptonic or hadronic) are heavily modulated by several additional and variable  aspects, such as e.g. the absorption of TeV photons by the companion photon field, or the system line of sight and inclination. Indeed, there are still unexplained discrepancies among the sources as far as their X-ray, GeV, and TeV emission and correlations are concerned \citep{dubus13}.
We chose to focus for an in-depth discussion on the microquasar source SS\,433 in Sect.~\ref{sect:ss433}.

The spectra of all the $\gamma$-ray binaries extend to the TeV range, although the spectral shape and the intensity of this emission is strongly correlated with the orbital phase (orbital periods vary from days to years) or with super-orbital modulations \citep[as in the case of the  binary LS I +61$^{\circ}$\,303,][]{ahnen16}. Five known $\gamma$-ray binaries can be easily observed from Teide and they are listed in Table~\ref{tab:sources_list}. In these sources a key goal is the study of their variability; we will show in Sect.~\ref{sect:ls5039} how, in the case of LS\,5039 (which has an orbital period of 3.9 days), the \astrima\ at this level of luminosity can trace with great accuracy its VHE modulation, and at the same time possible spectral changes \citep[see][for the spectral changes at different orbital phases]{pintore20}.  A peculiar and interesting case among the binaries is the source TeV J2032+4130, where the  compact object has recently displayed a flare in TeV emission when the source passed its periastron. Out of this phase the emission seems dominated by the pulsar's wind inflated nebula, so that the object displays characteristics more commonly attributed to the PWN class.
Given the next periastron passage will be around 2067, we studied this source in the context of PWNe as reported in Sect.~\ref{sect:j2032}. 

%% file: sect_ss433.tex
\subsection{VHE emission from Galactic microquasars: SS\,433} \label{sect:ss433}
\paragraph{Scientific Case}
SS\,433 is one of the most peculiar Galactic binary systems currently known and the prototype of micro-quasars \citep{1999ARA&A..37..409M}. It contains a supergiant star that provides super-critical accretion onto a compact object (neutron star or black hole) via Roche lobe overflow. It is extremely bright ($L_{\rm bol}$ $\sim$\,10$^{40}$ \ergsec, at 5.5 kpc) with the most powerful jets known in our Galaxy ($L_{\rm jet}$ $>$10$^{39}$ \ergsec, bulk velocity $\sim$\,0.26\,c).
The two jets, perpendicular to the line of sight, terminate inside the radio shell of W\,50, a large 2\degmark\,$\times$\,1\degmark\ nebula catalogued as SNR G39.7$-$2.0. The jets precess every 162.4\,d \citep{1989ApJ...347..448M}, while the orbital period of the system is 13.1\,d. 
Several X-ray hot-spots located west (\textit{w1}, \textit{w2}) and east (\textit{e1}, \textit{e2}, \textit{e3}) of the central binary are observed where the jets interact with the ambient medium \citep{safi97}. Radio lobes are also observed, east and west of the nebula. Radio to soft \gray\ photons are believed to be due to synchrotron emission from the relativistic electrons in the jets. 
\citet{2015ApJ...807L...8B} reported for the first time GeV emission tentatively associated to SS\,433/W\,50 using five years of \fermi-LAT data. More recent \fermi\ results (using ten years of data) reported GeV emission around the w1 region \citep[but not in the east region,][]{2019ApJ...872...25X}, while
\citet{2019A&A...626A.113S}, also using ten years of \fermi\ data, reported a spatially extended emission consistent with the extent of the nebula W\,50 (suggesting that the GeV emission may originate from the SNR). \citet{2019MNRAS.485.2970R} reported the first evidence for modulation at the jets precession period of 162.4\,d (but not at the orbital period) in \fermi\ data. 

More recently, based on accurate  background determination \citet{li20} was able to definitively confirm the w1 emission from the $w1$ region and the  associated precessional variability. 

No evidence of VHE \gray\ emission was found either from the central binary system or from the eastern and western interaction regions using MAGIC and \hess\ observations \citep{2018A&A...612A..14M}. \citet{2018Natur.564E..38A} reported VHE emission ($\sim$20\,TeV) from the eastern/western regions using HAWC. They characterised the \textit{e1}  and \textit{w1} emission as point-like with an upper limit on the size of 0.25\degmark\ and  0.35\degmark\ (90\% confidence), respectively (see Fig.1 of their paper).

The authors propose that the broad band (radio to TeV) emission of SS\,433/W\,50 is consistent with a single population of electrons with energies extending to at least 100\,TeV in a magnetic field of about 16 $\mu$G (synchrotron and IC scattering of ambient photon fields).
More recently, the third HAWC Catalog of VHE Gamma-ray Sources \citep{2020ApJ...905...76A} reported an updated spectrum of the \textit{e1} lobe as a power-law of index $\Gamma$\,=\,2.37 and a 7\,TeV prefactor $N$ = 6.8\,$\times$\,10$^{-15}$ $\phcmsectev$ in the 5.9--144 TeV energy range. 

SS\,433 is a unique laboratory to spatially resolve  particle acceleration in jets. Furthermore, sources with relativistic Galactic jets could contribute to the cosmic ray spectrum \citep{2002A&A...390..751H}. Indeed, it appears that SNR may not accelerate cosmic rays up to the knee of the spectrum \citep{2018AdSpR..62.2731A}. The highest energy of accelerated particles is limited by the requirements that the size of the accelerator must be larger than their Larmor radius \citep{2013A&ARv..21...70B}; hence it is important to constrain the size of the TeV emitting region.
Observations with the \astrima\  will address these two fundamental issues:
\begin{itemize} 
\item obtaining a meaningful broad-band (0.7--200 TeV) spectrum of the eastern source lobe, with the aim of constraining its slope and energy cut-off; 
\item investigating the radial extension of the TeV lobe emission. 
\end{itemize}

\paragraph{Feasibility and Simulations}
SS\,433 position is 1.4\degmark\ from the PeVatron candidate eHWC J1907$+$063 \citep[see Sect. 4.1.3 in][]{vercellone21} and the \astrima\ large field of view would allow a simultaneous investigation of these two important sources at once. 
The \textit{e1} and \textit{w1} regions are located at about 2\degmark\ from the Galactic Plane ($l$\,=\,-2.7\degmark\ and $b$\,=\,-2\degmark, respectively), so the contamination from the Galactic diffuse emission is expected to be negligible, especially in the \astrima\ energy range. The following results are presented for the East \textit{e1} region alone, since its position makes it also less contaminated by the nearby extended source MGRO J1908$+$06. The latter has an estimated size $<$0.44\degmark\ \citep{2014ApJ...787..166A}, it is 1.75\degmark\ away and undetected at the \textit{e1} position.\\
We considered the spectral model based on the 3HAWC publication: a power-law of index $\Gamma$\,=\,2.37 and a prefactor $N$ = 6.8\,$\times$\,10$^{-15}$ $\phcmsectev$ at 7\,TeV (that corresponds to a 0.7--200 TeV flux of about 3\,$\times$\,10$^{-12}$ $\ergcmsec$).  
We have performed 50 simulations (of 100\,h exposure each) for two configurations: point-like emission (as considered by HAWC) with radius of 0.001\degmark\ and extended emission. In the latter case, we have used a Gaussian model with width compatible with the maximum value allowed by HAWC: 0.25\degmark\ in size – 0.125\degmark\ radius – 90\% confidence.

\paragraph{Analysis and Results}  \mbox{} \\
\textbf{The VHE spectrum of the source}\\
In this part of the work, aimed at studying the spectrum, the spatial parameters (position and extension) have been kept frozen to the simulated values during the analysis. The final TS and best-fit values of the spectral slope, prefactor and flux are given in Table~\ref{tab:7.4.2.1} (columns 1 to 6). The final values are obtained as average from the simulations in which the source was detected (i.e. all of them in both point-like and extended configuration); the associated errors are one standard deviation of the distribution and are compatible with the 1$\sigma$ uncertainties in each realisation as given by the likelihood (\textsc{ctlike} in \ctools). 

\begin{figure*}
	\centering
	\includegraphics[width=0.43\textwidth]{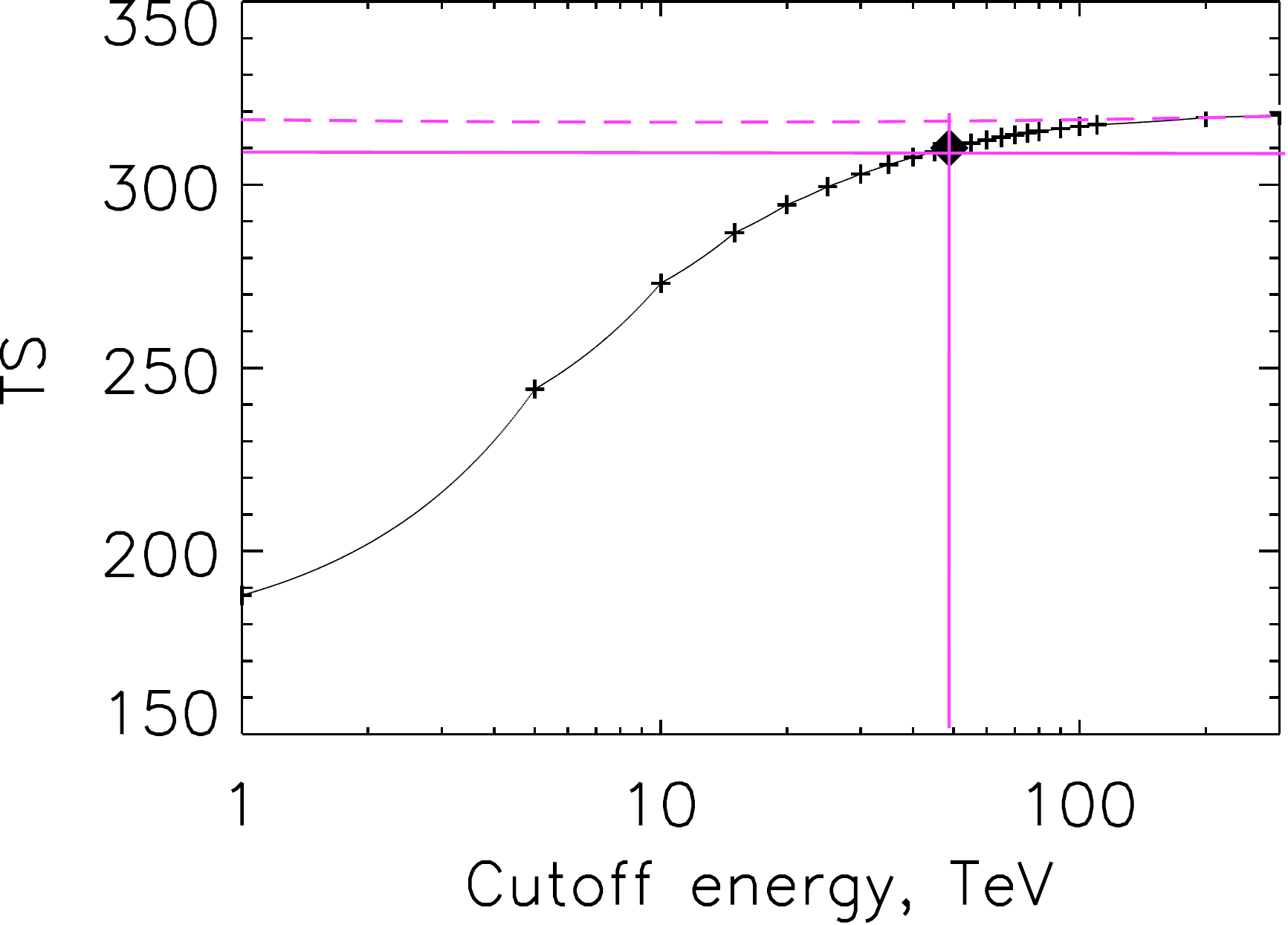}
	\includegraphics[width=0.41\textwidth]{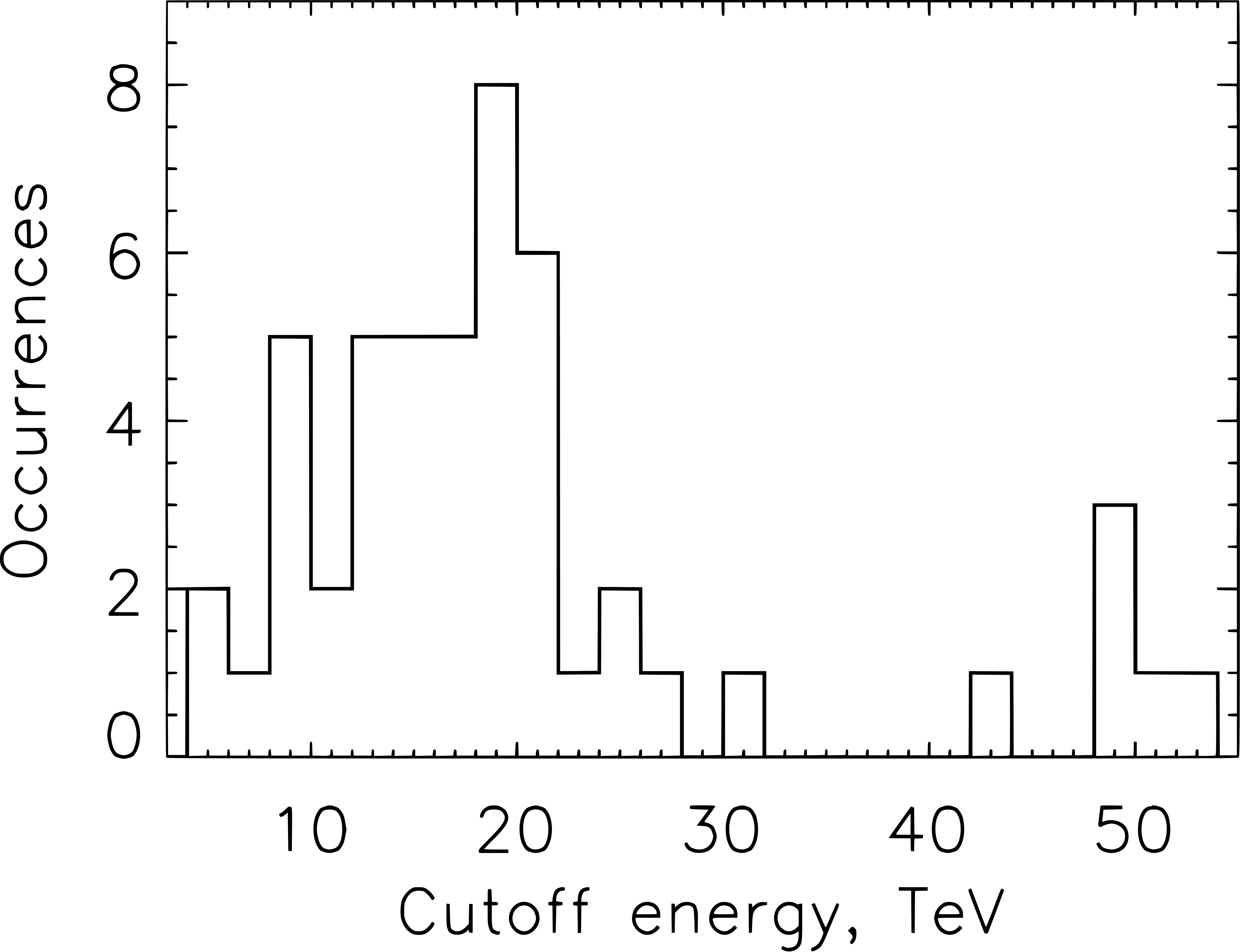}
	\caption{Cut-off energy investigation in SS\,433 (simulated as point-like). {\bf Left:} 
	TS versus high energy cut-off for a given simulation. The magenta horizontal dashed line marks the maximum TS value for the simulation. The magenta horizontal solid line indicates the value of  TS$_{\textrm{max}}$ - 9. The vertical magenta line is the corresponding value of $E_{\textrm{cut}}$ (in this case, $E_{\textrm{cut}}$\,=\,50 TeV). 
	{\bf Right:} distribution of all the 50 lower limits obtained for E$_{\rm cut}$ considering a TS decreases of 9 (i.e. 3$\sigma$) from its maximum (one per simulation). In 90\% of the cases, E$_{\rm cut}$ below 10\,TeV is excluded at a 3$\sigma$ level. This value is given as result in Table~\ref{tab:7.4.2.1}. }
	\label{FIG:Chap7_Fig02}
\end{figure*}

To explore possible constraints on the high-energy cut-off ($E_{\rm cut}$), we adopted the same approach outlined in Sect.\ref{sect:j2032}.
This resulted in plots like the one shown in Fig.~\ref{FIG:Chap7_Fig02} (left panel) where the trend of the TS with respect to $E_{\rm cut}$ is obtained (the plot relates to a single simulation). The grid of $E_{\rm cut}$ values used is shown with crosses, while the filled diamond symbol shows the level at which $\Delta$TS\,=\,9 from its maximum, (thus, roughly providing the relative 3$\sigma$ limit). The final values given in Table~\ref{tab:7.4.2.1}, column 7, are obtained as the ones containing 90\% of the results (see Fig.~\ref{FIG:Chap7_Fig02}, right). 

\begin{table*}[width=2.0\linewidth,cols=8,pos=htp!]
\centering
\caption{The VHE spectrum of SS\,433: results for the \textit{e1} hot-spot (100\,hr per configuration). SS\,433\,\textit{e1} has always been detected in both configurations. See text for details. \\ Notes: (a) Simulated sigma of the \emph{RadialGaussian} model in \ctools. (b) At 7\,TeV in units of 10$^{-15}$ ph cm$^{-2}$ s$^{-1}$ TeV$^{-1}$ (c) 10$^{-12}$ $\ergcmsec$ in 0.7-190\,TeV (d) Fixed in the analysis.}
\label{tab:7.4.2.1}
 \begin{tabular*}{\tblwidth}{@{}CCCCCCCC@{}}
\toprule
Config. & Simul. Radius (a) &  TS & Gamma & Prefactor (b) & Flux (c)   & $E_{\rm cut}$ & Reconst. Radius\\
 & [deg] &  &  &  &  & [TeV] & [deg]\\
\midrule
A1  & 0.001 & 321$\pm$43 & 2.4$\pm$0.1 &  6.3$\pm$0.5 & 2.8$\pm$0.2 & $>10$  & $<0.04$\\
   &  (d) & & & & & (90\% cases) & (90\% cases)\\ 
\\ 
A2 & 0.125  &  61$\pm$15 & 2.3$\pm$0.1 &  6.7$\pm$1.0 & 3.0$\pm$0.4 & $>2$  & $0.05<R<0.24$\\  
  &(d) &  & & & & (90\% cases) & (90\% cases) \\ 
\bottomrule
\end{tabular*}
\end{table*}

\begin{figure}
	\centering
	\includegraphics[scale=.49]{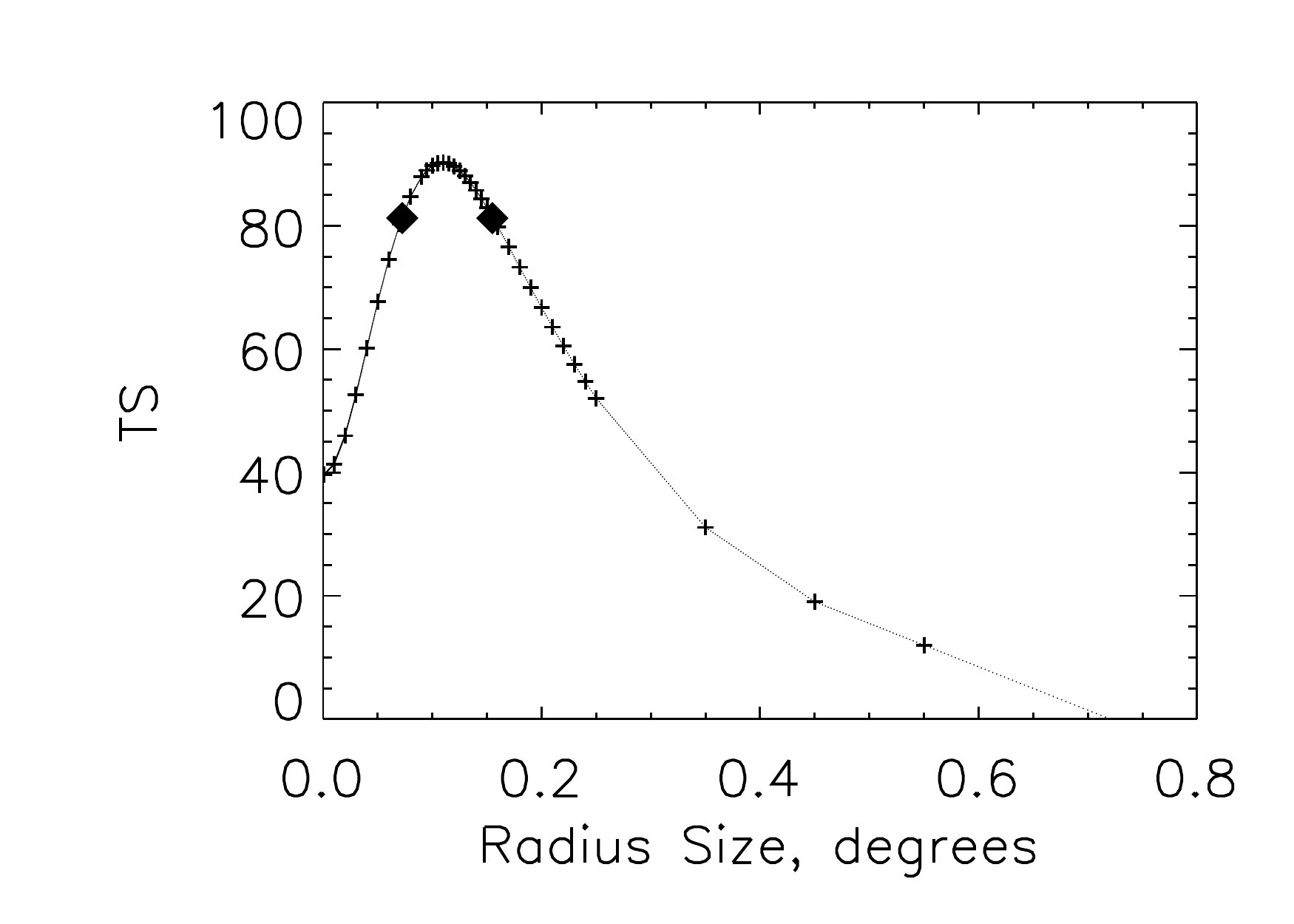}
	\caption{Example of a radial result for SS\,433. The source has been simulated with a 0.125\degmark\ radius and then fit with a grid of different fixed sizes (shown with '+'), each resulting in its own TS. A 3\,$\sigma$ (i.e. decrease of 9 from the TS maximum value) lower and upper limit on the source Gaussian radius are obtained in this simulation.}
	\label{FIG:Chap7_Fig03}
\end{figure}

\begin{figure*}
	\centering
	\includegraphics[scale=.5]{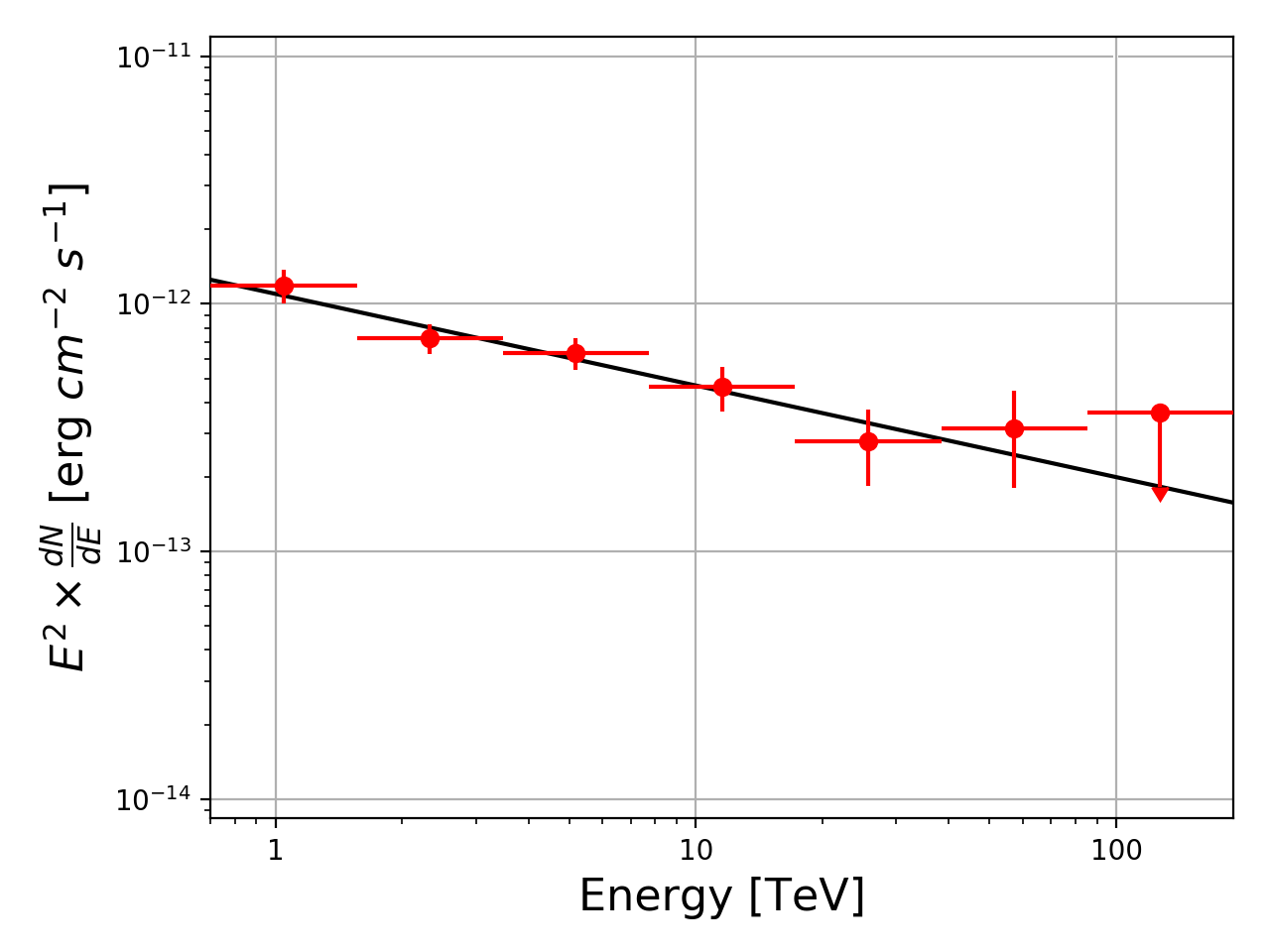}
	\includegraphics[scale=.5]{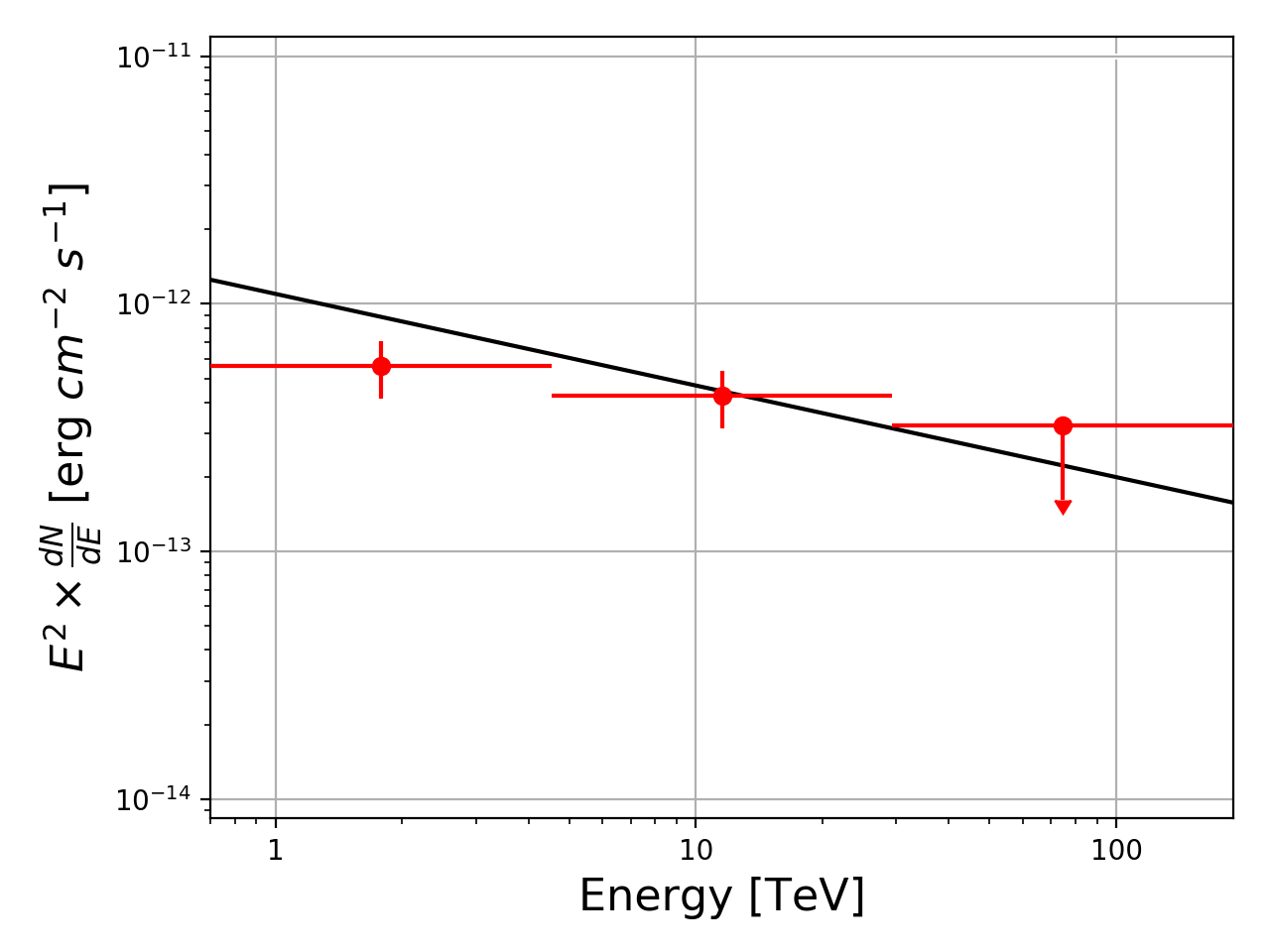}
	\caption{Examples of simulated spectra of the East \textit{e1} hot-spot of SS\,433 (100\,hr). Data points are at a 3\,$\sigma$ threshold.\\
 {\bf Left:} spectrum assuming source morphology as point-like.  {\bf Right:} assuming extended source morphology (0.25\degmark\ size). Best-fit power-law models shown as thick black lines.}
	\label{FIG:Chap7_Fig04}
\end{figure*}
\textbf{The radial extension of the VHE lobe emission}\\   
In the second part of our work, we investigated the extension of SS\,433\,\textit{e1}. The starting point is the same as for the former analysis: the 50 simulations with the power-law spectrum already described and two spatial configurations, point-like (R\,=\,0.001\degmark) and extended (R\,=\,0.125\degmark). In this case, no energy cut-off has been included in the fitting spectral model and the approach used is similar to what described above for the $E_{\rm cut}$ investigation: for each simulation, a grid of different fixed radii was used, each resulting in its own TS. 
Fig.~\ref{FIG:Chap7_Fig03} shows an example of the trend of TS with respect to the source extension when simulated as extended: a 3$\sigma$ lower and upper limit on the source Gaussian radius are obtained, i.e. indeed the simulated source is detected as extended.  Table~\ref{tab:7.4.2.1}, last column, shows the results obtained. 

Fig.~\ref{FIG:Chap7_Fig04} shows two realisations of the \astrima\ spectra obtained for SS\,433\,\textit{e1} in a 100\,hr simulation.\\ 

In summary, if the TeV extension of SS\,433\,\textit{e1} is below the \astrima\ angular resolution (configuration A1 in Table \ref{tab:7.4.2.1}), then a 100\,hr observation will allow us to study its spectrum, with particular attention to the spectral attenuation at higher energies that has not been investigated yet. Indeed, the source is consistently detected in the 0.7-200\,TeV energy range with a $\sim$\,4\% error on the spectral slope(1\,$\sigma$), and a 3\,$\sigma$ lower limit of $E_{\rm cut}$ $\sim$10\,TeV. The source will be reconstructed as point-like. 

In the case of an extended emission (A2, R\,=\,0.125\degmark), SS\,433\,\textit{e1} is characterised with a $\sim$\,4\% error on the spectral slope (1\,$\sigma$), and a 3\,$\sigma$ lower limit of $E_{\rm cut}\sim$2\,TeV. The source is indeed reconstructed as extended, allowing for larger radii than the instrumental angular sensitivity (0.05\,$<$ R $<$\,0.24\degmark, example shown in Fig.\ref{FIG:Chap7_Fig03} for one particular case). 

The \astrima\ will allow us to investigate the spectrum and source extension of SS\,433\,\textit{e1} in a 100\,hr observation, constraining the physical properties and size of the TeV emitting region, where the most energetic cosmic rays are accelerated.   

%% file: sect_ls5039.tex
\subsection{Periodic variability in the TeV range: the $\gamma$-ray binary LS 5039} \label{sect:ls5039}
\paragraph{Scientific Case}
LS 5039 was discovered as an X-ray source in 1997 \citep{motch97}. An early detection in the radio band \citep{marti98} unveiled a system in which particle acceleration occurs. The nature of the compact object orbiting a massive O-type stellar companion is still unknown, and currently the young non-accreting pulsar scenario (similarly to PSR~B1259$-$63 and PSR~J2032$+$4127) is preferred over the microquasar (i.e. black hole) one \citep[][and references therein]{dubus13}. Very recently, this scenario was reinforced by the results obtained by \citet{Yoneda+20}, who analyzed the HXD data of a \textit{Suzaku} observation performed in 2007 and revealed a pulsation with period $P \simeq$ 8.96 s; moreover, they also detected  a potential pulsation at $P \simeq$ 9.05 s in the data collected with \textit{NuSTAR} during an observation performed in 2016. The difference between the two periods would imply a fast spin-down of the pulsar, and suggests a magnetar nature for the compact object in LS 5039. However, a new analysis of the same data performed by \citet{Volkov+21} resulted in a very low statistical significance of this periodic signal, which casts doubts on its firm detection. This means that, if the compact object in LS 5039 is indeed a young neutron star, the intrabinary shock emission dominates the X-ray pulsations. Since the X-ray spectrum of this source can be described with a simple power-law model, without any hint of spectral lines or exponential cut off up to 70 keV, the candidate neutron star should not be accreting. The alternative scenario of black hole is less likely, due to the lack of flux variability and Fe K$\alpha$ lines.

The first very high energy ($>$\,0.1\,TeV) detection has been reported by \cite{aharonian05_binary} using \hess. Further \hess\ observations \citep{aharonian06_ls5039} revealed the first orbital modulation at TeV energies with a period of $\sim$\,3.9 days, the shortest currently known among $\gamma$-ray binaries \citep[e.g.][and references therein]{chernyakova19}. Using \hess\ data \citet{aharonian06_ls5039} showed that LS~5039 presented two different spectral states at TeV energies: a high-state in the orbital phase range 0.45--0.9 (luminosity L$_{0.2-10\rm{TeV}}$\,=\,1.1$\times$10$^{34}$~erg~s$^{-1}$, assuming a distance of 2.5 kpc), and a low-state for the remaining part of the orbit (L$_{0.2-10 \rm{TeV}}$ = 4.2\,$\times$\,10$^{33}$ erg\,s$^{-1}$). The high-state spectrum can be well described with a power-law model with exponential cut-off: $\Gamma$\,=\,1.85$\pm$0.06$_{\mathrm{stat}}$\,$\pm$\,0.1$_{\mathrm{syst}}$ and $E_{\rm cut}$\,=\,8.7\,$\pm$\,2.0 TeV. On the contrary, the low-state is consistent with a power-law ($\Gamma=2.53\pm0.07_{\mathrm{stat}} \pm 0.1_{\mathrm{syst}}$) without a cut off.

\paragraph{Feasibility and Simulations}
LS 5039 is located at ($l$, $b$) = (16.88, -1.29) and can be observed  from the Northern hemisphere at a minimum zenith angle of 43$^{\circ}$.
In the ACDC project we simulated a source observation with 300 h of total exposure, 250 of which in the dim low state and the remaining 50 in the high state \citep[for detailed results and considerations on the two average spectra we refer the reader to][]{pintore20}. We recall here that the \astrima\ can disentangle among the two spectral states in 300\,h: on one hand, the high state spectrum cannot be equally well fit with a simple power-law; on the other hand, the low state spectrum, once the slope is fixed to the \hess\ value, gives a lower limit $E_{\rm cut} >$ 46\,TeV on the energy cutoff. A plausible scenario could be that the energy of the cut off increases from the high-state (around 9\,TeV) to higher energies in the low state.

In this work, we show how the orbital modulation of LS~5039 can be constrained using the \astrima. We assumed the same, orbit averaged, spectrum for each phase bin: a power-law model with exponential cutoff with $\Gamma=2.06\pm0.05_{\mathrm{stat}}$ and $E_{\rm cut}=13.0\pm4.1$~TeV \citep{aharonian06_ls5039}, as more complicated models could not be constrained given the short exposure times. Starting from the orbital modulation detected by \hess\ in 2005, we simulated ten different spectra, one per orbital phase bin, by varying the normalization value as to reproduce the modulation obtained by \hess\footnote{The power-law pivot energy is fixed at 3\,TeV}. We considered a ﬁxed exposure time of 10 h for each phase bin. The new simulations and the analysis were performed as described in Sect.\ref{sect:intro}. 

\begin{figure}
\centering
\includegraphics[width=\columnwidth]{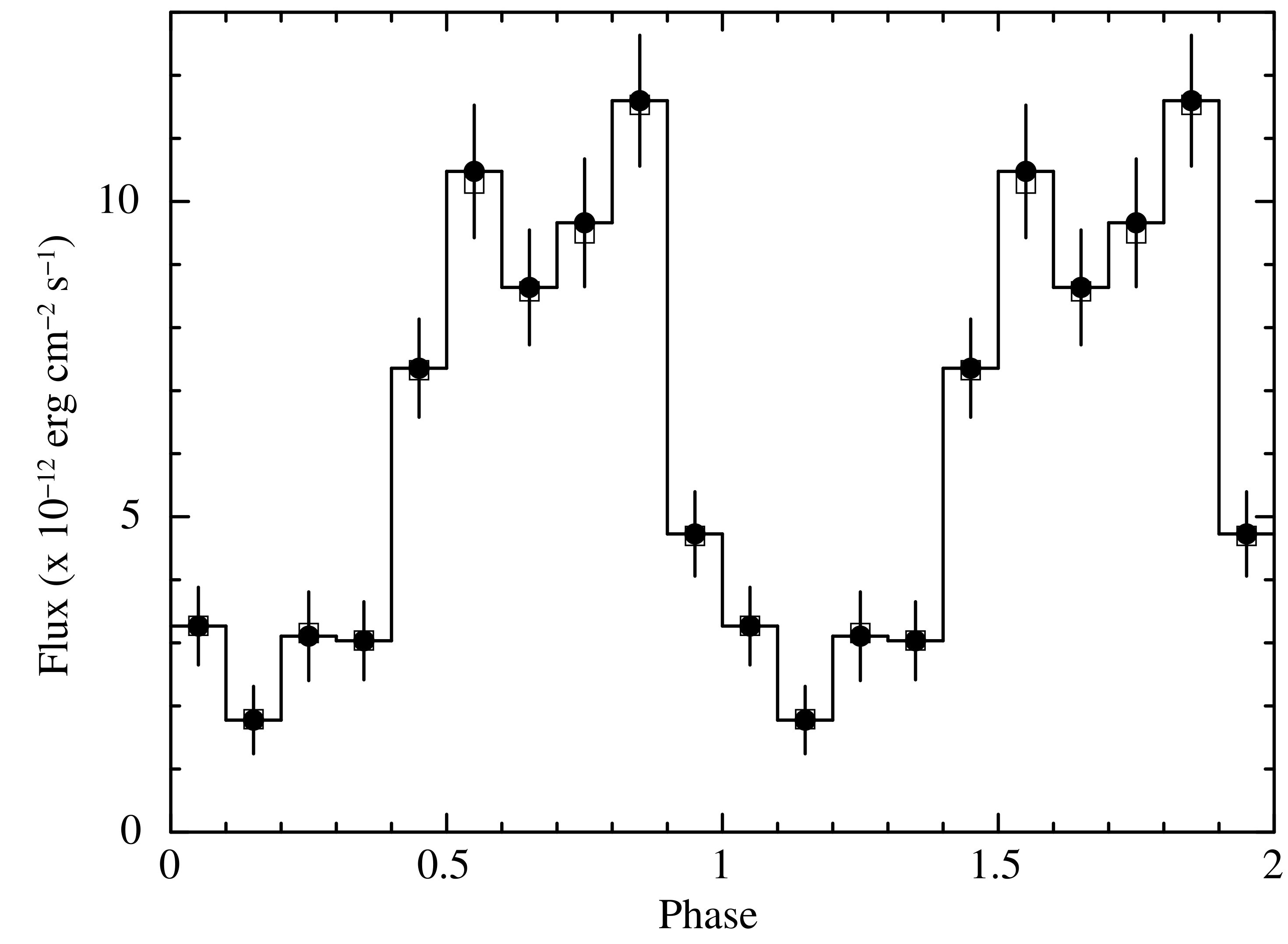}
\caption{LS~5039: orbital modulation obtained with 10\,h observation per bin. The open squares are the expected fluxes from the simulated models while the filled circles are the obtained flux in 0.8-200\,TeV with error bars at one sigma.}
\label{fig:LS5039_orbit_1TeV}
\end{figure}

\begin{figure}
\centering
\includegraphics[width=\columnwidth]{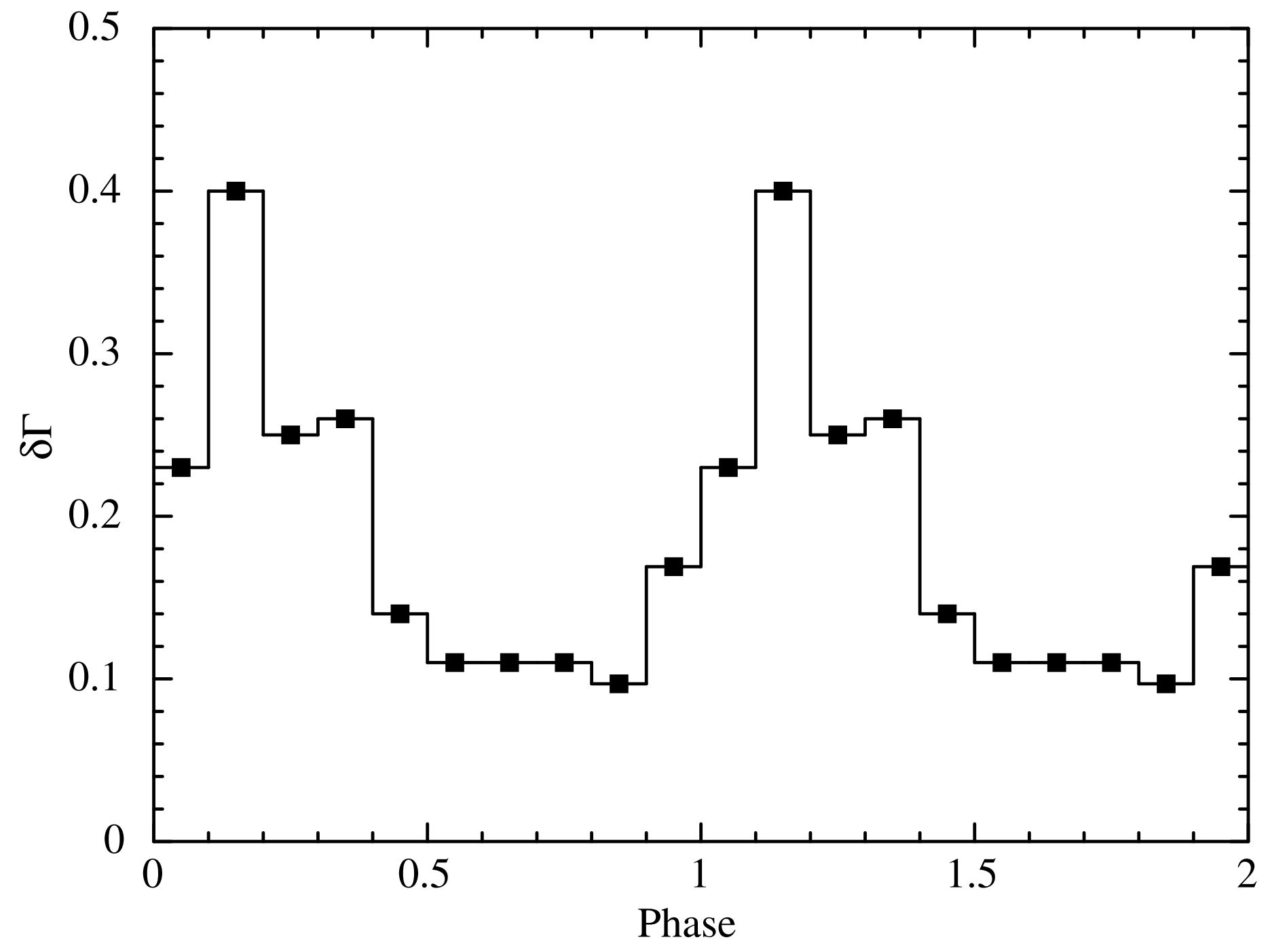} 
\caption{LS~5039: one sigma uncertainty on the spectral index obtained 
for 10\,h observations per orbital bin.}
\label{fig:LS5039_index}
\end{figure}

\paragraph{Analysis and Results}
In Fig.~\ref{fig:LS5039_orbit_1TeV} we report the orbital modulation obtained with the \astrima\ with 10\,h per orbital bin. It shows that, even during the low state of the source, the estimated source flux is fully consistent with the flux expected from the simulated model. Fig.~\ref{fig:LS5039_index} shows the one sigma uncertainty that is obtained on the spectral index. For 90 \% of the orbital phase this uncertainty is between 0.1 and 0.25, while only in the case of the lowest-flux bin (phase range 0.1-0.2) it value rises up to about 0.4. 

LS~5039 is among the few \gray\ binaries where TeV emission is detected along the whole orbital period. The MeV and TeV emissions appear to anti-correlate with the GeV one \citep{chang16}. A possible explanation for this behaviour could be that the X-ray/MeV emission is due to synchrotron emission from a highly relativistic particle population and the TeV one due to IC of stellar photons by the same particle population. In this scenario, the GeV emission would be the secondary result of TeV photons that are absorbed via pair-production when the compact object approaches the strong stellar wind of the companion \citep[e.g.][and references therein]{chernyakova19}. A pulsar-wind-driven interpretation of the HE and VHE emission of LS 5039 is also supported by the results obtained by \citet{Huber+21}, who reproduced the main spectral features of the observed multiband emission of LS 5039 with a numerical model for the non-thermal emission of shock-accelerated particles.

A new monitoring of the orbital modulation of LS~5039 at TeV energies is very important to address this point: it will allow us to evaluate the stability of the processes that accelerate the relativistic particles responsible for the TeV emission (acceleration in winds or jets, according to the scenario at play), together with the involved geometry (absorption, inclination, etc). The simulations performed show that with a total of 100 hours (10\,h for 10 orbital bins) it will be possible to build a good TeV orbital light-curve (Figure~\ref{fig:LS5039_orbit_1TeV}) to be compared to the ones presented by \hess\ in 2005 \citep[][simulated in this work]{aharonian06_ls5039} and in 2015 \citep{2015arXiv150905791M}. Concerning the phase resolved spectroscopy (Figure~\ref{fig:LS5039_index}), with a 10\,h bin time the spectral index can be constrained with a relative uncertainty of about 5\% in the high state and of about 10 \% in the low state, with the only expectation of the phase range corresponding to the minimum flux; in this case, the relative uncertainty increases up to about 20 \%. This implies that  a deeper coverage ($\sim$40\,h) in the low state is needed to reach a 10\% accuracy. We note that the bright phase of the modulation is of particular interest: the two-peaked shape (that has been observed only in LS~5039 up to now) can be interpreted in terms of the rotating hollow cone model \citep{neronov08}. In this model, an anisotropy pattern is produced when relativistic particles interact with a radiation field such as the one produced by a massive companion star in a binary system. Maxima in the light-curve (flux versus orbital phase or versus the true anomaly) are expected when the line of sight passes through the rotating hollow cone. A reliable constraint on the relative position and height of the two peaks would enable to reconstruct the system geometry (inclination angle, position of the TeV emitting region, etc). In the case of LS~5039, as observed by \citet{aharonian06_ls5039}, the difference between the peaks and $\phi=0.7$ (where the inferior conjunction occurs) is  $\sim$10\%. Should it be confirmed, it could be easily studied with the \astrima.

\subsection{Further observations  of \gray\ binaries and microquasars}
Regarding compact binaries, the \astrima\ will help unveil some intriguing issues like the TeV-detection of SS\,433 by HAWC. X-ray binary sources hosting a black-hole as a compact object show powerful jets as revealed in the radio band \citep[e.g. Cyg X-1 and Cyg X-3,][]{2017SSRv..207....5R}, and there are also clear examples of binary systems containing a neutron star, that have recently, and surprisingly, observed to launch extremely powerful jets  \citep[see e.g.][]{2018Natur.562..233V}.  Although jets seem a general ubiquitous phenomenon in many compact Galactic binaries, an issue which is yet to be explored is how efficiently the kinetic power of particles in jets is converted via IC in the VHE domain, and which is the overall contribution of these injections of relativistic particles in the Galactic medium.  

As can be noted from the Table~\ref{tab:sources_list}, the list of
potential \gray\ binary targets comprehends two more important 
objects: HESS J0632+057 and LS I 61+303. The former has 
an average VHE flux similar to that of LS 5039. As shown in Sect.~\ref{sect:ls5039}, the 
\astrima\ on the long-term is able to well track this TeV variability 
along the orbit; in the case of HESS J0632+057 the orbital period 
is 316.7\,$\pm$\,4.4 days with a pronounced peak at phase $\sim$\,0.3
when the source reaches 6\% of the Crab flux \citep{adams21}, thus allowing a 
possible detection and spectral characterisation in the TeV range.
The other important \gray\ binary is LS I 61+303, also high-mass binary hosting a 0.27 sec radio pulsar \citep{weng22}. 
The orbital period is 26.5 days and it shows jet emission, which is known 
to precess with a similar period, giving rise to a longer, super-orbital 
modulation of $\sim$\,4.5 years \citep{ahnen16}. The TeV light curve is strongly
modulated along the orbit, with over-imposed transient exceptionally
bright flares, which arrive at $\sim$\,16\% Crab units \citep{2016ApJ...817L...7A}
at certain orbital phases. The origin of these TeV flares is still to be 
fully understood.

%% file: sect_dmgc.tex
\section{Other possible Galactic targets} \label{sect:dm_gc}
\subsection{The Galactic Center: diffuse emission and dark matter search}
The Galactic Center (GC) is possibly the most interesting, and at the same time, the most complex  region of our Galaxy in the VHE domain. The presence of the super-massive black-hole Sgr A$^{\star}$ at its centre, the high density of star-forming regions, active pulsars and PWNe, SNRs and giant molecular clouds all contribute  to bright and diffuse patterns of gamma-ray emission. In \citetalias{vercellone21}, we examined how \astrima\ can well constrain the diffuse emission along the ridge up and beyond 100 TeV, thus giving us the opportunity to study the energetic hadron population and connect it to the CR population observed at Earth. We mention here two more important  scientific cases that can be at the same time addressed by deep observations of this  region: the possible detection of  dark matter signals and the study of the VHE emission from Globular Clusters.

Part of the diffuse emission from the GC and the halo could be due to self-interaction, through annihilation or decay \citep{bergstrom1998} of particle dark matter \citep[DM;][]{zwicky33}. Originally introduced to explain the flat rotation curves of spiral galaxies, DM is also required to justify $\sim$\,30\% of the Universe's energy content \citep[see e.g.][and refs. therein]{planck14}. The main physical property of DM is that it does not couple with radiation as baryonic matter. Such a feature is required in order to explain DM invisibility to traditional astronomical observations, which prevents astrophysicists to directly obtain data on its components. A framework for astronomical indirect DM searches arises from the possibility that weakly-interacting massive DM particles (WIMPs) annihilate or decay to produce standard model pairs. Such pairs subsequently annihilate into final-state products, among which $\gamma$-ray photons. A detailed description of the theoretical $\gamma$-ray flux produced from DM interactions is made in a companion paper for the extragalactic targets \citepalias{saturni21}; here, we simply recall that the average velocity-weighted cross sections of such processes are expected to be of the order of magnitude typical of electroweak interactions \citep[$\langle \sigma v \rangle \sim 3 \times 10^{-26}$ cm$^3$ s$^{-1}$; e.g.,][]{roszkowski2014}.

The possible detection of a significant $\gamma$-ray flux from DM interactions with the currently hypothesised physical parameters of the DM particles would require $\gg$1000 h observations even for the most favourable targets, i.e. the Galactic centre itself. Furthermore, the standard DM analysis takes advantage of a full-likelihood method \citep{aleksic12} derived from common VHE likelihood maximisation procedures, which require (i) a detailed knowledge of the astrophysical processes producing the foreground $\gamma$-ray emission (Fermi bubbles, unresolved point sources) to be subtracted from the GC exposures, and (ii) the production of dedicated 2D maps of spatial extension of the Galactic DM halo from Monte-Carlo analyses of the kinematic properties of Milky Way's baryonic matter \citep[e.g.,][]{bonnivard15,hayashi16}. 
Search of $\gamma$-ray signals from DM annihilation or decay is a challenging task, though worth to be pursued in multi-TeV deep observations of the GC and halo with the \astrima. This is a core-science case \citepalias[see][]{vercellone21}. However, given the large amount of observing hours needed to reach the sensitivities at which a theoretical $\gamma$-ray flux from DM would be detected, an upper limit is more likely to be drawn. Therefore, the immediate expected result will be a firmer constraint of the DM parameters (particle mass, annihilation cross section, particle lifetime) related to the limits on the expected $\gamma$-ray signal from DM annihilation or decay at $E \gtrsim 10$ TeV, at which the \astrima\ expected sensitivity reaches its maximum. Furthermore, the stacked exposures could in principle be combined with observations from other Cherenkov facilities to obtain deeper sensitivities to DM signals. Finally, synergy with observatories at shorter wavelengths is foreseen to produce updated estimates on the structure and DM content of the DM Galactic halo starting from observable astrophysical quantities (e.g., Galactic rotation curves, stellar kinematic samples, intra-galaxy medium distributions).

The observations of diffuse signals from the GC and halo will also take advantage from the very large zenith angle at which such sky regions are viewed from the \obsteide\ site (ZA $\sim 60^\circ$), as observations at larger zenith angles lead to a larger effective area and sensitivity in the highest energy band \citepalias[see][]{vercellone21}, allowing us to probe DM particle masses $m_\chi \gtrsim 10$ TeV. Furthermore, the \astrima\ large FoV 
will easily allow us to integrate wide regions of the Galactic halo, increasing the collected amount of signal from single pointed exposures. Finally, the combined good \astrima\ spectral and angular resolution can allow the search for exotic signals from DM interactions such as monochromatic \gray\ emission lines, whose expected flux can be enhanced by fundamental physics mechanisms \citep[see e.g. Sect. 4.1.5 of][and references  therein]{cta19} at the highest energies accessible to the \astrima.

\subsection{Serendipitous Science: the case of  Novae}

Observations of novae at high-energies is a field  of research  that is still in his infancy.  Nova stars,  and more generally nova-like objects, were discovered as  gamma-ray sources a few years ago \citep{abdo2010,ackerman2014}.  To date about a dozen of novae and symbiotic systems have been detected in gamma-rays at GeV levels.  The physical origin of this emission is not yet clear, but it is likely due to the inverse Compton scattering of the companion star radiation by electrons accelerated at the nova shock which originates when the dense nova ejecta collide with the interstellar medium or the wind from the secondary~\citep{martin2013}. As an alternative the shocks might be produced by the interaction of a fast wind, radiatively driven by the nuclear burning on the white dwarf \citep{martin2018} with inhomogeneity blobs formed into the expanding shells of the nova ejecta. 

However,~\citealt{sitarek2012} have hypothesized that also protons are accelerated in the shocks and their  interactions with the ambient medium or inhomogeneities into the nova shells  might be able to produce observables fluxes of TeV \grays\ (and neutrinos).  In general, the current  nova rate per year in the Milky Way is estimated to be  in the range 20-40 novae/yr~\citep{dellavalle2020}. This figure implies that about 3-6 novae/yr  are potential targets for the \astrima.  For example, 
very recently \hess\ and MAGIC detected  very high energy \grays\ fluxes (up to \,0.1--0.2 TeV) from the recurrent nova RS Ophiuchi, up to 1 month after its 2021 outburst \citep[][respectively]{hess22, magic22}. 



%% file: sect_terzan5.tex
\subsection{VHE emission from globular clusters: the case Terzan 5} \label{sect:terzan5}
\paragraph{Scientific Case}
Globular clusters (GlCs) have a density of millisecond pulsars (MSPs) per unit mass about 1000 higher than the one  present in the Galactic disk. This is due to the large stellar densities in the cores of GlCs that favour dynamical interactions,  such as exchange interactions and tidal captures.  The process lead to the formation of X-ray binary systems, where neutron stars are spun-up during the accretion process.
The GlC Terzan 5 contains 37 known MSP \citep{cadelano18} and it is by far the MSP richest GlC.  \citet{prager17} has recently argued that this  MSP over-population might be due to Terzan 5 being a past fragment of the Milky Way’s proto-bulge, and a much more massive GlC in the past.  Observations with the \textit{Fermi} satellite detected a \gray\  source consistent with the position of Terzan 5 in the 0.5--20 GeV range. The emission was explained as the convoluted effect from the emission of the whole MSP population \citep{kong10}. The GeV spectrum is  well described by a cut-off power-law of index 1.4 and cut-off energy  of 2.6 GeV. Similar GeV emission from many other GlCs has been detected  in the last years and the scenario of the MSP population as the source of this emission is now firmly established \citep{tam16}.

\citet{hess11} detected a  TeV source, HESS J1747-248, at a distance of 4$^\prime$ to the centre of Terzan 5. The positional \hess\ uncertainty translates to an offset of  $\sim$2\,$\sigma$ with respect to the GlC core centre.  The TeV source is extended  with an elliptical shape (9.6$^\prime$ $\times$ 1.8$^\prime$) and it only partially overlaps with the GlC, which has an inner half-core radius of only  0.15$^\prime$ and a tidal  radius of 4.6$^\prime$.
\citet{hess11} showed that the chance coincidence of the new source  with the GlC is only 10$^{-4}$ and favoured the association with one, or more, GlC sources. However, the significant displacement  of the TeV emission from the GlC centre opens an intriguing question about its origin. A possible explanation was proposed by \citet{bednarek14},  who hypothesised that the winds from normal stars and MSPs by interacting with  the more dense medium of the Galactic disk form a bow shock, that  is naturally misaligned with respect to the GlC. The shock traps high-energetic leptons escaping from the GlC, which in turn interact with  background photons through inverse-Compton producing the TeV emission. In a second scenario, \citet{domainko11} proposed that energetic leptons and hadrons, produced by the explosion  of a past supernova, or through a neutron star-neutron star  collision  in a kilonova explosion, decay after interaction with  ambient nuclei into other massive particles,  like the $\pi^0$, which eventually decay into two energetic $\gamma$ photons.  Although the first scenario appears more likely,   as it does not involve any ad-hoc event, the lack of TeV emission  in an extensive search with \hess\ for other GlCs \citep{hess13}  poses a difficult conundrum, as the Terzan\,5 TeV emission appears in this context a unique and unexplained case.

\paragraph{Feasibility and Simulations}
HESS J1747-248 is 4.2\degmark\ away from the Galactic Centre, 1.7\degmark\ above the Galactic plane, in a region where the central contribution of the Galactic Centre is negligible \citep{hess18_gc}. The Galactic Centre is part of the Science Core Programme in the first operational years of the \astrima\ \citepalias[see][]{vercellone21} and, for this reason, it is likely that HESS J1747-248 will benefit from long exposures, although at large offset angles. The TeV emission from HESS J1747-248 was simulated according to the morphology and spectral shape from \citet{hess11}; the spectrum  in the 0.8--50 TeV range  was consistent with a simple power-law of  photon index 2.5 and  integrated photon flux of (1.2\,$\pm$\,0.3) $\times$ 10$^{-12}$ cm$^{-2}$ s$^{-1}$, or 1.5\% of the Crab flux. We simulated three different event lists for 100, 200 and 400 hours of observations, assuming a pointing direction 3\degmark\ away from the source. 

\paragraph{Analysis and Results}
HESS J1747-248 is a good benchmark to test the detection efficiency  of the \astrima\ for a faint, slightly extended source, with a moderate hard spectrum.  We performed an unbinned likelihood analysis (\textsc{ctlike} task in \ctools) in the 0.7--200 TeV range for 100 independent realisations and for the three exposure times. The averaged best-fitting results are shown in Table~\ref{tab:results_terzan5}. Reported errors represent the standard deviation of the ensemble of the best-fitting parameter results.

\begin{figure}
\begin{center}
\includegraphics[width=\columnwidth]{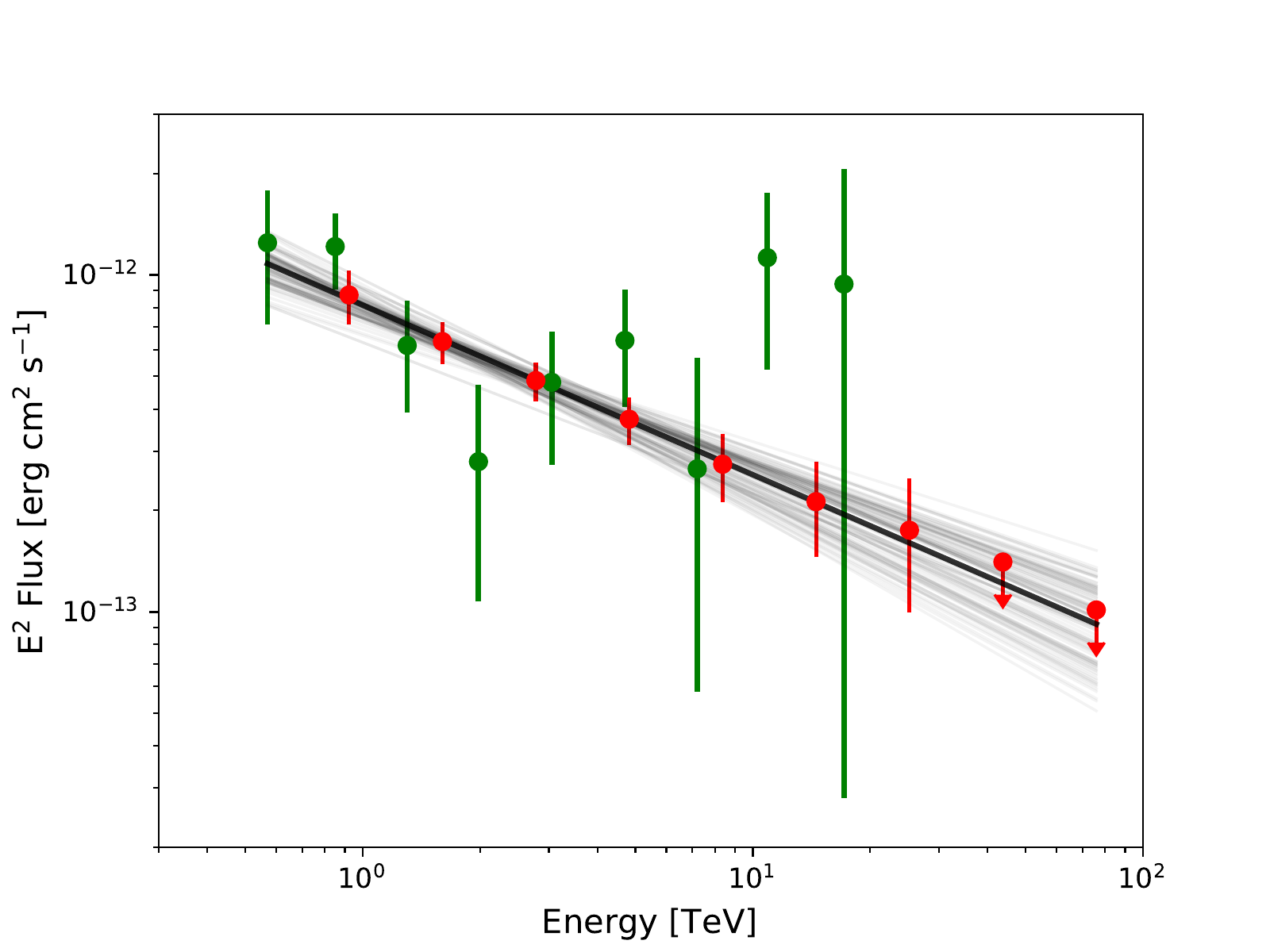}
\caption{A combined fit of the spectrum obtained by \hess\ \citep[][green points]{hess11} and spectral data points obtained with a possible 400\,hr observation with the \astrima\ (red points). A butterfly diagram, showing the envelope of all spectral models that are statistically compatible with the data, is marked by the grey lines.}
\label{fig:spec_terzan5}
\end{center}
\end{figure}

Finally we also show a combined analysis of the \astrima\ spectral points obtained for the 400\,hr exposure and the data published in 2011 by the \hess\ Collaboration \citep{hess11}. A power-law model fit of these data results in the determination of the photon index 2.51$\pm$0.08 and a normalisation value at 1 TeV of 5.1$\pm$0.4.
In Fig.~\ref{fig:spec_terzan5}, we show the simulated 400 h \astrima\ spectrum of HESS J1747-248 together with the \hess\  data \citep{hess11} and the combined best-fitting model (butterfly diagram).
\astrima\ would obtain a detection up to $\sim$\,20 TeV in case of no cut-off in the spectrum. For this specific case, the possibility to set a lower limit  on the cut-off energy could be of great importance for  distinguishing a leptonic scenario, where the effects of the reduction in the cross-section due to the Klein-Nishina correction predict a curvature in the energy range above 10 TeV, from  the hadronic one, which predicts no significant curvature above 10 TeV
\citep[see also discussion in][]{bednarek16}.
We adopted the same approach outlined in Sect.\ref{sect:j2032} and in Sect.\ref{sect:ss433} to estimate the sensitivity in detecting an energy cut-off in the spectrum for the 400\,hr exposure; for 10 random realisations, simulated with a power-law model with a corresponding $TS_{max}$ value, we found by fitting the data with a cut-off power-law model the cut-off energy that corresponded to a decrease of this value of nine by varying the cut-off energy in each fit. Taking the average threshold  value from this sample, we found we might be able to constrain the presence of a cutoff if it is below 12$\pm$2 TeV. 

\begin{table*}[h]
    \centering
     \begin{threeparttable}[t]
    \caption{Best-fitting results for HESS J1747-248 using three different exposures with the \astrima. Last column reports              \hess\ values and uncertainties for an immediate comparison.}
    \begin{tabular}{r|rrr|r} \hline 
         & \multicolumn{3}{c}{ASTRI-MA}  & \hess\tnote{1} \\
         & 100hr & 200hr & 400hr &  \\
         \vspace{0.2cm}
         TS & 73$\pm$19 & 144$\pm$28 & 282$\pm$37 & \\
         &\multicolumn{3}{c}{Spectral Model} \\
         $N_o$\tnote{2} & 5.0$\pm$1.3   & 4.9$\pm$1.0   & 4.8$\pm$0.6   & 5.2$\pm$1.1 \\ \vspace{0.2cm}
         $\Gamma$       & 2.47$\pm$0.18 & 2.47$\pm$0.14 & 2.47$\pm$0.09 & 2.5$\pm$0.3 \\
         &\multicolumn{3}{c}{Spatial Model} \\
         R.A. (deg) & 266.95$\pm$0.02 & 266.96$\pm$0.01  & 266.95$\pm$0.009 &  267.95$\pm$0.03\\
         Dec. (deg)  & -24.81$\pm$0.01 & -24.81$\pm 0.006$ & -24.81$\pm0.004$ & -24.81$\pm$0.01 \\
         PA\tnote{3} (deg)  & 90$\pm$46        & 83$\pm$64        & 92$\pm$5       & 92$\pm$6\\
         
         $R_{\textrm max}$ (arcmin) &  9.6$\pm$3.0  & 9.6$\pm$1.8 & 9.6$\pm$0.6 & 9.6$\pm$2.4 \\
         \vspace{0.2cm}
         $R_{\textrm min}$ (arcmin) & 2.0$\pm$2.0 & 1.9$\pm$1.8  & 2.0$\pm$1.8 &  1.8$\pm$1.2 \\
\hline
\end{tabular}
     \begin{tablenotes}
     \item[1] Value taken from \citet{hess11}.
     \item[2] The spectral best-fitting model is a power-law with 
     normalisation $N_0$ calculated at the reference energy of 1 TeV in units  
     of 10$^{-13}$ photons cm$^{-2}$ s$^{-1}$ TeV$^{-1}$.
     \item[3] Positional Angle  counted counterclockwise from North.
     \end{tablenotes}
\label{tab:results_terzan5}
\end{threeparttable}
\end{table*}

%% file: sect_conclusions.tex
\section{Conclusions} \label{sect:conclusions}
In this work, we have presented a list of scientific cases that the \astrima\ will be able to address in a long-term planning of observations of the Northern Galactic sky. A companion paper, \citetalias{vercellone21}, presented
the core-science program envisioned for the first years of operations. These campaigns will focus on a short list of targets that will be observed in depth  to answer \textit{primary and outstanding} scientific questions. Furthermore, the large FoV of the array will allow the simultaneous observations of many close-by VHE targets that will constitute the initial base for the \astrima\ observatory science programme, and we have shown how significant improvements can be obtained for a small subset of representative field targets.

Moreover, since, as shown in Table~\ref{tab:sources_list}, the number of expected detectable sources is of order a few tens, it is likely that most, if not all of them, will be observed at least for few hours, or few tens of hours, during the first years of operation. The sources that have been simulated and analysed in this paper illustrate how \astrima\ observations will allow to derive spatial and spectral constraints on a wide range of different classes of galactic \grays\ emitters.

%% file: sect_velax.tex
\subsection{A bright extended PWN: Vela X} \label{sect:velax} \vspace{0.3cm}
\paragraph{Scientific Case}
Vela X is a relatively evolved PWN with an estimated age of $\gtrsim$10$^4$ years \citep{lyne96}, emitting across all the electromagnetic spectrum. It is one of the brightest TeV sources detected with \hess~\citep{Aharonian_2006_VelaX, abramowski12}. In the VHE band Vela X has an extended morphology intermediate between that at radio wavelengths (see e.g. \citealt{frail97, bock98}) and that in the X-rays \citep{markwardt95}: the inner part of the TeV emission region coincides with the central X-ray cocoon, whereas the VHE extended wings (along the right ascension) are consistent with the radio morphology. \citet{abramowski12} showed that the VHE gamma-ray morphology of the Vela PWN can be accounted for by a \textit{radio-like} component emitting 65\% of the flux and by an \textit{X-ray-like} component emitting the remaining 35\% of the flux.

\paragraph{Feasibility and Simulations}
We simulated 100 hours of observations with the \astrima\ of Vela X  in the context of the ACDC project \citep{pintore20}, to probe the performance of this facility for a relatively large, complex and bright benchmark case. Although Vela X is not visible from Teide, this analysis is relevant to the investigation of other similar extended and bright TeV sources that, instead, might be observed. Following \citet{abramowski12}, we modelled the VHE morphology of the Vela PWN as a superposition of radio and X-ray brightness maps, accounting for the 65\% and 35\% of the TeV flux, respectively. 

To fully test the resolving capabilities of the \astrima\ , we simulated VHE emission from these components, creating radio and X-ray templates from archival high-resolution observations.

For the X-ray spatial template, we used the ROSAT image of Vela X in the 0.4--2.4 keV energy range \citep{voges99}, which has an angular resolution of $\sim$\,0.0167$^{\circ}$\footnote{The half-power radius corresponds to an encircled energy of about 50\%.}.

\begin{figure}
	\centering
	\includegraphics[width=0.5\textwidth]{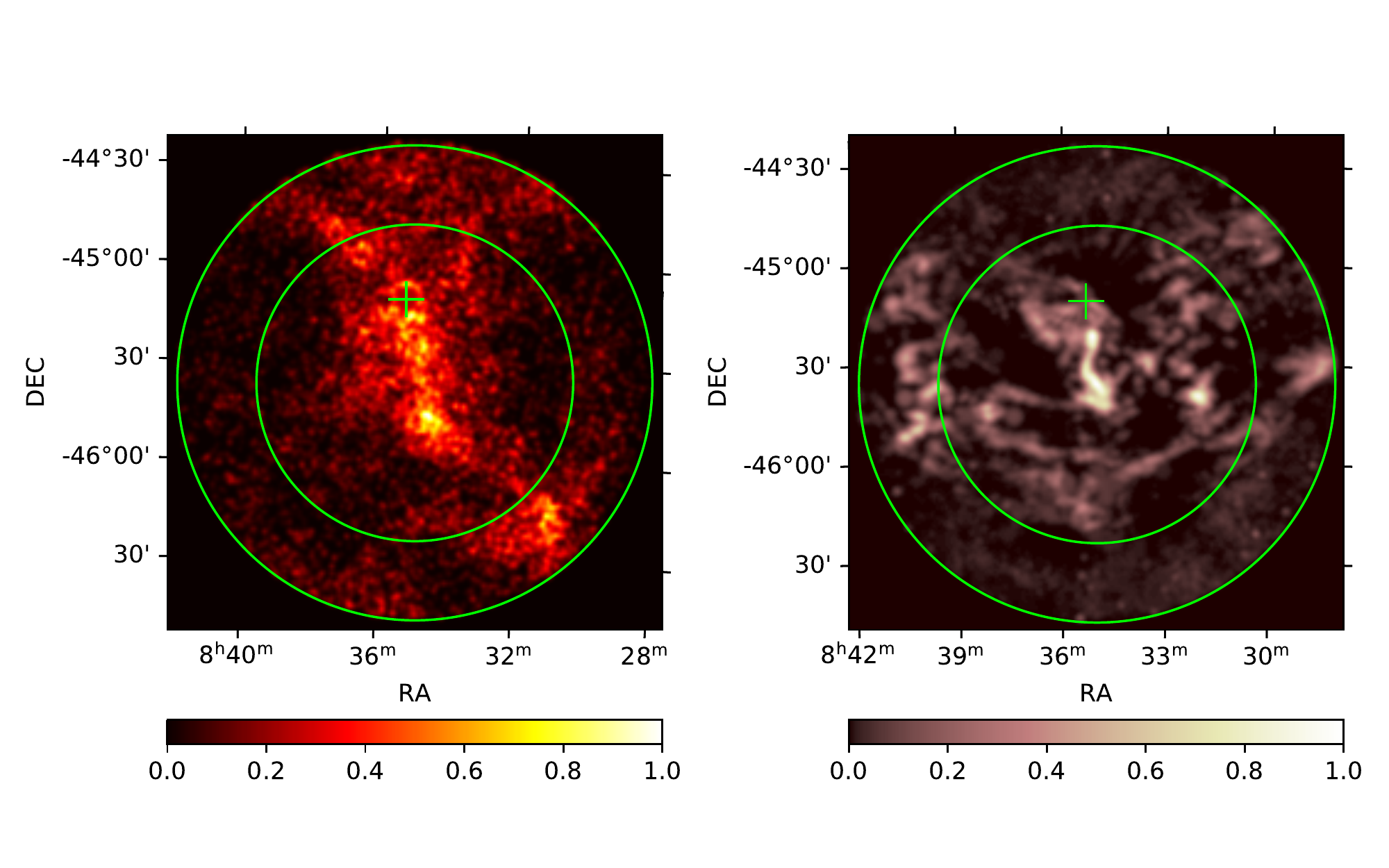}
	\caption{X-ray (\textit{left panel}) and radio (\textit{right panel}) templates of the extended Vela X emission. $x$- and $y$- axes represent right ascension and declination, respectively. Colour bars represent the normalised brightness. Circles have radii of 0.8$^{\circ}$ and 1.2$^{\circ}$, respectively. The cross marks the Vela pulsar position. Colour map units in counts/pixel.}
	\label{fig:Vela_Xray_Radio_Template_FULL}
\end{figure}

The radio spatial template was created adopting the exposures taken during the second Molongo Galactic Plane Survey (MGPS\,-\,2) at 843 MHz with the MOST radio telescope \citep{bock99,murphy07}, whose angular resolution is $\sim$ 0.0167 degrees. In order to convert both adopted images to an appropriate template for the VHE simulations, we subtracted the diffuse background unrelated to the Vela X extended source, and then we removed the contribution of bright point sources in the field (including the Vela pulsar). \citet{abramowski12} determined the spectral parameters of Vela X at VHE in two regions both centred at ${\rm RA} = 128.75^{\circ}$ and ${\rm Dec} = -45.6^{\circ}$. We selected in our spatial templates the same two regions: a circle with radius of 0.8$^\circ$ and a ring with inner and outer radii of 0.8$^{\circ}$ and 1.2$^{\circ}$, respectively. The final circular and ring-shaped X-ray and radio spatial templates of the extended Vela PWN emission are shown in Fig. ~\ref{fig:Vela_Xray_Radio_Template_FULL}.

For the VHE spectra of the circular and ring-shaped parts of the Vela PWN, we adopted the power-law models with an exponential cut-off, obtained for the same regions  in the analysis of the \hess\ data between 0.75--70 TeV \citep{abramowski12}, and extrapolated them up to 100 TeV. 
We also took into account emission from the Vela pulsar, simulating it as a point source with a power-law spectral model taken from \citet[][Table 4]{burtovoi17}. 

We extrapolated this spectrum to the VHE range as well, assuming the absence of a cut-off\footnote{Similarly to what was found in the spectrum of the Crab pulsar up to a few TeV \citep{ansoldi16}.}. 

We simulated the \astrima~observations which cover a circular area of radius $2.5^{\circ}$, centred at $\alpha_0=128.75^{\circ}$ and $\delta_0=-45.6^{\circ}$, with an overall exposure time of 100 h. In our analysis we considered two different energy ranges: E$>$1 and E$>$10 TeV. 
\paragraph{Analysis and Results}

\begin{figure}
	\centering
	\includegraphics[width=0.5\textwidth]{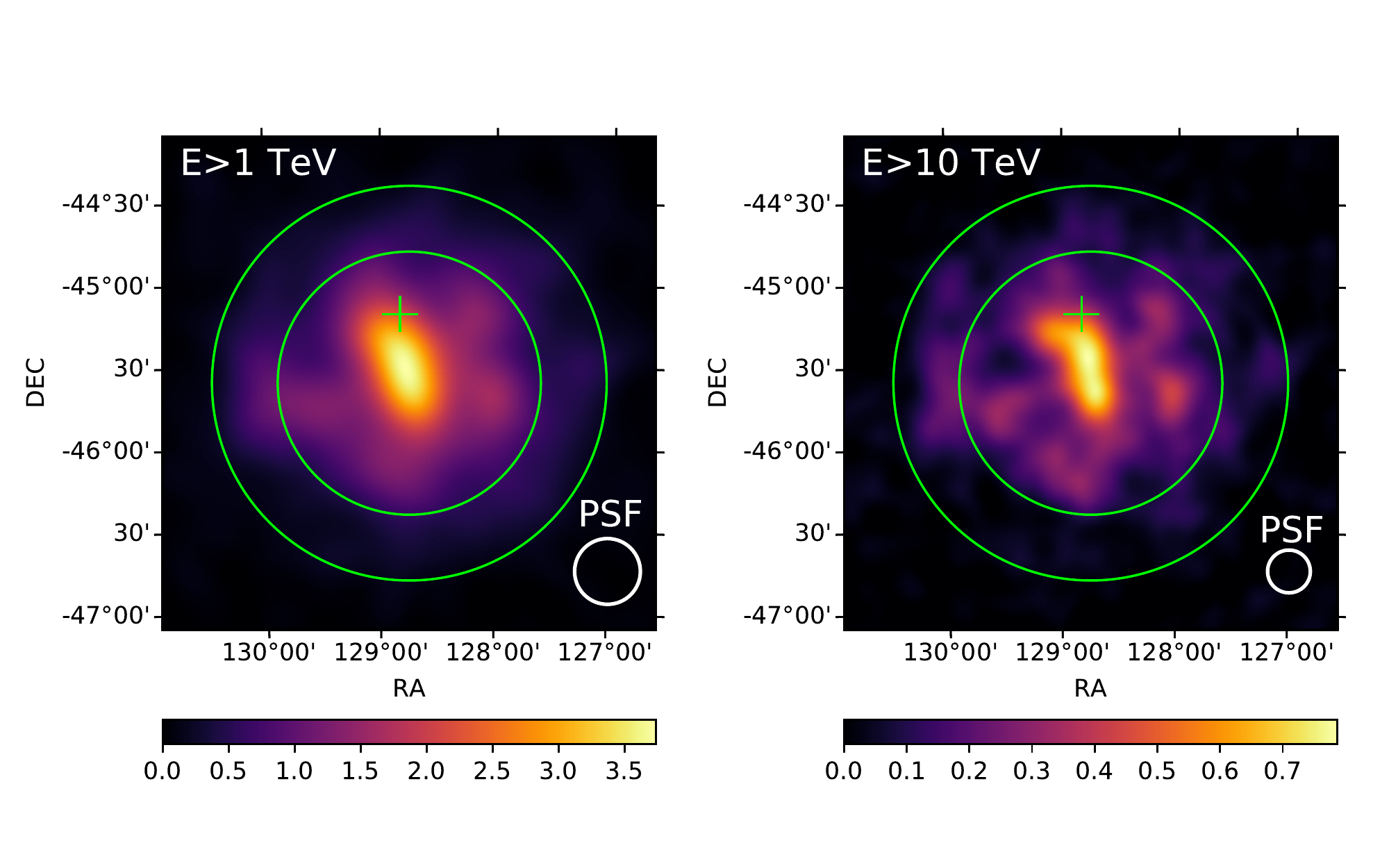}
	\caption{VHE Vela X background-subtracted maps simulated assuming 65\%-contribution from the radio template and 35\%-contribution from the X-ray template. The exposure time is 100 h. Energy ranges are $E>1$ TeV \textit{(left panel)} and $E>10$ TeV \textit{(right panel)}. $x$- and $y$- axes are right ascension and declination.  All maps are smoothed according to the size of the corresponding PSF (white circles). Green circles have radii of 0.8$^{\circ}$ and 1.2$^{\circ}$, respectively. The cross marks the Vela pulsar position. The colour bars represent residual number of counts per pixel.}
	\label{fig:VelaX_ASTRI_R_M_E}
\end{figure}

The resulting VHE background subtracted maps, obtained using the task \textsc{ctskymap}\footnote{The task \textsc{ctskymap} was used in the \textsc{IRF} mode, i.e. the background template provided with the Instrument Response Functions was subtracted from the map.}, are shown in Fig.~\ref{fig:VelaX_ASTRI_R_M_E}. All images are smoothed using the angular resolution of the \astrima\ at different energies. The smoothing radius adopted here is equal to the 68\% of the containment radius ($r_{\textrm{68}}$) of the gamma-ray PSF at the lower limit of the corresponding energy range\footnote{For the \astrima~$r_{\textrm{68}}$=\,0.20$^{\circ}$ at 1 TeV and $r_{\textrm{68}}$=\,0.13$^{\circ}$ at 10 TeV.}.

We fitted the simulated data using a multi-component model, which contains radio and X-ray circular and ring-shaped components of the extended Vela PWN emission, with free spectral parameters. We estimated the detection significance for these components using the task \textsc{ctlike} in binned mode. \textsc{ctlike} calculates the value of the Test Statistics\footnote{Test Statistics is twice the difference between the maximum log likelihood value for a model with an additional source and that for a model without the additional source.} (TS) for each source in the spectral model, performing a maximum likelihood analysis of the data. In order to account for the statistical fluctuations, we repeated the simulations 50 times, fitting the distribution of the TS values with a Gaussian function\footnote{To decrease the computational time, the energy dispersion was not considered in these simulations.}. 
The resulting mean values of TS, together with corresponding values of standard deviation, for different components of the Vela PWN, after 100 h of observation at $E>1$ TeV, are as follows:
\begin{itemize}
    \item for the radio circular component TS\,=\,2060 (100\,$\sigma$),
    \item for the radio ring-shaped component TS\,=\,719 (70\,$\sigma$),
    \item for the X-ray circular component TS\,=\,609 (50\,$\sigma$),
    \item for the X-ray ring-shaped component TS\,=\,168 (18\,$\sigma$).
\end{itemize}

In order to determine to what extent the contributions from the spectral components of the Vela PWN are distinguishable with the \astrima, we  repeated the simulations and analyses described above assuming different ratios between the radio and X-ray templates. Our goal was to determine the minimum contribution of radio/X-ray spectral components (assuming that the Vela X morphology is defined by the templates) that can be significantly detected with the \astrima. 

\begin{figure}
\begin{center}
\includegraphics[width=\columnwidth]{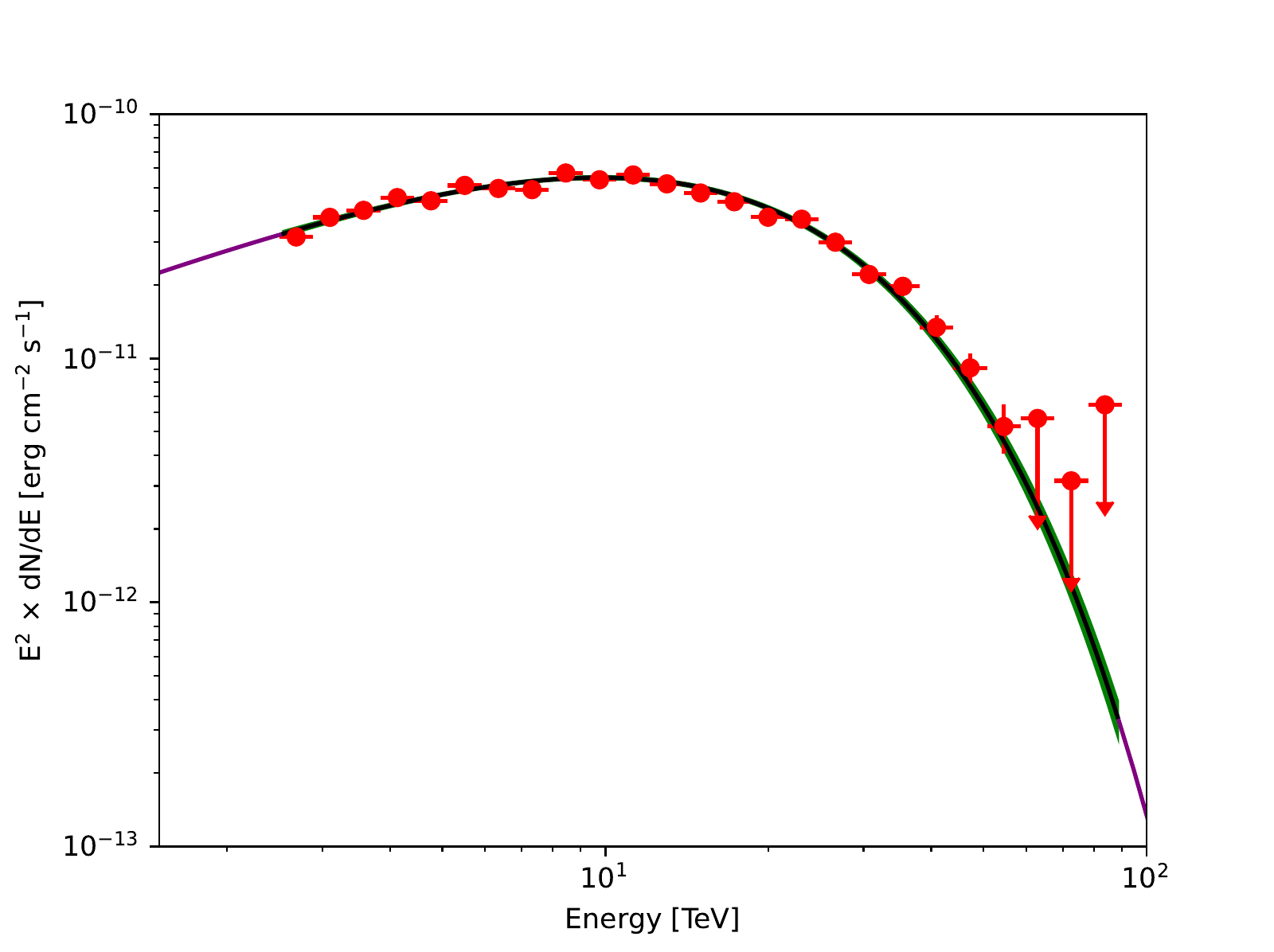}
\caption{
Spectral energy distribution (SED) of the Vela PWN for the $2.5 < E < 90$ TeV energy range. The solid black line shows the best-fitting model with an exponential cut-off power-law. The green zone shows the 1-sigma butterfly uncertainty of the best-fitting model.
}
\label{fig:sed_2-100}
\end{center}
\end{figure}

\begin{figure}
\begin{center}
\includegraphics[width=\columnwidth]{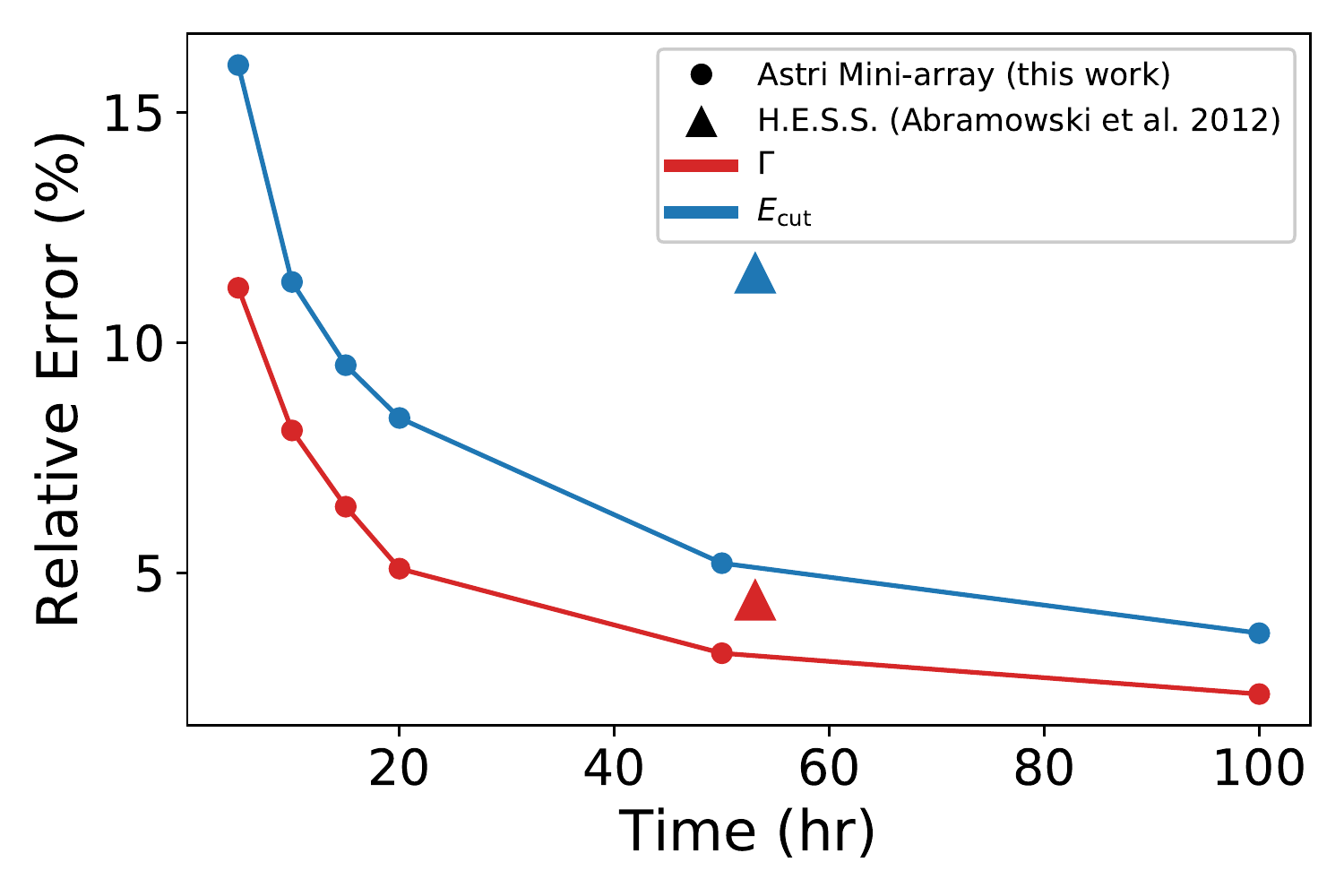}
\caption{
Relative error of the photon index $\Gamma$ (in red) and cut-off energy $E_{\rm cut}$ (in blue) as a function of the total observation time for the \astrima~simulations (dots). \hess~constraints are shown here for comparison as triangles.
}
\label{fig:scatter_t}
\end{center}
\end{figure}

Assuming that emission from the radio and X-ray spectral components are distinguished if both of them are detected with a significance TS$>$25, we found that, at energies $E>1$ TeV, the \astrima~would distinguish VHE emission from the circular radio and X-ray components, if the contribution from one of them is more than 8\% of the total flux. The ring-shaped components with a softer spectrum can be significantly detected if their contribution is more than $\sim$15\%. At higher energies ($E>10$ TeV) the components would be distinguishable with the \astrima~if their contribution is more than $\sim$\,16\% for the circular and $\sim$\,30\% for the ring-shaped components, respectively.
In addition, we evaluated how the statistical uncertainties for the spectral model parameters, as obtained for a 50 h and 100 h \astrima\ observations, compare with the results obtained in 50 hr by \hess\  \citep{abramowski12}. To this aim, we performed a maximum likelihood spectral analysis in the 2.5--90 TeV range, using 25 logarithmically spaced bins. In this analysis, we froze the spatial model adopting a uniform radial disk of radius of 0.8$^{\circ}$ centered at RA\,=\,128.75$^{\circ}$, DEC\,=\,-45.6$^{\circ}$. The resulting best-fitting spectral parameters are reported in Table~\ref{tab:vela_spectralresults}, whereas the spectral energy distribution of the Vela PWN obtained from a simulated 100 h exposure is shown in Fig.~\ref{fig:sed_2-100}. Upper limits are shown for those bins for which TS$<$9. The butterfly diagram is obtained with the task \textsc{ctbutterfly} which, for each energy bin, calculates an intensity envelope (the minimum and maximum values) of the best fitted spectral model compatible with the data. In Fig.~\ref{fig:scatter_t} we show how the relative errors on the photon index (in red) and on the cut-off energy $E_{\rm cut}$ (in blue) decrease as a function of the exposure time. The \hess~measurements are also shown for comparison (see triangles in Fig.~\ref{fig:scatter_t}). We note that only statistical uncertainties are taken into account here.

\begin{table}[!htbp]
\caption{Best-fitting spectral parameters of the Vela PWN for the simulated \astrima~observations.}
  \begin{threeparttable}[t]
  \centering
       \begin{tabular}{c|ll}
\hline
       Exposure (h) & 50  & 100        \\
       TS         & 9372& 18922      \\
        \hline
            &  \multicolumn{2}{c}{Spectral model parameters}\\
         $N_0$           & 9.6$\pm$0.4   & 9.5$\pm$0.3 \\ 
         $\Gamma$        & 1.25$\pm$0.04 & 1.24$\pm$0.03 \\ 
         $E_{\rm cut}$ (TeV) & 12.6$\pm$0.6  & 12.4$\pm$0.4  \\
         \hline
  \end{tabular}
     \begin{tablenotes}
     \item Normalization $N_0$ is in units of 10$^{-12}$ photons/cm$^{2}$/s/TeV. Reference energy is 1 TeV.
   \end{tablenotes}
    \end{threeparttable}%
  \label{tab:vela_spectralresults}%
\end{table}%
Finally, we noted that 100 h observations are not sufficient for a significant detection of the Vela pulsar at energies $E>1$ TeV even assuming no VHE cut-off in its spectrum.

%% file: sect_j1303.tex
\subsection{Energy-dependent morphology: the PWN HESS J1303-631} \label{sect:j1303}

\paragraph{Scientific Case}
We consider here the case of the PWN HESS J1303-631 whose energy dependent morphology constitutes an important test-case to probe the capabilities 
of the \astrima\  to disentangle softer or  harder emission zones from the same nebula.  Although this particular source cannot be observed from Teide, the aim is to show the feasibility of such studies for similar PWNe. We also show, comparatively, the constraints that can be obtained on the averaged spectral emission. The PWN HESS J1303-631 was one of the first so-called  \emph{dark sources} discovered by \hess\ \citep{hess05a}, namely a source which lacked an obvious counterpart at longer wavelengths. It was later realised that the extended morphology  of such TeV sources could be explained within the PWN scenario, 
but with a parent, runaway pulsar at the edge of the nebula, or, alternatively, with the expanding bubble of energetic relativistic particles, shaped by strong density gradients in the surrounding ISM, so that the emission becomes strongly asymmetrical. For HESS J1303-631, the parent pulsar \citep[PSR J1301-6305, ][]{manchester05} is located at 0.6$^\circ$ North of the nebula, it is relatively  young (11 kyr) with a spin period of 184 ms and a spin-down luminosity $\dot{E}$\,=\,1.7$\times 10^{36}$ erg s$^{-1}$.  Based on an estimated distance of 6.6 kpc, the overall TeV luminosity amounts to $\sim$\,4\% of the total  spin down luminosity. Observations with the Fermi/LAT instrument  constrained with some difficulty the MeV/GeV region of the spectrum, due to contamination of the nearby source SNR Kes 17 \citep{acero13}. It appears, however, that a single one-zone leptonic model based on observed fluxes in the radio, X-ray and TeV bands \citep{hess12} underpredicts the observed \fermi\ flux. Recently, \citet{sushch17}, using deeper observations in the radio bands at 5.5 GHz and 7.5 GHz with ATCA, did not find evidence  for an extended radio nebula around the pulsar, contrary to what could be expected based on the X-ray and HE/VHE maps.

\paragraph{Feasibility and Simulations}
For the \astrima\ simulations we adopted the morphology and the spectral shape based on the \citet{hess12} data
(parameter values and their uncertainties are listed in Table~\ref{table:results_j1303}) 
for a total exposure time of 300 hours.
In our simulation we also took into account the presence 
of other TeV sources in the field of view \citep[e.g. the 
source PSR B1259–63/LS 2883, at a distance of only 0.75$^{\circ}$ 
from HESS J1303-631,][]{hess20}.  
In \citet{hess12}, an analysis of the events selected by 
energy range revealed an energy-dependent shape of 
the emission, where events of higher energy, above 10 TeV, 
were found preferentially closer to the 
present pulsar's position, whereas events of lower 
energy showed an offset of $\sim$ 0.14$^{\circ}$. 
This result was obtained based on the distance distribution
of the events from the pulsar position. To test the capability of 
the \astrima\ to detect such energy-dependent morphology, 
we also made an ad-hoc simulation. In this case, we 
simulated the extended emission of HESS J1303-631 as 
the superposition of two point sources at a distance 
of 0.14$^{\circ}$ from one another. 
One point source (labelled \textit{PSRsource}) is placed at the 
coordinates of PSR J1301-6305 and has a power-law
spectrum with no cut-off;  the second point 
source (labelled \textit{Nebula}) is placed at the 
centre of the nebula, and has a spectral cut-off at $E_{\textrm cut}$\,=\,4 TeV. In both cases, we
assumed the power-law photon index to be -1.5, 
adjusting the normalization of each component so that each
point source contributes half of the 1--100 TeV flux
observed from \hess\ \citep{hess12}.

\begin{figure*}
    \centering
    \begin{tabular}{ccc}
    \includegraphics[width=0.6\columnwidth]{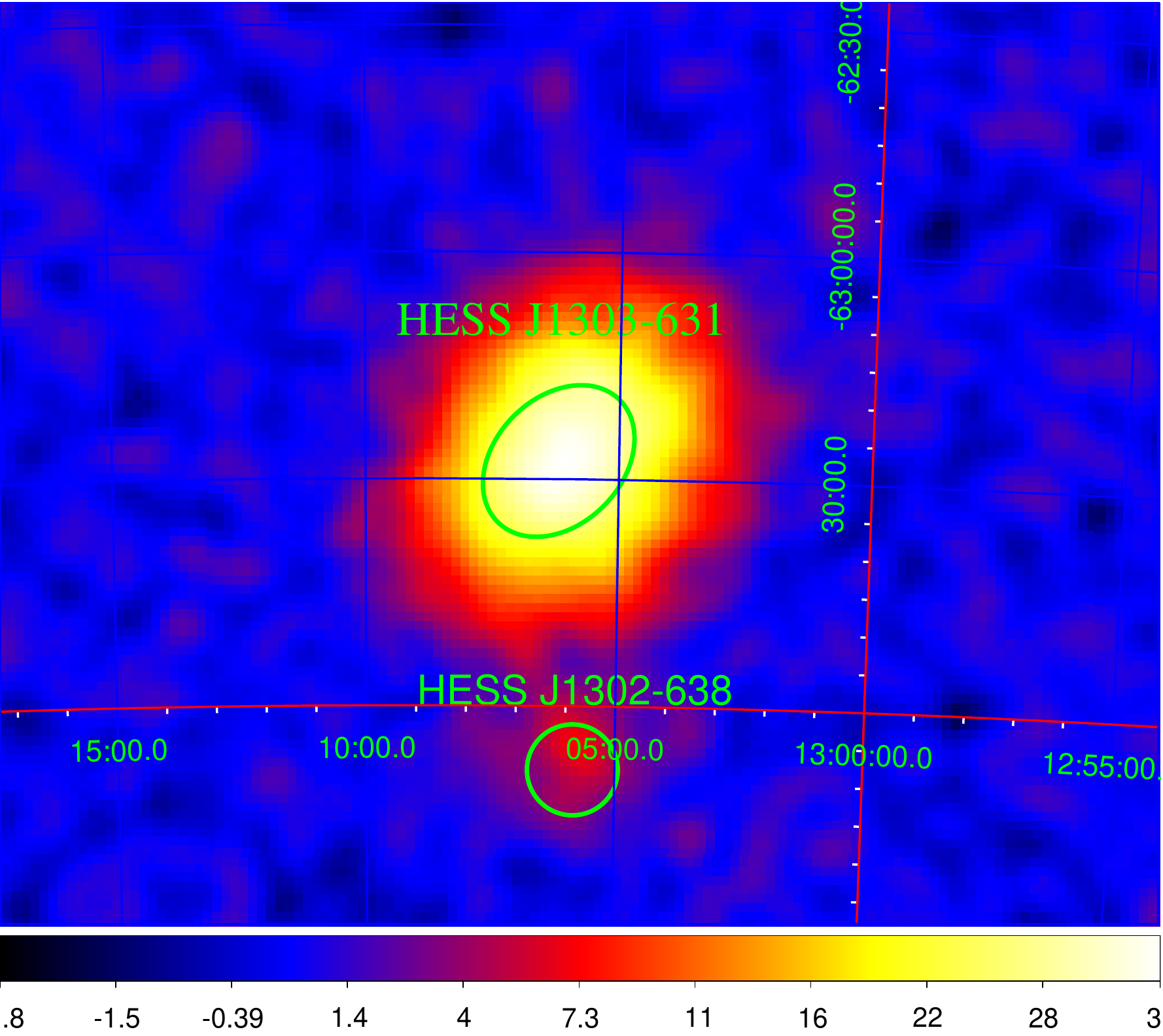} &
    \includegraphics[width=0.6\columnwidth]{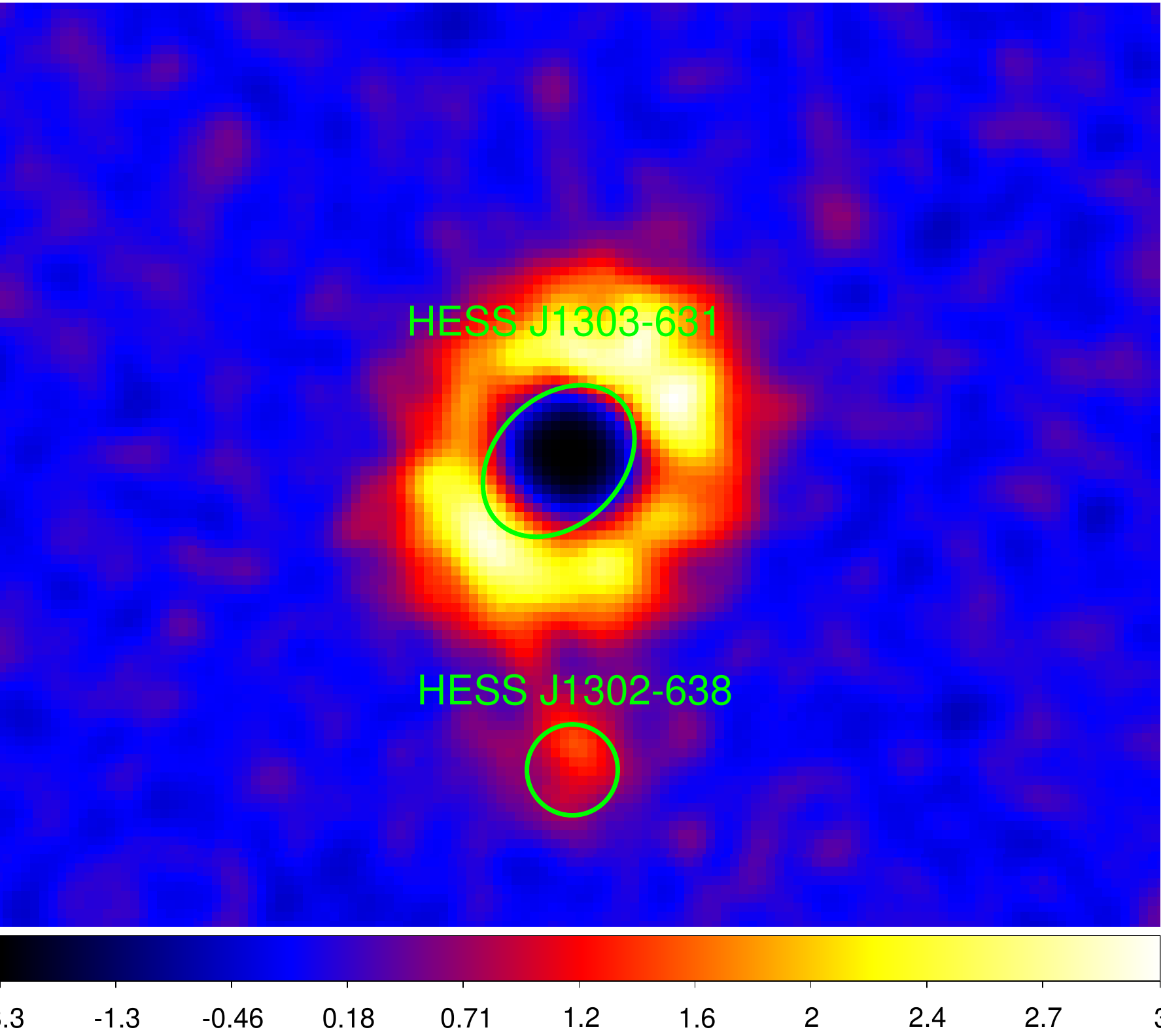} &
    \includegraphics[width=0.6\columnwidth]{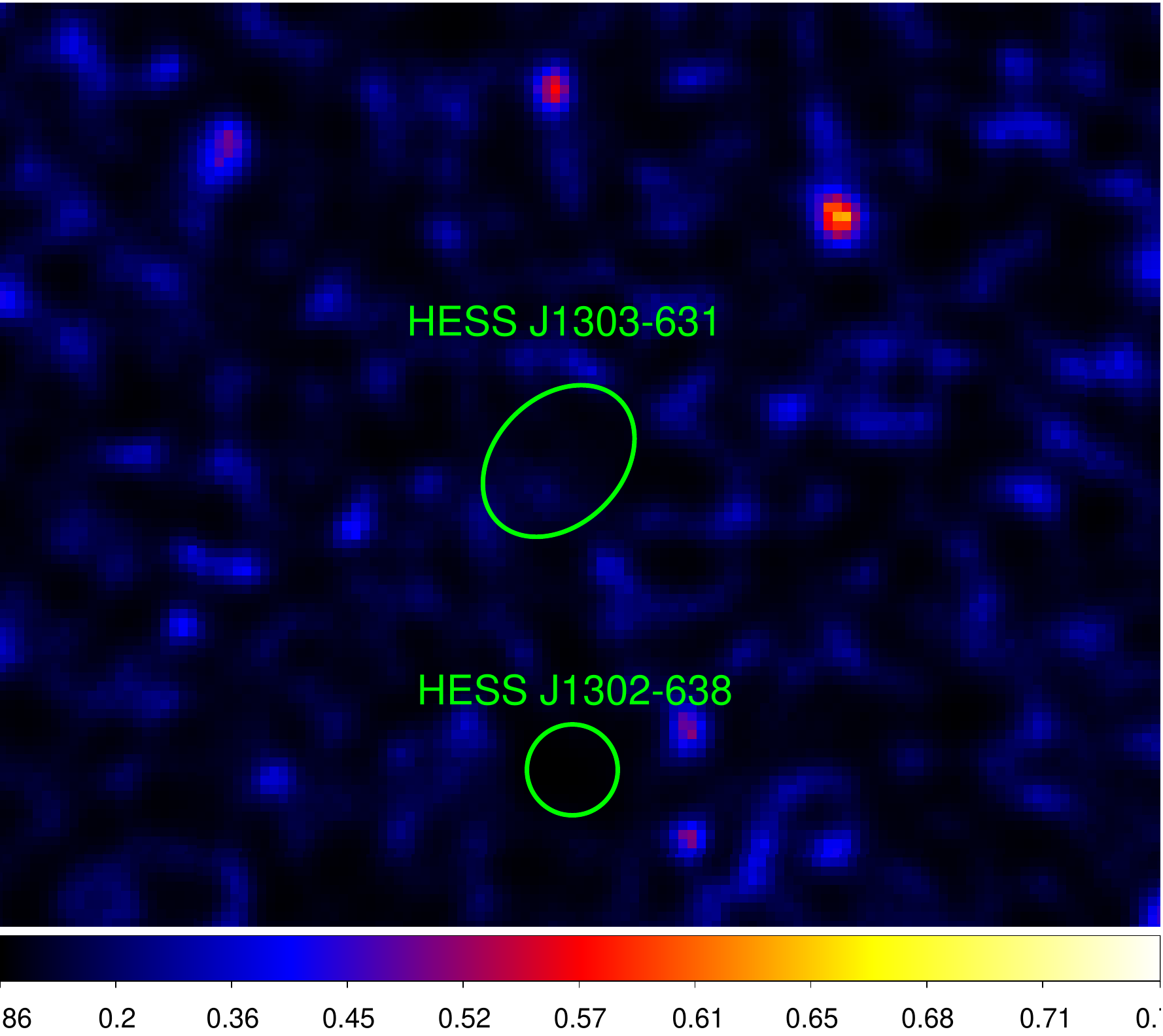} \\
    \end{tabular}
    \caption{ \textit{Left panel}: background-subtracted sky map of the region around
    HESS J1303-631 from a 300h \astrima\ simulation in the 1--100 TeV band. 
    The only other source in the field
    is the $\gamma$-ray binary HESS J1302-638, 
    that we modelled as a point source.
    \textit{Centre panel}: residual map of the HESS J1303-631 region when a point-source 
    spatial model is adopted. The large residuals in the map clearly indicate 
    that the extended emission can be resolved by the \astrima. \textit{Right panel}: 
    residual map around HESS J1303-631 adopting the template model (see results 
    in Table~\ref{table:results_j1303}). Sky maps units are counts/pixel. }
    \label{fig:resmaps_j1303}
\end{figure*}

\paragraph{Analysis and Results}
We performed a binned likelihood analysis using 
15 logarithmically spaced energy bins in the 2.5--90 TeV 
energy band and a pixel size of 0.015$^{\circ}$. 
The source has an extension which is slightly larger than 
the overall \astrima\ PSF 
(e.g. 68\% PSF at a reference value of 5 TeV is 0.15$^{\circ}$). 
If we assumed a point source for the spatial model
of the source, significant residuals would appear in the 
residual map, as it can be comparatively observed 
in Fig.~\ref{fig:resmaps_j1303}.\\ 
The statistical uncertainties of the spatial and 
spectral parameters are comparable with the values 
given by \cite{hess12}, and, as expected, they become better for our assumed exposure time of 300 hr (see Table~\ref{table:results_j1303}). 

\begin{figure}
    \centering
    \includegraphics[width=\columnwidth]{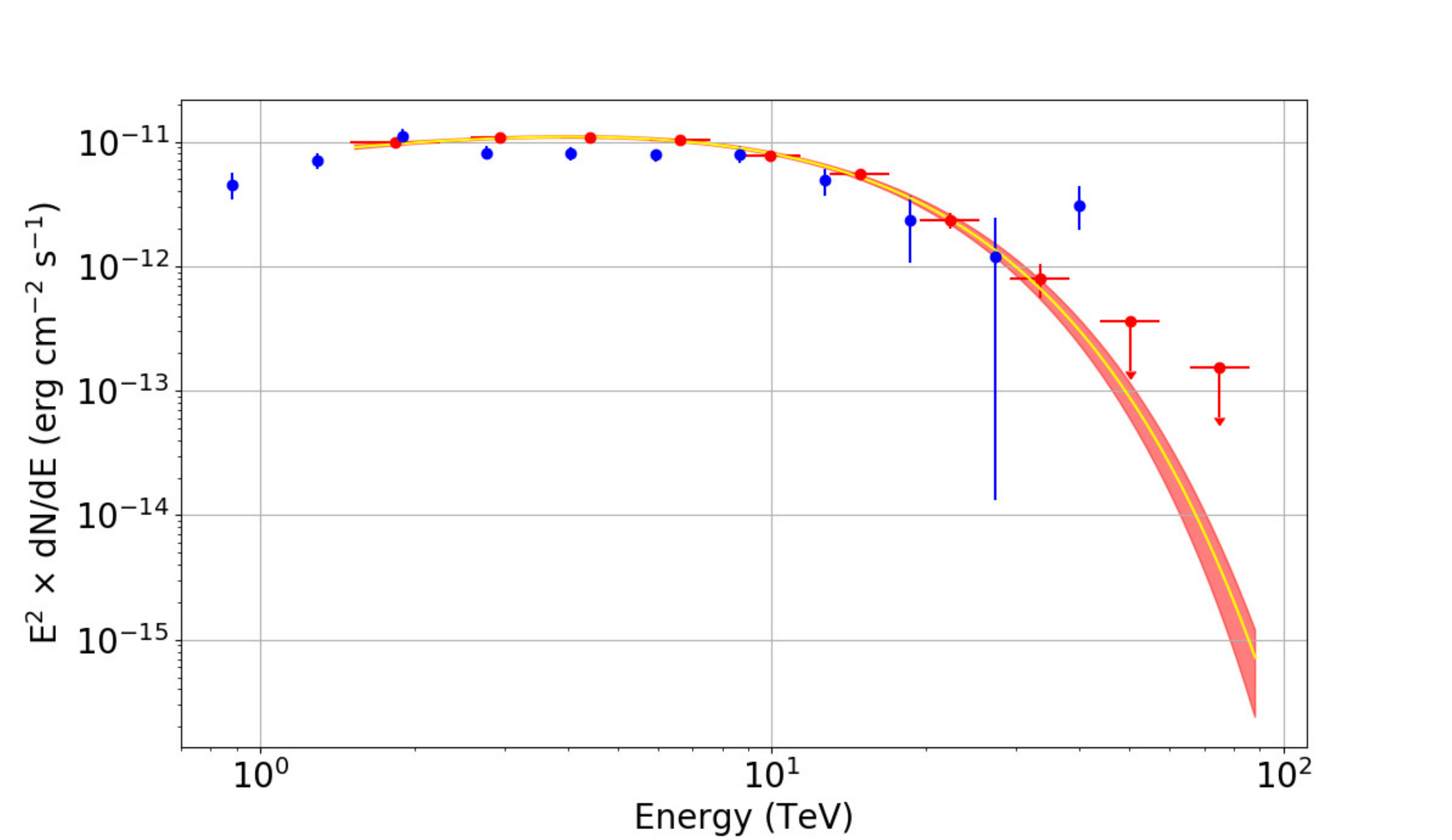} 
    \caption{HESS J1303-632 spectrum: \astrima\ data-points in red; \hess\ data from 
    \citet{hess12} in blue. The butterfly diagram shows the best-fit 
    and the 68\% confidence contour as derived from the minimum likelihood
     estimate of 110 hours \astrima\ simulations.}
    \label{fig:spectrum_j1303}
\end{figure}

In Fig.~\ref{fig:spectrum_j1303}, we show the \hess\ spectral 
data points and the \astrima\ simulated spectrum based 
on a sub-sample of data corresponding to 110 hours of observations. 
Using the same observing time, the quality of the data is comparable in the 2--10 TeV range, while the larger collecting 
area above 10 TeV for \astrima\ allows for a better characterisation 
of the spectral curvature beyond the cut-off energy.

\begin{table}[h]
    \centering
     \begin{threeparttable}[t]
    \caption{Likelihood results for HESS J1303-631. Comparison of \astrima\
    results with \hess\ values and uncertainties. The spectral model is an 
    exponentially cut-off power-law model. The spatial model is an elliptical 
    disk model, which assumes uniform intensity distribution within the ellipse.}
    \begin{tabular}{r|rrr|r|r} \hline \hline
                      & \astrima  & \hess\tnote{1}  \\
         Exposure (hr)      & 300           & 108.3  \\
 
         \vspace{0.2cm}
        
              & \multicolumn{2}{c}{Spectral Model} \\
         $N_o$\tnote{2}       & 4.6$\pm$0.1 &  5.6$\pm$0.5 \\ \vspace{0.2cm}
         $\Gamma$             & 1.52$\pm$0.04 &  1.5$\pm$0.2 \\
         $E_{\textrm{cut}}$ (TeV) & 8.5$\pm$0.5 &  7.7$\pm$2.2 \\ 
         &\multicolumn{2}{c}{Spatial Model} \\
         R.A. (deg)   & 195.703$\pm$0.005  & 195.7000$\pm$0.0008  \\ 
         DEC (deg)   &  -63.176$\pm$0.002&  -63.178$\pm$0.007  \\
         PA\tnote{3} (deg)    & 146$\pm$2 & 147$\pm$6  \\
         
         $R_{\rm max}$ (arcmin) &  0.198$\pm$0.003 & 0.194$\pm$0.008 \\
         \vspace{0.2cm}
         $R_{\rm min}$ (arcmin) & 0.149$\pm$0.002 & 0.145$\pm$0.006   \\
         
\hline
\end{tabular}
     \begin{tablenotes}
     \item[1] Value taken from \citet{hess12}.
     \item[2] $N_0$ calculated at the reference energy of 1 TeV in units  
     of 10$^{-12}$ photons cm$^{-2}$ s$^{-1}$ TeV$^{-1}$.
     \item[3] PA is the positional angle, measured counterclockwise from North.
   \end{tablenotes}
\label{table:results_j1303}
\end{threeparttable}
\end{table}

\begin{figure}
    \centering
    \includegraphics[width=\columnwidth]{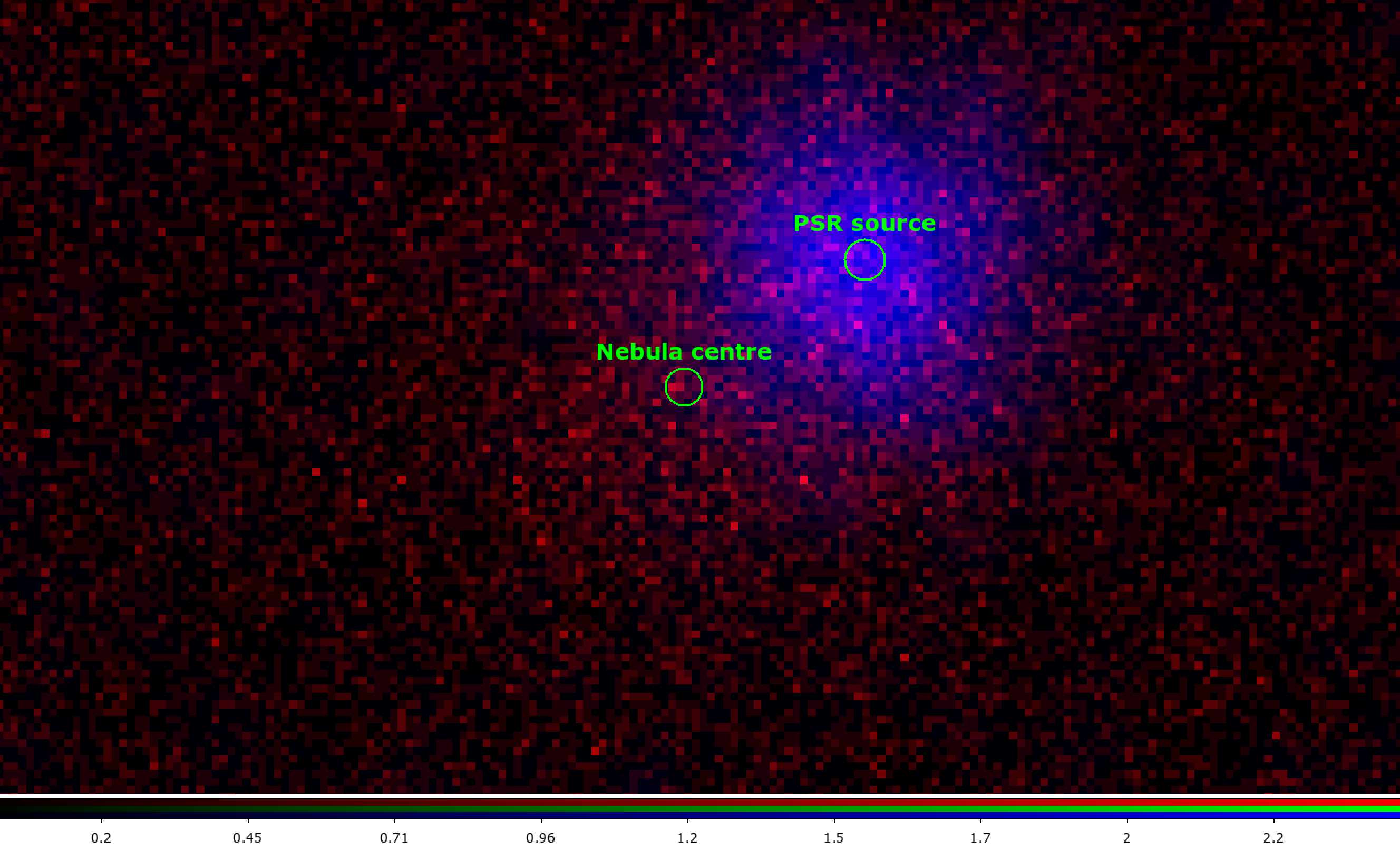} 
    \caption{Colour map for the energy-dependent (red and blue, respectively) are over-imposed. 
    Circles around the position of the \textit{PSR\,source} and for the extended \textit{Nebula} show the positional 
     uncertainties as derived from a maximum likelihood fit. Colour map units are counts/pixel.}
    \label{fig:j1303_rgbmap}
\end{figure}

To show the capabilities of the \astrima\ to  detect possible spatial-dependent spectral variations,  where the variation length is of the same order of the PSF for 
energies of few TeV, we adopted the very simple baseline model 
illustrated in the previous paragraph, as two point sources separated 
by $\sim$\,8.5$^{\prime}$ and with different spectral shapes; however,  we 
reproduced as closely as possible the expected rate
and flux from HESS J1303-631.
Adopting the template model of the two-source emission, we performed 
an unbinned likelihood analysis. For the  source \textit{Nebula}
we obtained the following best-fit values for its position: RA\,=\,195.699$\pm$0.012, Dec.\,=\,-63.17$\pm$0.05, 
while for \textit{PSR source} RA\,=\,195.44 $\pm$ 0.03, Dec.\,=\,-63.088 $\pm$ 0.013. 
The \textit{Nebula} and the \textit{PSR source} positions have an average uncertainty 
less than 3$^{\prime}$. Considering the assumed 
radial distance, both sources can be therefore clearly resolved. 
The composite image using two IRF-background subtracted sky maps in 
two energy ranges (a soft one, below  10 TeV and a hard one above this 
energy range) in Fig.~\ref{fig:j1303_rgbmap} shows the two simulated
sources together with their error regions (circle radii correspond 
to the largest value between the error in R.A. and in
declination).

